# Iron-Based Superconductors: current status of materials and pairing mechanism

Hideo Hosono [a,\*] and Kazuhiko Kuroki [b]

[a] Materials and Structures Laboratory & Materials Research Center for Element Strategy, Tokyo Institute of Technology, 4259 Nagatsuta, Midori-ku, Yokohama 226-8503, Japan

[b] Department of Physics, Graduate School of Science, Osaka University, Toyonaka, Osaka 560-0043, Japan

## Abstract

Since the discovery of high Tc iron-based superconductors in early 2008, more than 15,000 papers have been published as a result of intensive research. This paper describes the current status of iron-based superconductors (IBSC) covering most up-to-date research progress along with the some background research, focusing on materials (bulk and thin film) and pairing mechanism.

\*Corresponding author: TEL +81-45-924-5009, FAX +81-45-924-5134, E-mail hosono@msl.titech.ac.jp

# 1. Introduction

The dynamic formation of electron pair is prerequisite for emergence of superconductivity, while (anti)ferromagnetism emerges by long range static spin ordering. This is the reason why it is widely believed both compete with each other. Iron is a typical magnetic element with a large magnetic spin moment, and had been believed to the most harmful for emergence of superconductivity. However, the situation was totally changed since the discovery of an iron oxypnictide superconductors $LaFeAsO1_{-x}F_x$ with Tc=26K[1] in early 2008. This discovery sparked intense research activity on superconductivity in this system. As a consequence, more than 15,000 papers have been published to date along with several comprehensive review articles[2] and monographs[3].

What is the impact of iron-based superconductors? There will be two answers, i.e., the first is the breaking of a widely accepted belief that "iron is antagonistic against superconductivity", which led to the opening of a versatile frontier in superconducting materials. It has become clear through intense research in the last 7.5 years that iron can be a good friend for high Tc-superconductors under certain conditions. The second is a rich variety in candidate materials and in pairing interaction. It has turned out that there are many material varieties in iron-based superconductors such as 7 parent materials,1111, 122, 111, 112, 245, 11 and thick-blocking layer bearing materials (where the number denotes the atomic ratio in constituting the compound, see Fig.2 for crystal structure of each compound). Each type has rather different electrical and magnetic properties including anti-ferromagnetic semimetal, Pauli para metal and antiferromagnetic Mott insulator.

Iron-based superconductors (IBSCs) have several unique properties such as robustness to impurity, high upper critical field and excellent grain boundary nature. These properties are advantageous for wire application. Recent progress in the performance of superconducting wires of IBSC is wide eyed, i.e., the maximal critical current has reached the level of commercial metal-based superconducting wires.

In this article we review the current status of IBSC focusing on materials and pairing mechanism along with a brief research background in order to give a comprehensive view of this rapidly growing superconductor to relevant researchers.

## 2.Materials : bulk

### 2.1. 3d transition metal oxypnictides

IBSC was first discovered in LaFePO [4] with Tc =~4K in 2006. Subsequently superconductivity was found in LaNiAsO [5] with Tc=2.4K in 2007 and then Tc jumped to 26K in early 2008 for LaFeAsO$_{1-x}$F$_x$ [1]. The electromagnetic properties of 3d transition metal (TM) oxypnictides vary drastically with TM [6]. Figure 1 summarizes the properties of LaTMPnO, where TM= 3d transition metal, Pn=P or As). One may see that electric and magnetic properties of layered TM oxynictides strongly depend on TM. The synthesis of early TMs and Cu oxypnictides was tried but unsuccessful even using high pressure up to ~9GPa and no distinct correlation was found between the stability of the 1111-type compound and the kind of TM. Bulk superconductivity appears in TM=Fe$^{2+}$and Ni$^{2+}$, both of which have even number of 3d electrons but no superconductivity has found in TM=Cr$^{2+}$ [7] with 3d$^4$ electronic configuration. Undoped LaFeAsO is an antiferromagnetic metal but does not exhibit superconductivity. For TM=Mn, an exceptionally high electron doping is possible by applying H$^-$[8] in place of F$^-$ as a substituent for the oxygen site and transition of antiferromagnetic insulator to ferromagnetic metal was observed but no Tc appeared. LaCoAsO is an itinerant ferromagnetic metal [9]. No Tc exceeding 10K has been reported in the 1111 system for except the iron oxyarsenides up to date.

### 2.2. Parent materials

Since the paper reporting Tc=26K in LaFeAsO$_{1-x}$F$_x$, several ten superconducting materials have been reported in layered iron pnictides or chalcogenides. These materials contain a common building block of square lattice of Fe$^{2+}$ ions which take tetrahedral coordination with Pn (where P and/or As) or chalcogenide ions. Figure 2 (A) summarizes crystal structure of 10 parent compounds and a simplified classification of these parent compounds is shown in fig.2 (B). Since the Fermi level of each parent compound is primarily governed by Fe five 3d-orbitals, iron plays the central role of superconductivity. These compounds have tetragonal symmetry in the superconducting phase, are Pauli para metals in the normal state and undergo crystallographic/magnetic transition to orthorhombic or monoclinic anti-ferromagnetism at low temperatures. Exception is a 111-type compound with Pauli paramagnetism even at lower temperature and 245 compounds having antiferromagnetic insulating properties. Table 1 summarizes the various properties of the parent materials. Superconductivity emerges when anti-ferromagnetism disappears or diminishes by carrier doping or structural modification by applying external pressure or by chemical pressure induced by isovalent substitution. In any case, the parent materials are metal having itinerant carriers and how to remove the obstacles for emergence of superconductivity is an experimental approach.

## 2.3. Carrier doping to induce Tc

Most of the parent materials listed in section 2.2 is antiferromagnetic metal and superconductivity is induced by appropriate carrier doping or structural modification. Although some of the parent phases exhibit superconductivity without doping, the Tc value of such a material was low as exemplified by LaFePO [4] with Tc=~4K, implying the occurrence of close relationship between magnetic ordering in the parent phase and resulting Tc. Metal iron with bcc structure is a ferro-magnet with a Curie temperature of 1043K, but high pressure phase with hexagonal structure loses ferromagnetism and exhibits superconductivity of Tc= ~0.4K under 15-30GPa[10]. We may understand in a sense that high Tc IBSCs are obtained by eliminating long range spin ordering in layered iron pnictides using carrier doping in place of applying high pressure to metal iron. Three types of carrier doping described below are possible for the parent compounds of IBSCs.

## 2.3.1 Aliovalent doping

The first high Tc IBSC was discovered by a partial replacement of $F^-$ ion at the oxygen site in La-1111 compounds. The1111 type compounds have a 2-dimensional electronic structure and a metallic conducting FeAs layer are sandwiched by insulating and LaO layers. When the $O^{2-}$ site is replaced by an $F^-$ ion, an electron generated is transferred to the FeAs layer due to the energy offset. Figure 3 shows the schematic phase diagram of the 1111 and 122 system. For the 1111 system, the Tc appears when the anti-ferromagnetism (AFM) disappears. On the other hand, the AFM and superconductivity coexist in the 122 system and the optimal Tc appears to be obtained at a doping level where the Neel temperature ($T_N$) reached 0K, suggesting the close relationship between the optimal Tc and quantum criticality. Electron doping into RE-1111 compounds (where RE= rare earth metal) by this substitution was very successfully, i.e., the max.Tc was increased from 26K to 55K by replacing La with other RE ion with smaller ionic radius [11].

However, complete experimental data on the shape and width of the Tc-dome in the 1111 system with the highest Tc were not obtained until 2011[1(a)], i.e., electron-doping level was insufficient to observe the over-doped region. The primary origin is poor solubility of F ions at the oxygen site (10-15%) due to the preferential formation of stable REOF crystal. This restriction was removed by use of hydride ion H⁻ in place of F⁻[12]. Hydrogen is the simplest bipolar element, taking +1 and -1 charge state depending on its local environment [13]. The ionic radius of H⁻(~110pm) is not so different from that of F⁻ (133pm) or $O^{2-}$(140pm). The H-substituted RE-1111 compounds, $REFeAsO_{1-x}H_x$, were successfully synthesized with an aid of high pressure. This synthesis is based on an idea that hydride substituted state is more stable than oxygen vacancy in the charge blocking layer REO with fluorite structure (an oxygen ion occupies tetrahedral site) [14]. Following this idea, the mixture of starting materials with $REFeAsO_{1-x}H_x$ was heated with a solid hydrogen source which release $H_2$ gas at high temperature under 2GPa. This procedure gives the targeted materials and the location and charge state of hydrogen at the oxygen sites are confirmed by powder neutron diffraction [8] and DFT calculations [12], respectively. Figure 4 shows the electronic phase diagrams of $REFeAsO_{1-x}H_x$ with different RE (La, Ce, Sm and Gd)[15]. Three new findings are evident from the figure. First is that La-1111 has two-dome structure in which the first dome is the same as that reported previously for $LaFeAsO_{1-x}F_x$? The second dome newly found by H-doping has a higher optimal Tc (36K) and a larger width. The temperature dependence of conductivity at the normal state (150K>T>Tc) just above Tc follows $T^2$ (Fermi liquid like) for the first dome but $T^1$ (non- Fermi liquid like). The double dome structure is not unique for the La-1111 system and is seen for the chemical compositions with ~30K >Tc of $SmFeAs_{1-x}P_xO_{1-x}Hx$ [16].

Second is that although Tc has a single dome for other RE systems, its range is much wider than that reported in the F-substituted case for the other lanthanide ion systems. Third is that the optimal doping level is decreased with deceasing the size of RE ion.

Hole-doping to the 122 system is possible by substitution of an alkaline earth ion site with an appropriate alkali ion such as potassium substitution of Ba site [17]. On the other hand, hole doping effect into RE1111 system is not so clear to date. The solubility of Ca ion to RE site is so restricted to several% that the Tc is not observed [1]. Although substitution by Sr ion can be doped to more than 10%, the shielding volume fraction is small (<10%) [18].

### 2.3.2. Electron-doping by introduction of oxygen vacancy

Another electron doping way reported was by introduction of oxygen vacancy to REO layers in $ReFeAsO_{1-x}$ samples [19]. These samples are synthesized by heating the batch of oxygen deficient compositions under high pressure. If an oxygen vacancy substitutes the oxygen ion site, 2 carrier electrons should be generated. However, the optimal Tc reported is the almost the same as the value obtained for $REFeAsO_{1-x}F_x$.

### 2.3.3. Isovalent doping

A unique characteristic of doping into IBSCs is isovalent doping. Two typical examples are introduced. One is partial substitution of $Fe^{2+}$ site by $Co^{2+}$ and another is replacement of As site by P. The former is understood in term of electron doping because the $Co^{2+}$ ($3d^7$) has an excess electron compared with $Fe^{2+}$ ($3d^6$) [20]. Figure 5 shows the correlation between Tc and excess electron count on Fe site in Ba $(Fe_{1-x}TMx_2)$ $As_2$ [21]. It is evident that Tc is scaled by excess electron number on the Fe sites. This finding makes a sharp contrast to the results of impurity effects in high Tc cuprates for which Tc is easily degraded by partial replacement of $Cu^{2+}$ site. Wadati et al.[22] pointed out by DFT calculations that doped electron is not distributed uniformly but is concentrated within the Muffin-Tin sphere at the substitute sites compensating for the increased nuclear charge. Nakamura et al. [23] reported that these TM ions do not work as a strong scattering center, forming an alloy with Fe. Robustness of Tc to impurity is closely related to the pairing mechanism to be discussed in the pairing mechanism. This type of substitution is often called *direct doping* because the TM replaces the iron sites in which superconductivity emerges. It is natural to consider that the Tc induced by the direct doing is rather lower than that by indirect doping.

However, the experimental difference in Tc between them is not clear for the 122 system [24] but distinct (~20K) for the 1111 system [25]. This marked difference may be understood by that in dimensionality of electronic structure between these two systems, i.e., the 2D-nature of electronic structure is much weaker in the 122 system [26] than that in the 1111 system and in the former system effects of both doping modes would give almost the same effect on the FeAs layers.

Another effective isovalent substitution is seen in the 122 system [27] such as $BaFe_2(As_{1-x}P_x)_2$ [28]. As the $T_N$ of the parent phase is reduced by x, the Tc appears and reaches the maximum of ~30K around x=0.35 which looks to correspond to the quantum critical point. The shape of this phase diagram is similar to that obtained by electron doping using Co-substitution. Emergence of superconductivity by the similar isovalent substitution of anions directly bonding with iron is observed for $FeSe_xTe_{1-x}$ [29]. Since isovalent anion substitution does not generate carriers unlike Co-substitution, it is understood that the anion substitution modifies the local geometry around irons, which in turn leads to weakening of AFM order competing with emergence of superconductivity. Since the parent materials of IBSC are metals containing carriers enough to induce superconductivity, the primary effect of isovalent anion substitution is to weaken the AFM.

2.3.4. Doping by intercalation

The parent materials of IBSC have layered structure. Insertion of ions and/or molecules is possible without keeping the original FePn(Ch) layers in some parent materials. Metal-superconductor conversion has been reported to date for 11 and 122 compounds. The FeSe intercalates obtained from low-temperature alkali metal and $NH_3$ co-intercalation exhibit higher Tc of 30-46K compared with the samples obtained by conventional high temperature methods [30]. A unique feature of this process is that a small sized-alkali cation such as Li and Na combined with the $NH_2^-$ anion or $NH_3$ molecules can be intercalated into the FeSe layers[31] because the formation of ion intercalates is restricted to large-sized monovalent cations such as Cs and Tl [32] by conventional high temperature methods.

When $SrFe_2As_2$ thin films are placed in an ambient atmosphere, this film converts into superconductor accompanying shrinkage of the c-axis [33]. Based on an observation that this conversion does not occur in a dry atmosphere, it was suggested that the intercalation of $H_2O$-relevant species into a vacant site in the Sr-layers [34]. Such a conversion is not observed for $BaFe_2As_2$ [35] with a vacancy with smaller space than that in $SrFe_2As_2$. This finding led to the shift of thin film research from $SrFe_2As_2$ to $BaFe_2As_2$ which is less sensitive to ambient atmosphere [35]. A similar conversion was reported by immersing the parent compounds into polar organic solvents including wines [36]. Interestingly, it is reported that strain can induced the similar effect in the bulk single crystal [37].

.

2.4. Correlation between Tc and local structure

Tables 2-7 enlist data of Tc in various type IBSCs along with the doping modes, local structure around iron and pressure dependence. It is a general trend that the optimal Tc is higher in the order 1111>122>11. This result implies that the optimal Tc is enhanced by the interlayer spacing of FeAs layers. However, this view does not valid as shown in Fig.6. Instead, it is now a consensus that the Tc of IBSCs is sensitive to the local geometry of $FePn(Ch)_4$ tetrahedron. Lee et al. [38] first reported that the *optimal* Tc is obtained when the bond angle ($\alpha$) of Pn(Ch)-Fe-Pn(Ch) approaches that (109° 5') of a regular tetrahedron. Figure 7 plots the most of data including *non-optimal* Tc in various types of IBSCs. The phenomenological correlation between the Tc and $\alpha$ becomes worse compared with that between the *optimal* Tc and $\alpha$, but the tendency still remains. However, data on the 11 system and the first dome in $LaFeAsO_{1-x}H_x$ are far from this empirical rule. This discrepancy is due to that the Tc is not determined only by the local structure of $FePn(Ch)_4$. Kuroki et al. [39] proposed a model that the pnictogen (chalcogen) height ($h$) from the iron plane is a good structural parameter associated with strength of spin fluctuation and the Tc is enhanced by increasing $h$. The correlation between the $h$ value and Tc is comparable to that between Tc and $\alpha$. The factors controlling Tc will be discussed later.

2.5. Electronic phase diagrams

There is a clear difference between 1111 and 122 systems as shown in Fig.3. A striking difference is whether the AF phase deriving from the non-doped parent compound coexists with superconducting phase or not. It is consensuses that both do not distinctly coexist in the 1111 system, while the two phases do in the 122 system. It is of interest to note that there is a distinct separation between magnetic (PM-AF) and structural (tetra-ortho) transitions in the 1111 system. There are two major discoveries associated with the phase diagram in last two years. One is the discovery of electronic nematic ordering phase which appears in $BaFe_2(As_{1-x}P_x)_2$ at a temperatures higher than the structural/magnetic transition [40]. Magnetic torque measurements revealed that the electronic nematic phase has 2-fold symmetry notwithstanding that crystal lattice still keeps 4-fold symmetry (tetragonal). The presence of such an electronic nematic phase with $C_2$ symmetry in the crystalline phase with $C_4$ symmetry is suggested in various superconductors and was clearly demonstrated in this system. Although a similar electronic nematic phase has been reported recently in FeSe [40], it is still unknown whether the existence of electro nematic phase is universal for each IBSC and the relation with superconductivity.

Second is the discovery of bipartite phases in La-1111 system [41]. It was described in 2.1 that the Tc-dome in RE-1111 systems is rather extended than that one thought before the over-doped region is elucidated by heavy electron-doping using hydrogen anion as the dopant instead of F. As a result, the two dome structure was elucidated in $LaFeAsO_{1-x}H_x$ and each Tc dome has a different parent compound, i.e., the parent compound of the first dome (0.05<x<0.18) is the non-doped LaFeAsO with $T_N$ of ~140K, which was already known, and that of the second dome (0.2<x<0.45) is $LaFeAsO_{0.5}H_{0.5}$ with $T_N$ of ~90K. Figure 8 shows the phase diagram elucidated along with the comparison between these two parent phases. The latter parent phase has a magnetic moment as large as twice and lower space symmetry (1.2 $\mu_B$/Fe,$Aem2$, non-Centro symmetric) than the former (0.6 $\mu B$/Fe, $Cmme$, Centro symmetric). The temperature of structural and magnetic transitions is distinctly separated (by ~20K) for the former but this difference becomes small (by ~5K). A similar difference between these two parent phases is seen in the relation of magnetic phase and superconducting phases. The inelastic neutron scattering [41] revealed that the resonance energy and scattering vector are distinctly different between the first dome and the second dome. The former vector (1.1A$^{-1}$, 15meV) is the same as that in the 122 compounds, while the latter (1.27A$^{-1}$, 17 meV) is unique for the second dome of La-1111 system. The observed scattering vectors may be understood in term of nesting, the vector for the first dome is attributed to nesting between the hole pockets at $\Gamma$-point (0,0) to the electron pockets at M-point ($\pi$,0) of the Fermi surface of Fe $d_{xz,yz}$ orbitals, while that for the second dome does to nesting between electron pockets (0,-$\pi$) and a hole pocket ($\pi$,$\pi$) on the Fermi surface of Fe $d_{xy}$ orbital. The primary point is the dominant orbital participating to nesting is switched between these two Tc domes.

A series of the experimental results strongly suggest that there are two factors controlling the superconductivity of the first and two domes and higher Tc is obtained when the two domes merge to form a single dome. The unification of the two dome structure and a marked Tc increase up to 52K have been found for LaFeAsO$_{1-x}$H$_x$ upon applying high pressure as shown in Fig.9 [42]. This finding is a good support for the above view and provides a guide to approach higher Tc.

## 3. Materials : thin film

Research on thin film growth of IBSC and device fabrication using the thin films such as Josephson junction has much advanced in last 8 years [43, 44,45,46]. Figures 10 (a), (b), and (c) summarize the important progress chronologically for three representative systems. The fabrication of epitaxial thin films was first reported [47] by pulsed laser deposition for the 122 system at the early stage. Since then, this material system has been most extensively studied among IBSCs. The reason is two, first is easy fabrication compared with the 1111 system and second is a small anisotropy and relatively high Tc [48]. These features are favorable for wire application.

### 3.1. 122 system

The extensive works led to the realization of high and isotropic Jc was reported utilizing the super lattice structure in Ba122/Ba122:Co on SrTiO$_3$ substrates [49] or introduction of a metal iron buffer [50], and the highest Jc of 1MA/cm$^2$ at 10T and 4.2K in BaFe$_2$(As$_{1-x}$P$_x$)$_2$ thin films [51]. The critical titling angle at the twin grain boundary to keep high Jc is of primarily importance to examine the grain boundary nature toward the wire application [52,53]. This critical angle ($\theta_\chi$) was determined employing epitaxial thin films grown on twinned (LaAlO$_3$)$_{0.3}$-(SrAl$_{0.5}$Ta$_{0.5}$O$_3$)$_{0.7}$ (LSAT) substrates with varied tilting angles. Figure 11 shows the results in comparison with those of high Hc cuprates (YBCO) [54]. The $\theta$c determined is 4-5 degree, which is twice as large as that of the YBCO. In addition, the current does not suddenly drop for the IBSC because the grain boundary of IBSCs has metallic nature. This finding encouraged the experts of bulk superconducting wires. Powder-in-tube (PIT) method has been almost exclusively applied to bulk wire fabrications. Three research groups in US [55], Japan [56] and China [57] have been competing to reach the practical Jc level of 0.1MA/cm² at 4K, which is comparable to that of commercial metal superconducting wires of Nb-based intermetallic compounds and each group has realized this Jc in early 2014. Figure 12 summarizes the magnetic field dependence of Jc in various superconducting wires. It is obvious that the Jc of superconducting wires of the 122-type IBSC is pretty robust to magnetic field and exceeds the metallic superconductors at higher fields. Although Tc of IBSC (122-type) is rather lower than that of cuprates, easy formation reflecting tetragonal symmetry of the materials (note high Tc cuprates have orthorhombic symmetry in superconducting state) and robust grain boundary nature makes PIT-wires of IBSC favorable in cost for application.

3.2. 11 system

Research on the thin films of the 11 system started after the reports on the 122 thin films [58]. Tc of the bulk of the 11 compounds has markedly increased by modification of the isovalent substitution, high pressure (see Table 7) or intercalation. In the thin films, effects of strain on the Tc were first reported on this system utilizing lattice mismatch between the substrate and the superconductor [59]. The value of max Jc was improved to 0.4MA/cm² at 9T and 4K in 2013 as well [60]. In 2012, a striking finding was reported on FeSe single unit cell epitaxial thin films grown on $SrTiO_3$ substrates [61]. In situ STM measurements revealed the opening of 20meV gap below ~45K and the onset Tc of ~50K (zero resistivity was attained at 32K). Subsequently, the gap closing temperature was determined to be 65±5K [62]. In July 2014 a paper reporting zero resistivity temperature of ~100K was posted on Arxiv and published in November, 2014 [63]. Since no data on Meissner effects were reported, it still remains incomplete that the sample has a Tc higher than 77K. Simultaneously Shen et al. [64] suggested the oxygen optical phonon in STO couples to the FeSe electrons through the high resolution angle resolved photoemission measurements. Although further effort is obviously needed to clarify whether this single layer epitaxial thin film has a Tc higher than 77K or not, we may expect to obtain a clue to reaching higher Tc from this system.

### 3.3. 1111 system

Last is the thin film of the 1111 system. The thin film growth of the materials in this system is extremely difficult compared with 122 and 11 systems by pulsed laser deposition or sputtering method. The primary reason is that since the precipitation temperature is so high that decomposition and/or preferential evaporation of a component including F as a dopant tends to occur. Successful deposition method was molecular beam epitaxy (MBE) which enabled to fabricate the epitaxial thin films with Tc comparable to that of the bulk [65, 66].

### 3.4. Field effect

It is a novel approach foe exploration of new superconductors by applying intense electric field to insulators utilizing an electric double layer transistor structure [67]. The 234 phase such as $K_{0.8}Fe_{1.6}Se_2$ (see Fig.2b) is the only one insulating parent compound with high $T_N$ and a distinct band gap [68]. Katase et al. reported a distinct field effect on the conductivity of epitaxial thin films of $TlFe_{1.6}Se_2$ employing this method but superconductivity was not attained yet[69]. Very recently, Wang et al .[70] reported that the Fe-vacancy ordered $K_2Fe_4Se_5$ is the magnetic, Mott insulating parent compound of the superconducting state. According to this conclusion, it is essential to realize superconductivity in this material that antiferromagnetism due to the Fe-vacancy ordering is eliminated by field effect.

### 4. Band structure and modeling

### 4.1 Basic band structure

We now turn to the theoretical aspects of the study on IBSC [71, 72, 73, and 74]. In the early stage of the theoretical study of IBSC, it was shown that phonon-mediated pairing mechanism cannot account for the experimentally observed high $T_c$ [75]. Hence, the superconductivity most likely occurs due to some kind of electron correlation effects. In order to theoretically investigate the electron correlation effects, it is necessary to obtain a many-body model Hamiltonian that correctly describes the low energy physics. The kinetic energy part of the effective model can be described by a tightbinding model whose hopping integrals reproduce the first principles band structure near the Fermi level. This downfolding procedure can be performed by constructing Wannier orbitals from the band calculations, which can be accomplished by adopting formalisms such as the "Maximally localized Wannier orbital" [76]. In ref [77], a five orbital model of LaFeAsO was constructed by this downfolding procedure from a first principles band structure obtained using the experimentally determined lattice structure. Due to the tetrahedral coordination of As, there are two Fe atoms per unit cell, so ten maximally localized Wannier orbitals are obtained. The two Wannier orbitals with the same symmetry in each unit cell are equivalent in that each Fe atom has the same local arrangement of the surrounding atoms. Therefore, a unit cell that contains only one orbital (for each orbital symmetry) can be taken by unfolding the Brillouin zone, ending up with an effective five orbital model as shown in Fig.13. Here, x and y axes of the reduced unit cell are in the direction of the nearest neighbor iron-sites. The Fermi surface consists two hole pockets around $(k_x, k_y) = (0, 0)$ , two electron pockets around $(\pi, 0)$ or $(0, \pi)$ , and one hole pocket around $(\pi,\pi)$. The hole Fermi surfaces around $(0,0)$ and portions of electron pockets near the Brillouin zone edge have mainly $d_{xz}$ and $d_{yz}$ orbital character, while the hole Fermi surface around $(\pi,\pi)$ and the portions of the electron Fermi surfaces away from the Brillouin zone edge have mainly $d_{xy}$ orbital character. A five orbital model was also constructed in ref.[78] using the lattice parameters determined by theoretical optimization.

There are also models that involve fewer bands/orbitals. Two band model in refs.[79], three band models in refs.[80], and four band models in refs.[81,82,83] are among those models. These models have similar shape of the Fermi surface as those in the five orbital model, but in some cases have different orbital characters on some of the Fermi surfaces. There are also some models that explicitly considers the 4p orbitals of the arsenic atoms. [84, 85, 86, 87, 88]

4.2 Local lattice structure dependence

At this point, it is instructive to show how the band structure is affected by the change in the local lattice structure since the correlation between the Pn-Fe-Pn bond angle[38] or the pnictogen height[90] and $T_c$ has been experimentally suggested. Let us take LaFeAsO as a reference, and consider first the As→P partial substitution. As the phosphorous content increases, it is known that the average Pn-Fe-Pn bond angle increases (the pnictogen height $h_{Pn}$ decreases), and this results in a loss of the $d_{xy}$ Fermi surface around $(\pi,\pi)$ in the unfolded Brillouin zone [91]. On the other hand, replacing

La by smaller atoms such as Nd or Sm in LaFeAsO reduces the average Pn-Fe-Pn bond angle (the pnictogen height $h_{Pn}$ increases), and will make the $d_{xy}$ hole Fermi surface larger. Simultaneously, the $d_{xy}$ portion of the band around $(0,0)_{unfold}$ above the Fermi level comes down, and eventually sinks below the $d_{xz/yz}$ bands for sufficiently small bond angle [92, 93, 94, 95, 96]. This is accompanied by a band reconstruction among $d_{xy}$,$d_{xz/yz}$ bands, as shown in Fig.14; namely, in the original band structure, the $d_{xz/yz}$ bands degenerate at $(0,0)$ have concave curvatures, but after the reconstruction, one of them has a convex curvature. Such a band structure reconstruction can be clearly seen in a band calculation of materials such as $Ca_4Al_2O_6Fe_2As_2$ [92, 97] or $LiFeO_2Fe_2Se_2$ [98].

This large variation of the $d_{xy}$ portion of the band, being closely linked to the Pn-Fe-Pn bond angle or the pnictogen height, can be understood as follows. In Fig.14 (left), we show the $d_{xy}$ portion of the bands. If we approximate this $d_{xy}$ portion by a single band tightbinding model that considers electron hoppings up to third nearest neighbors, i.e.,

$$\varepsilon(\boldsymbol{k}) = 2t_1(\cos k_x + \cos k_y) + 4t_2 \cos k_x \cos k_y + 2t_3(\cos 2k_x + \cos 2k_y) \quad (1)$$

the energy difference between $(0,0)$ and $(\pi,\pi)$ is proportional to the nearest neighbor hopping $t_1$. Therefore, the rising of the $d_{xy}$ band at $(\pi,\pi)$ and the lowering at $(0,0)$ both originate from the decrease of $t_1$. In fact, the nearest neighbor hopping $t_1$ within the $d_{xy}$ orbital is extremely sensitive to various factors because this consists of two components, the direct Fe-Fe hopping and the indirect hopping via As 4p orbitals, which have opposite signs (see Fig.21). The relatively large $t_1$ in LaFeAsO is due to the indirect component dominating over the direct one. When La is substituted by Nd or Sm and the Pn-Fe-Pn bond angle decreases, the indirect contribution becomes smaller, so that $t_1$ is reduced and can even change its sign. On the other hand, LaFePO has a much larger $t_1$ because the indirect component strongly dominates. .

4.3 Consideration of many-body interaction

Many body interactions are considered on top of the above mentioned multiorbital band structure. In the itinerant spin picture, on-site electron-electron interactions are the intra-orbital Coulomb U, the inter-orbital Coulomb U', the Hund's coupling J, and the pair hopping J'. Off-site interactions are also considered in some models as we shall in the next section. Values of the Coulomb interactions were evaluated from first principles in some studies [99, 100].

In the picture in which the electron correlations are considered to be strong, the localized spin model is often adopted. There, the nearest and next nearest neighbor antiferromagnetic interactions are taken into account, as explained in section 5.2. The strong electron correlation effects are also captured in some other ways. For instance, in ref.[101] the itinerant five orbital model has been studied using the variational Monte Carlo method, and it is concluded that orbital selective Mottness of the $d_{xy}$ orbital plays an important role in this system. Dynamical mean field studies suggest that the Hund's coupling rather than the Mottness is important as the origin of the electron correlation in IBSC [102, 103]. In ref.[ 104], a model that considers both itinerant and localized spins was studied, and there also the importance of the Hund's coupling was pointed out.

## 5. Pairing Mechanism

In this chapter, we describe theoretical studies on the pairing mechanism and the pairing states. There are various issues under debate regarding the pairing mechanism as summarized in Fig.15. We will discuss these issues in the following sections.

### 5.1 Spin and/or orbital fluctuation mediated pairing

As mentioned in section 2.2, in many of IBSC, structural transition that breaks the original four fold rotational symmetry takes place. There is now growing consensus that this structural transition is electronically driven, but there has been intensive debate as to whether it is driven by spin [105,106,107] or orbital [108, 109, 110, 111, 112, 113] degrees of freedom. In the spin-driven scenario, the lattice lowers its symmetry from $C_4$ (four fold rotational symmetric) to $C_2$ (two fold symmetric) through magneto-elastic coupling in order to lift the degeneracy of the stripe-type antiferromagnetism (explained later) in x and y directions. In the orbital scenario, the orbital ordering (also explained later) that lifts the degeneracy of $d_{xz}$ and $d_{yz}$ orbitals, as observed experimentally[114,115], is the driving force of the structural transition. In this article, we will not go further into the problem of the origin of the structural transition. Rather, we will put more focus on superconductivity, such as the competition between different pairing states and their possible origins. In fact, as we shall conclude, the    origin of the structural transition/nematicity and the pairing mechanism or the pairing symmetry are not in one to one correspondence. On the other hand, one should keep in mind that there is a possibility of   nematicity enhancing the spin fluctuation by lifting the degeneracy [116, 117], and this may affect superconductivity provided that the pairing is mediated magnetically.

The superconducting gap equation can be written in the form in momentum space,

$$\Delta(k) = \sum_{k'} V(k-k') \frac{\tan[E(k')/2k_BT]}{2E(k')} \Delta(k') \quad (2)$$

$V$(k-k') is the pairing interaction which is mediated by bosonic modes and $E(k)$ is the dispersion of the quasiparticle excitation. In the case of conventional superconductors, the pairing interaction is mediated by phonons, and is usually attractive ($V$<0). On the other hand, when the spin fluctuation mediates pairing, the pairing interaction is repulsive ($V$>0), and is often peaked around a certain wave vector $Q$ at which the spin fluctuation develops. The electronic charge fluctuations (using this term in a broad sense to include orbital fluctuations, plasmons, etc...) can also mediate pairing, and the pairing interaction is attractive as in the case of phonons. When the system is close to some kind of charge ordering with a wave vector $Q$, $V(Q)$ can be large and attractive. From eq.(2), it can be seen that when V($Q$)<(>)0, $\Delta(k)$ and $\Delta(k+Q)$ tend to have the same (opposite) signs. A typical example is the case of phonon-mediated pairing with $V$(q)<0 for all q, where the gap has a constant sign on the entire Fermi surface. Another well-known example is the case of antiferromagnetic spin fluctuations with $Q$=(π,π), where $\Delta$(k)*$\Delta$(k+Q)<0 gives the d-wave gap form proportional to cos(k$_x$)-cos(k$_y$) as in the high $T_c$ cuprates.

In the itinerant spin picture of IBSC, the Fermi surface nesting between electron and hole Fermi surfaces gives rise to spin fluctuation around the wave vector (π,0)(0,π) in the unfolded Brillouin zone. The corresponding spin ordering configuration is shown in Fig.16(left). The (π,0)(0,π) spin fluctuation mediates pair scattering between electron and hole pockets, as shown in Fig.17(left), and this favors a superconducting gap with opposite signs between electron and hole Fermi surfaces. (This kind of explanation often leads to a misunderstanding that the better the nesting, the higher the superconducting $T_c$, but this is not necessarily correct, as we shall see later.) This state is often called the s± state[77, 118], and has been considered as one of the prime candidates of the pairing states. Many purely electronic, itinerant models have been analyzed in different manners, showing the robustness of the s± pairing state. Those include fluctuation exchange [119], functional renormalization group [120, 121, 122], third order perturbation [123], variational Monte Carlo [101,124], and two particle self-consistent studies [125]. There are also studies on models that explicitly consider the As 4p orbitals, which also obtain the s± state[84, 88]. S± pairing state can also be obtained in the localized spin picture, which will be discussed in the next section.

s± is indeed a pairing state which exploits peculiar features of IBSC. One such feature is the presence of the spin fluctuations. As mentioned above, the spin fluctuation can act as a glue for Cooper pairing, but the pairing interaction is repulsive, so that the gap function is accompanied by sign changes, which usually results in nodes intersecting the Fermi surface. However, when the Fermi surface consists of disconnected pieces, the sign change can take place without nodes intersecting the Fermi surface, and this can enhance $T_c$. This kind of s± state had already been discussed in various contexts[126, 127, 128, 129, 130]. Ref. [128] focused particularly on the $T_c$ enhancement owing to the absence of nodes, and this possibility has been analyzed more recently using dynamical cluster approximation on the same model [130]. In the actual IBSC, there are cases where the nodes do intersect the Fermi surface, and this will be discussed in section 5.5.

We will now turn to the orbital fluctuation pairing. In the early days, there were several studies that proposed orbital fluctuation enhanced by the electron-phonon coupling [131, 132, 133]. Those theories proposed that the electron-phonon coupling enhances the orbital fluctuation at wave vectors $(0,0)$ and $(\pi,0)/(0,\pi)$, which correspond to ferro-orbital and antiferro-orbital fluctuations, respectively. The former is a fluctuation that arises near an orbital ordering shown schematically in Fig.16 (right). This order breaks the $C_4$ symmetry present in the original lattice structure, and hence is often called "nematic". Since this fluctuation mediates attractive pair scattering with small momentum transfer (Fig.17 lower right), it can enhance all kinds of pairing states, regardless of the sign reversal in the gap function. On the other hand, the antiferro-orbital fluctuation around $(\pi,0)/(0,\pi)$ competes with the antiferromagnetic spin fluctuation around the same wave vectors (Fig.17 upper right). When the spin fluctuation dominates, the pairing state is $s\pm$ as mentioned above, but when the orbital fluctuation dominates, the sign of the gap remains the same between the electron and hole pockets. This state is called the s++ state.

Later on, Nomura and Arita estimated the electron-phonon coupling from first principles [134]. It has turned out that the electron-phonon interactions are an order of magnitude smaller than those assumed in the early orbital fluctuation pairing theories because there are a large number of bands (orbitals) contributing to the coupling, and they basically cancel with each other. Using these realistic values of the electron-phonon pairing interaction, it was shown that the orbital fluctuation is not strong enough to dominate over the spin fluctuation to result in the s++ state.

In the meantime, the orbital fluctuation pairing theories have developed in several different ways. Onari and Kontani revisited the purely electronic five orbital Hubbard model, taking into account Asmalazov-Larkin type vertex correction that are not taken into account in RPA type calculations[112]. There again, it was suggested that orbital fluctuation is enhanced around the wave vectors around $(0,0)$ and $(\pi,0)/(0,\pi)$, but this time not due to electron-phonon coupling, but due to the coupling among spin fluctuation modes. Although the orbital fluctuation mediated pairing interaction in this case originates from the spin fluctuations, the resulting pairing interaction competes with the spin-fluctuation-mediated pairing interaction around the wave vector $(\pi,0)/(0,\pi)$ in the same sense as in the case when the orbital fluctuation is enhanced by the electron-phonon interaction. In ref. [112], it is concluded that s++ pairing can again take place in the purely repulsive five orbital model. This conclusion is in contrast to those studies on the same model but with different many body techniques. In particular, the variational Monte Carlo technique should take into account the higher order electron correlation effects, but it is found that the sign changing gap state is always found to be favored [101,124]. Hence, it remains an interesting open issue as to whether the purely repulsive five orbital model can actually have s++ pairing dominating over s$\pm$. Even when the s++ state is realized from antiferro-orbital fluctuation, to obtain "high $T_c$" can be another challenge because the pairing interaction arising from the *difference* between orbital fluctuation and spin fluctuation has to strongly dominate over the spin fluctuation in order to overcome the large self-energy effect that originates additively from both spin and orbital fluctuations.

Alternatively, Zhou et al considered an eight orbital model that explicitly takes into account the As 4p orbitals, and studied the effect of the electron repulsion between neighboring Fe3d and As 4p orbitals [87]. Along this line of study considering the interaction between Fe 3d and As 4p orbitals, Ono et al. later introduced an orbital polarization interaction that was not considered in ref.[87] as shown in Fig.18(left)[111]. This orbital polarization interaction enhances $x^2$-$y^2$ quadrupole fluctuation around the wave vector (0,0), namely, ferro-orbital fluctuation. As mentioned in the above, the pairing interaction mediated by this fluctuation enhances all kinds of pairing states, regardless of the sign change. Hence the leading pairing state remains to be s$\pm$ with an enhanced $T_c$ as shown in Fig.18(right). In other words, the spin and orbital fluctuations naturally cooperate to enhance this pairing state. A more general and phenomenological study on the nematic fluctuation mediated pairing has been performed by Yamase and Zehyer [135].

An important conclusion here is that the presence of strong ferro-orbital fluctuation can go hand-in-hand with antiferromagnetic spin fluctuations to enhance s$\pm$ superconductivity, so that the problem of the origin of the electronic nematicity and the issue regarding s$\pm$ vs. s++ are not in one to one correspondence, as mentioned in the beginning of the theoretical part of this review. The ferro-orbital fluctuation simply "boosts" the $T_c$ of s$\pm$ pairing, so that many aspects of s$\pm$ pairing studied in the context of purely spin-fluctuation-mediated pairing (some of them reviewed below) basically hold even when the ferro-orbital fluctuation is present.

## 5.2 Correspondence between real space and momentum space

From the early stage of the study, there has been a debate regarding the origin of the spin fluctuation. In the itinerant spin picture, the spin fluctuation is attributed to the Fermi surface nesting in momentum space as mentioned in the previous section. On the other hand, several studies proposed models with localized spins, where nearest and next nearest antiferromagnetic interactions $J_1$ and $J_2$ are introduced ($J_1$-$J_2$ model, Fig.19) [136,137, 138, 139, 140]. Antiferromagnetic interactions can act as glues of singlet pairing in real space, so when $J_2$ dominates over $J_1$, next nearest neighbor pairing superconductivity is likely to take place. This is reminiscent of the nearest neighbor d-wave pairing in the t-J model for the high $T_c$ cuprates [141, 142], where J is the nearest neighbor antiferromagnetic interaction. In the present case, the pairing is dominant in the s-wave channel, which implies that the pair wave function in real space does not change sign (Fig.20 left). The reason for s-wave dominating over d-wave can be clearly understood if the pairing form factor in real space is Fourier transformed to momentum space as

$$\sum_{\Delta r} \exp(i\boldsymbol{k} \cdot \Delta \boldsymbol{r}) \sim \cos(k_x)\cos(k_y), \quad (3)$$

where the next nearest neighbors are $\Delta \boldsymbol{r}$=(1,1),(1, -1),(-1,1)(-1,-1) in units of the lattice constant. In momentum space, this is nothing but the s$\pm$ wave superconducting gap, where the absolute value of the gap is maximized at $(0,0),(\pi,0)(0,\pi)(\pi,\pi)$ around which the Fermi surfaces exist, and nodes of the gap do not intersect the Fermi surfaces because of their disconnectivity. If it were the next nearest neighbor d-wave ($d_{xy}$) pairing, the gap in momentum space would be $\sin(k_x)*\sin(k_y)$, whose vertical and horizontal nodes would all be intersecting the Fermi surface. In the case of the cuprates, the nearest neighbor d-wave pairing has a superconducting gap with $\cos(k_x)$-$\cos(k_y)$ dependence in momentum space, whose absolute values become large around the wave vectors $(\pi,0)(0,\pi)$ with large density of states at the Fermi level, as shown in Fig.20(right). Namely, in both iron-pnictides and cuprates, the sign of the pair wave function in real space is determined so as to maximize the superconducting gap around the wave vectors with large density of states near the Fermi level.

In the case of the curates, the $t$-$J$ model can be derived as an effective model of the single band Hubbard model in the large $U/t$ limit, where t and $U$ are the nearest neighbor electron hopping and the on-site repulsion, respectively. In this case $J$ is proportional to $t^2/U$ in the second order perturbation with respect to $t/U$, namely, larger t leads to larger J as far as the perturbational treatment is valid. On the other hand, there is some ambiguity in deriving the $J_1$-$J_2$ model microscopically, starting from the five orbital Hubbard model. Nonetheless, there can be some microscopic relation between the $J_1$-$J_2$ model and the five orbital Hubbard model, as we shall see in section 5.3. The fully gapped s± state is an ideal pairing state in the case of spin-fluctuation-mediated pairing, but it competes with other sign reversing gap states when $J_1$ is comparable with $J_2$. This is related to the nodes intersecting the Fermi surface, briefly mentioned in the previous section. These states will be discussed in detail in section 5.5.

Finally, it is intriguing to look at the s± vs. s++ competition from the real space point of view. The s++ state has a superconducting gap that has the same sign on all of the Fermi surfaces. In a rough approximation where the gap is taken as a constant, this corresponds to on-site pairing in real space. In this rough sense, s± vs. s++ is a competition between next nearest neighbor and on-site pairings. While the off-site pairings such as s± and d-wave pairing can avoid the strong on-site Coulomb interaction that is characteristic of strongly correlated systems, the on-site pairing has to suffer from this interaction. Then, in order to have the s++ state, a large pairing interaction originating from the difference between the orbital and the fluctuations is necessary to overcome this problem.

5.3 $d_{xy}$ or $d_{xz/yz}$ orbital : which is important ?

As mentioned in section 4.1, $d_{xz}$, $d_{yz}$ and $d_{xy}$ orbitals mainly contribute to the Fermi surface of IBSC. In the spin-fluctuation-mediated pairing in particular, the pairing interaction basically arises within each orbital component (Cooper pairs are scattered within portions of the Fermi surface having the same orbital character, so that one can distinguish the contributions coming from each orbitals [91]. Then, a question arises as to which one of the orbitals is the main player in the occurrence of superconductivity. In principle, they can all cooperate, but when the pairing is mediated by repulsive pairing interactions, different orbitals can in general favor different kinds of sign change in the gap function, so they compete with each other. In fact, such a problem of different orbitals favoring different pairing symmetries was studied in the two orbital model of the cuprates [143]. Cuprates are usually viewed as single orbital systems with only $dx^2\text{-}y^2$ orbital being relevant, but actually in some materials such as $(La,Sr)_2CuO_4$, the $dz^2$ orbital hybridization is not negligible. Since the two orbitals favor different pairing symmetries, the $dz^2$ hybridization degrades the d-wave superconductivity. Hence, in multiorbital systems, it is important to determine which orbital plays the main role. In this section, we shall see that the situation is rather special for IBSC when considered as a spin-fluctuation-mediated superconductor ; both $d_{xy}$ and $d_{xz/yz}$ orbitals can by themselves play an important role in the occurrence of superconductivity depending on the material, and more surprisingly, the three orbitals can cooperate to enhance superconductivity because they favor  pairing states with basically the same form of the order parameter, namely $s\pm$.

In this context, the 1111 family is very instructive to study since the Fermi surface configuration changes by chemical substitution, or doping, as described in chapter 4.2. Let us now see the consequences of these band structure variations to spin-fluctuation-mediated superconductivity. As the phosphorous content increases and hence the Pn-Fe-Pn bond angle increases, it is expected that the $d_{xy}$ hole Fermi surface first shrinks, then it is lost, and finally the top of the $d_{xy}$ band moves away from the Fermi level. Therefore, the $d_{xy}$ band  contribution to the spin fluctuation decreases monotonically as the phosphorous content increases, especially after the $d_{xy}$ Fermi surface is lost. Theoretical calculations show that the spin-fluctuation-mediated pairing is strongly degraded when the $d_{xy}$ hole Fermi surfaces is absent, suggesting that the $d_{xy}$ band plays an important role in the occurrence of high $T_c$ superconductivity. Nonetheless, a very recent theoretical calculation shows that $T_c$ of the spin fluctuation mediated pairing shows a non-monotonic behavior even after the $d_{xy}$ hole Fermi surface is lost; it exhibits a local maximum around a certain phosphorous content (i.e., certain Pn-Fe-Pn bond angle) greater than that at which the $d_{xy}$ Fermi surface disappears as shown in Fig.22(a) [144]. The origin of this is traced back to the fact that the loss of the $d_{xy}$ hole Fermi surface leads to a better nesting within the $d_{xz/yz}$ portion of the Fermi surface, thereby enhancing the low lying spin excitations. It is also found in the calculation that the Pn-Fe-Pn bond angle at which the $T_c$ is locally maximized decreases when the electron doping rate is increased. These results are in fact consistent with recent experimental observations [145, 146, 147]. This indicates that the $d_{xz/yz}$ orbitals by themselves can play the main role in the occurrence of spin-fluctuation-mediated superconductivity, although the $T_c$ in that case is not very high. The resulting superconducting gap in this case is not expected to be a fully gapped $s\pm$ form, as we shall see in section 5.5.

Doping electrons by partially substituting O by F or H in LaFeAsO is also very interesting in this context [139]. As mentioned in 2.5, electron doping rate can exceed 50 % in LaFeAsO$_{1-x}$H$_x$, and the T$_c$ phase diagram against the doping rate $x$ exhibits a double dome structure, where the second dome with higher doping concentration has the higher T$_c$. In a rigid band picture, such a large amount of electron doping would wipe out the hole Fermi surfaces, so that the nesting scenario is no longer valid in the higher T$_c$ second dome. First principles band calculation that takes into account the band structure variation with chemical substitution reveals that the band structure rapidly changes with doping, and the rigid band picture is not valid[149, 150, 151]. In momentum space, around (0,0) the d$_{xz/yz}$ hole Fermi surfaces shrink monotonically and are eventually lost with sufficient electron doping, and in turn an electron Fermi surface appears. On the other hand, an interesting point is that the d$_{xy}$ hole Fermi surface around ($\pi,\pi$) is barely changed with the doping rate $x$, which is clearly a non-rigid band feature. Quite surprisingly, this is found to be due to a rapid decrease of t$_1$ within the d$_{xy}$ orbital upon increasing x as shown in Fig.21, which pushes up the d$_{xy}$ band top at ($\pi,\pi$) to follow the increase of the Fermi level. The t$_1$ reduction is reminiscent of the case of La→Sm substitution as described in section 4.2, but an important point here is that it is largely due to the increase of the positive charge within the blocking layer by O(2-)→H(1-) substitution, which in turn reduces the As 4p electronic level and leads to the suppression of the indirect component of t$_1$ (Fig.21) [151].

As seen in the above, the d$_{xy}$ hole Fermi surface remains even at large doping rate, while the d$_{xz/yz}$ hole Fermi surfaces are lost, so that the importance of the d$_{xy}$ orbital increases with doping. Interestingly, a fluctuation exchange study of these non-rigid band models show that the spin fluctuation and the s$\pm$ pairing are both enhanced in this largely doped regime, exhibiting a double dome feature of the superconducting T$_c$ as a function of doping. Moreover, the two domes are merged into a single dome when the Pn-Fe-Pn bond angle is reduced (a change that takes place when the rare earth is varied as La→Ce→Sm→Gd), as shown in Fig.22(b), in agreement with the experiment. Although the d$_{xy}$ hole Fermi surface remains unchanged in the highly doped regime, the Fermi surface nesting (in its original sense of the term) is monotonically degraded because the volume of the electron Fermi surfaces increases, so the origin of the second (higher T$_c$) dome in LaFeAsO$_{1-x}$H$_x$ cannot be attributed to a good Fermi surface nesting. Here one should recall the discussion of section 5.2. s$\pm$ pairing is a next nearest neighbor pairing, which is favored by $J_2 > J_1$, corresponding to $t_2 > t_1$. In fact, as mentioned above, $t_2$ dominating over $t_1$ is what is happening in the second $T_c$ dome regime. Hence, intuitively speaking, $t_2 > t_1$ can be considered as the origin of the $T_c$ enhancement in the largely doped regime. To be precise, however, the fluctuation exchange approximation is a weak coupling method based on the itinerant spin model, so using the $J_1$-$J_2$ term of the localized spin model is not conceptually correct. What is actually happening in the above FLEX calculation is that the entire $d_{xy}$ portion of the band structure (including that away from the Fermi level) is strongly modified in a manner that it favors the second nearest neighbor pairing. In any case, this example shows that a good Fermi surface nesting is not necessary for the spin fluctuation mediated pairing even in the itinerant spin picture. On the other hand, we have mentioned that the nesting of the $d_{xz/yz}$ Fermi surface can enhance the pairing, so the bottom line is that the nesting and s$\pm$ superconductivity can in some cases be correlated with the Fermi surface nesting, while in other cases not. This also explains why the strength of the low energy spin fluctuation probed by NMR is in some cases correlated with $T_c$[139,152,153,154], while in other cases not[2(e), 139], because the low energy spin fluctuation is largely governed by the Fermi surface nesting. This theoretical interpretation has recently been summarized schematically as shown in Fig.22(c).

Quite recently, it has been shown theoretically[155] that the strong reduction of $t_1$ within the $d_{xy}$ orbital by electron doping occurs also in alkali metal/ammonia intercalated FeSe, where a large enhancement of the $T_c$ is observed, as mentioned in section 2.3.4. There again, the increased positive charge in the layers inserted between FeSe layers lowers the Se4p level, thereby reducing the indirect Fe->Se->Fe hopping. This issue is also closely related to $K_xFe_{2-y}Se_2$, another electron doped material, which will be discussed in the next section. These examples show that the $t_1$ reduction within the $d_{xy}$ orbital occurs rather universality in largely electron doped materials with high $T_c$.

In the above, we discussed the $d_{xz/yz}$ to $d_{xy}$ orbital crossover in LaFeAsO$_{1-x}$H$_x$ in the context of the spin fluctuation mediated pairing. Such a crossover of the dominating orbital from $d_{xz/yz}$ to $d_{xy}$ has also been discussed in the context of the orbital fluctuation mediated pairing [156]. There it has been proposed that $x^2$-$y^2$ and $3z^2$-$r^2$ quadrupole fluctuations lead to the s++

state in the lightly doped and the heavily doped regimes, respectively [157].

## 5.4 Hole or electron Fermi surfaces only

In the previous sections, we saw that good Fermi surface nesting is not important for the enhancement of s± superconductivity as far as $t_2 > t_1$ is satisfied within the $d_{xy}$ orbitals. Then, it is of interest to investigate as to whether the superconductivity is maintained even when all the hole bands sink below the Fermi level, owing to finite-energy pair scattering processes between the electron Fermi surfaces and the top of the hole bands. This problem was phenomenologically studied in detail in ref.[158]. Microscopically, further calculations within the fluctuation exchange approximation (i.e., weak coupling theory) on a five orbital model show that the top of the hole band has to sit very close to the Fermi level, within few 10 meV; if the hole band sinks below the Fermi surface more than that, s± superconductivity is no longer enhanced even if $t_2 > t_1$ [159]. In fact, the pairing symmetry in the absence of the hole Fermi surface was already studied in the paper that first introduced the five orbital model, and d-wave pairing was found to dominate within the weak coupling theory.

In this context, $K_xFe_{2-y}Se_2$ is an interesting material [160] since the ARPES experiments indicate absence of the hole Fermi surfaces; the hole bands have been observed to be lying about 50~100 meV below the Fermi level [161, 162, 163, 164]. As just mentioned, s± pairing state is not enhanced in such a situation within the weak coupling scheme, and indeed a number of theories have obtained a d-wave superconducting state for models with only electron Fermi surfaces[165, 166]. As explained in detail in the next section, this turns out to be inconsistent with the ARPES experiments. In ref.[167, 168], an alternative state with sign change was proposed, but this was not supported by a microscopic calculation based on a realistic ten orbital model [169]. We will come back to this point also in the next section.

The strong coupling approaches to the problem of $K_xFe_{2-y}Se_2$ based on localized spin picture do obtain an s-wave superconducting state even in the absence of the hole Fermi surfaces [170, 171, 172], which is consistent with a fully open gap. Orbital fluctuation mediated pairing theories also predict sign conserving s-wave states [173]. However, these sign conserving superconducting gap scenarios are in contradiction with a neutron scattering experiment, in which the observation of the spin resonance suggest the sign change in the superconducting gap. Therefore, existing theories all confront some kind of difficulties.

Quite recently, ARPES study of $K_xFe_{2-y}Se_2$ has been revisited in ref.[174], and a hole band that intersects the Fermi level has been observed for the first time. The hole band seems to have eluded detection in previous studies because it is observed only for a certain photon energy and polarization probably due to matrix element problem. It is interesting to look at this material from the viewpoint of electron hoppings within the $d_{xy}$ orbital ; a first principles study of the stoichiometric $KFe_2Se_2$ gives $t_1=$-0.008eV, $t_2=$0.056eV, namely, $t_1$ is very small. Therefore, the $d_{xy}$ electron band around $(0,0)_{unfold}$ comes down and the $d_{xy}$ hole band around $(\pi,\pi)_{unfold}$ rises up. The electron band has already been seen around the $\Gamma$-Z line in previous studies, and the newly observed hole band most likely corresponds to the $d_{xy}$ hole band. If this is indeed the case, the s$\pm$ pairing state is obtained within the weak coupling theory, and this may give explanation to the existing experiments.

We will now turn to the opposite situation, where only the hole Fermi surfaces are present. In the 122 material $BaFe_2As_2$, holes can be doped by substituting Ba by K, and in the end compound $KFe_2As_2$, sufficient number of holes is doped so as to wipe out the electron Fermi surfaces. It is worth mentioning that a first principles estimation of the electron hoppings of $KFe_2As_2$ gives $t_1=$0.165eV, $t_2=$0.113eV, so that in this case the nearest neighbor $t_1$ strongly dominates [174]. Hence, the s$\pm$ state is not so favorable as in the cases with more electrons, both from the viewpoint of the Fermi surface configuration and the hoppings in real space. A random phase approximation study has looked into the problem of the gap symmetry based on a five orbital model [175]. S-wave pairing in this case occurs due to the spin-fluctuation-mediated interaction between the hole Fermi surface around $(0,0)$ and the states around $(\pi,0)/(0,\pi)$, although there is no electron pocket there. On the other hand, d-wave pairing is in very close competition with s-wave pairing. In the next section, we will come back to this problem on the gap nodes of $KFe_2As_2$.

5.5 Various sign changes in the superconducting order parameter

s$\pm$ and s++ pairings usually refer to fully gapped states consistent with a number of experiments [176,177,178,179,180,181,182], but on the other hand in some of IBSC, experiments suggest presence of nodes on the Fermi surface. One of the examples found in the early days is LaFePO[183, 184]. As mentioned in section 5.3, LaFePO has larger Pn-Fe-Pn bond angle compared to other 1111 arsenides, and therefore does not have the $d_{xy}$ hole Fermi surface around $(\pi,\pi)_{unfold}$. Hence, the spin-fluctuation-mediated pairing interaction between electron and hole Fermi surfaces mainly come from the $d_{xz/yz}$ orbitals. On the other hand, the $d_{xy}$ portion of the two electron Fermi surfaces interact with each other with spin fluctuation around the wave vector $(\pi,\pi/2)(\pi/2,\pi)$ and therefore these portions tend to change signs. This can be accomplished either in the s-wave or in the d-wave form as shown in Fig.23(a). In both cases, electron-hole interaction coming from the $d_{xz/yz}$ orbitals is the main driving force of the superconductivity (as mentioned in section 5.3), but the superconducting gap is accompanied by nodes on the Fermi surface owing to the interaction within the $d_{xy}$ portion of the electron Fermi surfaces. This kind of sign change was first shown in ref.[78], where the band structure of LaFeAsO was obtained using the lattice structure determined by first principles optimization, which now is known to give smaller estimation for the pnictogen height, or equivalently, larger estimation of the Pn-Fe-Pn bond angle. Functional renormalization group studies [120, 122] also obtained similar results.

The modification of the s± state has also been studied from the localized spin picture as well[136, 137]. As mentioned in section 5.2, a fully gapped s± state is favored when $J_2$ strongly dominates over $J_1$, but when $J_1$ is comparable to $J_2$, the nearest neighbor s-wave and d-wave pairings becomes competitive against the s± state. The nearest neighbor s-wave state has a gap form of cos(kx)+cos(ky) in momentum space, whose nodes intersect the electron Fermi surfaces. Therefore, the $J_2$ vs. $J_1$ competition in real space corresponds to $(\pi,0)(0,\pi)$ vs. $(\pi,\pi/2)(\pi/2,\pi)$ spin fluctuation competition in momentum space.

As the experimental studies on the gap structure proceeded, nodes of the superconducting gap were observed in some other materials. In some of those materials such as $BaFe_2(As,P)_2$[185, 186, 187], the pnictogen height is not so small and the $d_{xy}$ hole Fermi surface is expected to be present. Within the random phase approximation, gap nodes of the type discussed in the previous paragraph are not expected. Instead, some theoretical studies have suggested presence of horizontal nodes on one of the hole Fermi surfaces [188, 189]. This sign change occurs close to the Z point, where the hole Fermi surface exhibits a strong warping peculiar to the 122 materials. There, the $dz^2$ orbital character is strong, and the superconducting gap is fragile within the spin fluctuation-mediated pairing because there are no counterparts originating from the $dz^2$ orbital in the electron Fermi surfaces. On the other hand, nodes can arise from other origins, such as the intra-Fermi surface repulsive interaction [190]. Another possibility is a competition between spin and orbital fluctuations. Since these two fluctuations mediate pairing interaction with opposite signs, the competition gives rise to a frustration in momentum space, and can lead to nodes in the superconducting gap [191].

As mentioned in section 5.4, previous ARPES (except ref.[174]) observed absence of hole Fermi surfaces in $K_xFe_{2-y}Se_2$ [161, 162, 163, 164]. In such a case, the d-wave pairing state is a natural pairing state that can explain the neutron scattering experiment [192]. In the unfolded Brillouin zone, where the two electron Fermi surfaces are located around $(\pi,0)$ and $(0,\pi)$, the diagonal nodes of the d-wave state do not intersect the Fermi surfaces, and therefore results in a fully gapped state, as was first pointed out in ref.[77]. At first glance, this seems to be consistent with the ARPES experiments suggesting a fully open gap. In the 122 compounds including $K_xFe_{2-y}Se_2$ the two electron bands are actually hybridized in the original folded Brillouin zone (the "unfolding" of the band can only be done approximately ) so as to give rise to inner and outer Fermi surfaces (see Fig.23(b)). This leads to nodes intersecting the Fermi surfaces around those points where the hybridization takes place as shown in Fig.23(b). D-wave pairing is also inconsistent with the ARPES experiments suggesting absence of nodes in the electron Fermi surface observed around the $\Gamma$-Z line. In ref.[167], an alternative pairing state was proposed, where the superconducting gap changes sign between inner and outer electron Fermi surfaces in the folded Brillouin zone, as shown in Fig.23(b). In ref. [168], it has been shown that d-wave gives way to this bonding-antibonding $s\pm$ pairing state when the hybridization between the two electron bands is large enough. However, a microscopic study based on a realistic ten orbital model shows that *d*-wave pairing always dominate over the bonding-antibonding $s\pm$ wave pairing when the hole Fermi surfaces are absent[169].

The sign change in the gap of other particular materials has also been discussed in detail. In ref.[193], the gap function of LiFeAs has been studied within the spin-fluctuation-mediated s±pairing mechanism. There, the most inner Fermi surface was found to have the smallest magnitude of the gap, in contradiction to the experimental observation [194, 195, 196]. Some other new pairing states with a gap sign reverse among the three hole Fermi surfaces have also been proposed [197, 198, 199], which give the correct (the experimentally observed) relative magnitude among the gaps. It has also been noticed recently that the correct magnitude relation may be obtained even in the conventional s± state by taking into account the self-energy correction (not taken into account in ref.[193]) within the fluctuation exchange approximation [200].

Another material for which the sign change of the gap function has been under debate is $KFe_2As_2$. Functional renormalization study shows d-wave dominating over s-wave [201], while some other studies suggest close competition between d-wave and s-wave [175, 202]. The possible existence of horizontal nodes was pointed out in ref. [175]. There is also a proposal of a new s-wave gap with a sign reverse of the gap among the whole Fermi surfaces [203]. Among the recent experiments, a Laser ARPES experiment shows an anisotropic s-wave gap [204], while experiments under pressure suggest a possibility of the change in the pairing state with increasing the pressure [205, 206].

5.6 Theoretical proposals of detecting the sign change in the order parameter

One of the aspects of IBSC that have led to the debate of the s+- vs. s++ pairing is the sensitivity of $T_c$ against impurities. Namely, in refs.[207, 208, 209] it was shown that the suppression of $T_c$ by impurity doping or by particle irradiation is too weak for a pairing state with sign changes in the order parameter. The suppression of $T_c$ by impurities was calculated within the five orbital model , which showed that the experimentally observed suppression rate of $T_c$ is small compared to the expectation of the s± state[210]. Quite recently, the problem of this disorder effect has been studied in detail combining electron irradiation experiment on $BaFe_2(As,P)_2$ and a theoretical analysis. As the disorder is increased by irradiation, the temperature dependence of the penetration depth shows a non-monotonic variation, suggesting that the nodes of the superconducting gap are lifted by the introduction of the disorder, and the resulting fully gapped state has a sign reversing superconducting gap. This is taken as a strong evidence for the s± state [211].

It was proposed at the early stage [212, 213] that one of the ways to determine the presence/absence of the sign change in the superconducting order parameter is the resonance in the neutron scattering experiment. Neutron scattering experiments have indeed observed a peak like structure in the superconducting state around the nesting vector of the Fermi surface, and this has been taken as an evidence for the sign change in the superconducting gap [214, 215], along with other phase sensitive experiments [216, 217]. However, ref.[218] later showed that the quasiparticle damping can give rise to a peak like enhancement of the dynamical spin susceptibility over the normal state values in the s++ state, which is due to the suppression of the normal state susceptibility originating from the damping. In a later paper by Nagai and Kuroki[219], it was shown that the experimental results are close to the calculation results of the s± state when the superconducting gap and the quasiparticle damping is ~ 10meV, while they are closer to those of the s++ state for larger gap and damping. It was further proposed that the two possible candidates can be distinguished by investigating experimentally the wave vector $(\pi,\pi)$ along with $(\pi,0)$ [220]. Alternatively, in ref. [221], it was proposed that the realization of s++ state can be detected by probing the orbital susceptibility through measuring the phonon spectra in the neutron scattering experiments. Also, a Bogoliubov-de Gennes equation approach for the single impurity problem shows that the bound state peak in the local density of states can be used to distinguish between s± and s++[222].

## 6. Summary and Perspective

Table 8 summarizes the features of IBSC in comparison with $MgB_2$ and high Tc cuprates. The unique properties such as high $Hc^2$ and robustness to impurity and huge diversity of superconducting materials originate from multiband and bonding nature of Fe and pnictogen (chalcogen), respectively. Irrespective of intense research globally, the maximum temperature remains at ~55K for past 6 years but understanding of the paring mechanism is much advanced by a combination of advanced measurement with theoretical treatments incorporating spin, orbital, charge and phonon. What are next challenges in IBSCs? We pick up the following subjects based on the recent

progress:

a. Searching higher Tc by optimizing the two factors controlling the two Tc domes.

b. Elucidating the origin of high Tc in single layer FeSe epitaxially grown on SrTiO$_3$

c. Searching new As-free IBSCs with less anisotropy and relatively high Tc for wire application

d. Clarifying the role of spin and orbital fluctuations in the pairing mechanism, and if they are coexisting, whether they are cooperating or competing

e. Establishing the presence/absence of the sign reversal in the superconducting gap, especially for those materials with Tc exceeding 50K.

All the major actors are now ready in this system associated with charge, spin, and orbitals. We expect "*iron age*" in superconductors will come in near future by successful collaboration between experimentalists and theoreticians.

## Acknowledgements

HH was supported by the MEXT Element Strategy Initiative to from a core research center  and  KK acknowledges the Grants-in-Aid for Scientific Research No. 24340079. HH appreciates Soshi Iimura and Hidenori Hiramatsu for their efforts for collecting data and helpful discussions. KK acknowledges Katsuhiro Suzuki and Hidetomo Usui for collaboration on the theoretical problems on IBSC.

## References

[1] Y. Kamihara, T. Watanabe, M. Hirano, and H. Hosono, J. Am. Chem. Soc., 130 (2008) 3296.

[2] (a) D.C.Johnston, Adv.Phys.59(2010)1063,(b) G.R.Stewart, Rev. Mod. Phys. 83(2011)1589, (c)A.A.Kordyuk, Low Temp.Phys.38(2011)888, (d)X.Chen et al. Nat.Sci.Rev.1(2014)371, (e)  K.Ishida, Y.Nakai, and H.Hosono, J. Phys. Soc. Jpn. 78 (2009) 062001,  (f) Special issue, Sol.Stat.Commun.152(2012)631, (g) J. Paglione and R.L.Greene, Nature Phys 6(2010) 645, (h)  E.Dagotto, Rev.Mod.Phys,85(2013)849, (i) J.E.Hoffman, Rep. Prog. Phys. 74 (2011) 124513, (j) P.J.Hirschfeld, Rep. Prog.

Phys. 74 (2011) 124508, (k) T.Shibauchi, A.Carrington and Y,Matsuda, Annu. Rev.Condens.Matter Phys. 5(2014)113, (l) L.Malavasi and S.Margadonna, Chem. Soc. Rev.41(2012) 3897.

[3] (a) N.L.Wang, H.Hosono, and P.Dai (ed.) Iron-based superconductors: materials, properties and mechanism, Pan Stanford(2013)., (b) P.D.Johnson, G.Xu, W-G.Yin (ed), Iron-based superconductors, Springer (2015).

[4] Y. Kamihara, H. Hiramatsu, M. Hirano, R. Kawamura, H. Yanagi, T. Kamiya and H. Hosono, J. Am. Chem. Soc. 128 (2006) 10012.

[5] T.Watanabe, H.Yanagi, T. Kamiya, Y. Kamihara, H. Hiramatsu, M. Hirano, and H.Hosono, Inorg. Chem. 46 (2007) 7719.

[6] H.Hosono, Physica C, 469(2009) 314.

[7] S. W. Park, H. Mizoguchi, K. Kodama, S. Shamoto , T. Otomo, S. Matsuishi, T. Kamiya, and H. Hosono, Inorg. Chem. 52 (2013) 13363.

[8] T. Hanna, S. Matsuishi, K. Kodama, T. Otomo, S. Shamoto, and H. Hosono, Phys. Rev. B 87 (2013) 020401(R)

[9] H.Yanagi et al. Phys. Rev. B 77(2008)224431.

[10] K.Shimizu et al. .Nature 412(2001)316.

[11] Z. A. Ren et al., Chin. Phys. Lett. 25 (2008) 2215.

[12] S. Matsuishi et al.Phys. Rev. B85 (2012) 014514.

[13] K. Hayashi, P. V. Sushko, Y. Hashimoto, A. L. Shluger, and  H. Hosono, Nat. Commun. 5 (2014) 3515.

[14] H. Hosono, Physica C469(2009) 314.

[15] S. Iimura et al. Nat.Comm.3 (2012) 943.

[16] S.Matsuishi , T. Maruyama, S. Iimura, and H.Hosono Phys. Rev. B 89, 094510.

[17] M.Rotter, M.Tegel, and D.Johrendt, Phys. Rev. Lett. 101(2008)107006.

[18] H-H. Wen, G. Mu, L.Fang, H.Yang and X.Zhu,  Euro Phys.Lett. 82 (2008)17009.

[19] (a)Z.A.Ren et al. Euro.Phys.Lett.83(2008)17002.
(b)H.Kito,H.Eisaki and A.Iyo, J.Phys.Soc.Jpn.77(2008)063707.

[20] A.S.Safat et al.Phys.Rev.Lett.101(2008)11704.

[21] P.C.Canfield and S.L Bud'ko, Annu.Rev.Conds.Matter.Phys.1(2010)27.

[22] H. Wadati, I. Elfimov and G. A. Sawatzky, Phys.Rev.Lett.105(2010) 157004.

[23] K.Nakamura, R.Arita and H.Ikeda, Phys.Rev.83(2011)144512.

[24] T.Katase, S.Iimura S, Hiramatsu, T.Kamiya and H.Hosono, Phys. Rev. B

85(2012) 140516(R)

[25] C. Wang et al., Phys. Rev. B 79(2009)054521.

[26] T.Yoshida et al. Phys. Rev. Lett. 106(2011)117001.

[27] Z. Ren et al. Phys.Rev.Lett. 102,(2009)137002.

[28] S.Jiang et al., J. Phys.: Condens. Matter 21 (2009) 382203.

[29] F.-C.Hsu et al. Proc.Natl.Acad.Sci.USA, 105(200)14262.

[30] M.Burrard-Lucus et al.. Nat.Mat.12(2013)15.

[31] J.B.Guo, H.Lei, F.Hayahi and H.Hosono, Nat.Comm.5(2014)4756.

[32] H-H. Wen, Rep.Prog.Phys.75(2012)112501.

[33] H.Hiramatsu, T.Katase, T. Kamiya, M.Hirano,and H.Hosono, Phys. Rev. B 80, (2009)052501.

[34] T.Kamiya et al. 173(2010) 244.

[35] T. Katase et al. Solid State Commun. 149(2009)2121.

[36] K.Deguchi et al. Supercond. Sci. Technol. 24(2011) 055008.

[37] S. R. Saha, N. P. Butch, K. Kirshenbaum, J.Paglione, and P. Y. Zavalij, Phys. Rev. Lett. 103(2009) 037005.

[38] C-H.Lee et al. J. Phys. Soc. Jpn. 77(2008) 083704.

[39] K.Kuroki, H.Usui, S.Onari, R.Arita, and H.Aoki, Phys.Rev.B79(2009) 224511.

[40] S. Kasahara, et al. Nature 486(2012) 382.

[41] S-H.Baek et al.Nat.Mat. (2014) in press.

[42] M. Hiraishi et al. Nat. Phys. 10(2014)300.

[43] S.Iimura et al. Phys. Rev. B88(2013) 060501(R).

[44] H.Takahashi et al. Sci. Rep., 5(2015)7829.

[45] H.Hiramatsu, T.Katase, T.Kamiya, and H.Hosono, J. Phys. Soc. Jpn., 81 (2012) 011011.

[46] S. Haindl et al.Rep. Prog. Phys. 77(2014) 046502.

[47]X.Ma, X.Chen and Q-K.Xue, Chap.3 in Iron-Based Superconductivity, Ed.by P.D.Johnson, G.Xu and W-G.Yin, Springer (2015).

[46] Q. Li, W. Si and I. K Dimitrov, Rep. Prog. Phys. 74(2011) 124510.

[47] H.Hiramatsu et al, Appl. Phys. Express, 1(2008)101702..

[48] S. A. Baily et al. Phys. Rev. Lett., 102(2009) 117004.

[49] S.Lee et al. Nat.Mat.12(2013)392.

[50] T.Thetsleff et al. Appl.Phys.Lett.97(201)022506.

[51] H.Sato et al. Appl. Phys. Lett.104(2014)182603.

[52] J.H.Durrell et al. Rep. Prog. Phys. 74(2011) 124511.

[53] T.Katase et al Nat.Commun. 2(2011) 409.

[54] D.C.Larbalestier et al. Nature Materials 13 (2014)375.

[55] Z. Gao, K. Togano, A. Matsumoto and H. Kumakura, Sci.rep. 4 (2014) 4065.

[56] H. Lin et al. Sci. Rep.4(2014)4465.

[57] J.Shimoyama, Supercond.Sci.and Technol. 27(2014) 044002.

[58] M. K. Wu et al. Physica C 496 (2009) 340; Y. Han et al . J. Phys.: Condens. Matter 21 (2009) 235702.

[59] M. J. Wang et al. Phys. Rev. Lett. 103 (2009) 117002.

[60] E. Bellingeri et al. Appl. Phys. Lett. 96 (2010) 102512.

[61] Q.-Y. Wang et al. Chin. Phys. Lett. 29 (2012) 037402.

[62] S.Y.Tan et al.Nat.Mater. 12(2013) 634.

[63] J-F. Ge et al. Nat.Mater.   on line published.

[64] J. J. Lee et al. Nature 515(2014)245.

[65] S. Ueda, Appl. Phys. Lett. 99 (2011) 232505

[66]T. Kawaguchi et al. Appl. Phys. Lett. 97 (2010) 042509.

[67] K.Ueno et al. J. Phys. Soc. Jpn. 83(2014) 032001.

[68] B.C.Sales et al. Phys Rev B 83(2011)224510

[69]T.Katase et al. Proc.Natl.Acad,Sci,USA. 111(2014) 3979.

[70] C-H.Wang, T-K.Chen, C-C.Chang, C-H.Hsu, Y-C. Lee, M-J.Wang, P.M. Wu, M-K.Wu, arXiv:1502.01116.

[71] For a review on itinerant model based studies, see also "Iron-based Superconductors: Materials, Properties and Mechanisms",   ed by Nan Lin Wang, Hideo Hosono, Pengcheng Dai, Pan Stanford publishing (2013) K.Kuroki, chap.8

[72] A. V. Chubukov, Annu. Rev. Cond. Mat. Phys. 3 (2012) 57.

[73] For a review on the gap symmetery, see also P. J. Hirschfeld, M. M. Korshunov, I. I. Mazin, Rep. Prog. Phys. 74(2011)124508.

[74] Fore a review on the spin fluctuation pairing in unconventional superconductors, see also D. J. Scalapino ,   Rev. Mod. Phys. 84, 1383 (2012).

[75] L. Boeri, O.V. Dolgov, and A.A. Golubov, Phys. Rev. Lett. 101(2008) 026403.

[76] N. Marzari and D. Vanderbilt, Phys. Rev. B 56 (1997) 12847; I. Souza, N. Marzari and D.Vanderbilt, Phys. Rev. B 65(2002)035109.

[77] K.Kuroki, S.Onari, R.Arita, H.Usui, Y.Tanaka, H.Kontani, and H.Aoki, Phys. Rev. Lett.101(2008)087004.

[78] S. Graser, T. A. Maier, P. J. Hirschfeld, and D. J. Scalapino, New J. Phys. 11 (2009)025016.

[79] S. Raghu, X.-L. Qi, C.-X. Liu, D.J. Scalapino, and S.-C.Zhang, Phys. Rev. B 77(2007) 220503(R)

[80] P.A. Lee and X.-G.Wen, Phys. Rev. B 78(2008) 144517.

[81] M. Daghofer, A. Nicholson, A. Moreo, and E. Dagotto, Phys. Rev. B 81 (2010) 014511.

[82] R. Thomale, C. Platt, J. Hu, C. Honerkamp, B.A. Bernevig, Phys. Rev. B 80, 180505(R) (2009).

[83] R. Arita and H. Ikeda, J. Phys. Soc. Jpn. 78(2009)113707.

[84] Y. Yanagi, Y. Yamakawa, and Y. Ohno, J. Phys. Soc. Jpn. 77(2008) 123701.

[85] T. Miyake, L. Pourovskii, V. Vildosola, S. Biermann, and A. Georges, J. Phys. Soc. Jpn. 77 (2008) Suppl. C 99.

[85] T. Miyake, K. Nakamura, R. Arita, and M. Imada, J. Phys. Soc. Jpn. 79, (2010)044705.

[86] T. Miyake, K. Nakamura, R. Arita, and M. Imada, J. Phys. Soc. Jpn. 79(2010), 044705.

[87] S. Zhou, G. Kotliar, and Z.Wang: Phys. Rev. B 84 (2011) 14505(R).

[88] J. Lichtenstein, S. A. Maier, C. Honerkamp, C. Platt, R. Thomale, O. K. Andersen, L. Boeri,  Phys. Rev. B 89(2014) 214514.

[89] J. Zhao, Q. Huang, C. de la Cruz, S. Li, J. W. Lynn, Y. Chen, M.A. Green, G.F. Chen, G.Li, Z. Li, J.L. Luo, N.L.Wang, and P. Dai, Nature Mater. 7(2008)953.

[90] Y. Mizuguchi and Y. Takano, J. Phys. Soc. Jpn. 79, 102001 (2010) and references therein.

[91] K. Kuroki, H. Usui, S. Onari, R. Arita, and H. Aoki, Phys. Rev. B 79(2009) 224511.

[92] T. Miyake, T. Kosugi, S. Ishibashi, and K. Terakura, J. Phys. Soc. Jpn. 79, (2010)123713.

[93] O.K. Andersen and L. Boeri, Annalen der Physik, 523(2011)8.

[94] Z. P. Yin, K. Haule and G. Kotliar, Nat. Mater. 10(2011) 932.

[95] H. Usui and K. Kuroki, Phys. Rev. B 84(2011) 024505.

[96] H. Usui, K. Suzuki, and K. Kuroki, Supercond. Sci.Technol.25(2012) 084004.

[97] H. Usui, K. Suzuki, K. Kuroki, N. Takeshita, P. M. Shirage, H. Eisaki,

and A. Iyo, Phys. Rev. B 87(2013)174528.

[98] C. Heil, L. Boeri, H. Sormann, W. von der Linden, M. Aichhorn, Phys. Rev. B 90(2014) 165122.

[99] K.Nakamura, R.Arita, and M.Imada, J. Phys. Soc. Jpn. 77(2008) 093711.

[100]T. Miyake, K. Nakamura, R. Arita, and M. Imada, J. Phys. Soc. Jpn. 79(2010) 044705.

[101] T. Misawa and M. Imada, Nat.Commun.5(2014)5738.

[102] K. Haule and G. Kotliar, New J. Phys.11(2009) 025021.

[103] A. Georges, L. de' Medici, J. Mravlje Annual Reviews of Condensed Matter Physics 4(2013) 137.

[104] W.-G. Yin, C.-C. Lee, and W. Ku,  Phys. Rev. Lett. 105(2010)107004.

[105] R. M. Fernandes, L. H. VanBebber, S. Bhattacharya, P. Chandra, V. Keppens, D. Mandrus, M. A. McGuire, B. C. Sales, A. S. Sefat, and J. Schmalian, Phys. Rev. Lett. 105(2010)157003.

[106]I. Eremin, J. Knolle, R.M. Fernandes, J. Schmalian, A.V. Chubukov , J. Phys. Soc. Jpn. 83 (2014) 061015.

[107]R. M. Fernandes, A. V. Chubukov, J. Schmalian, Nature Phys. 10(2014)97.

[108]F. Kruger, S. Kumar, J. Zaanen, J. van den Brink, Phys. Rev. B 79, (2009)054504.

[109]  W. Lv, F. Kr̈uger, and P. Phillips, Phys. Rev. B 82(2010) 045125.

[110]  C.-C. Chen, J. Maciejko, A. P. Sorini, B. Moritz, R. R. P. Singh, and] T.P. Devereaux, Phys. Rev. B 82(2010) 100504.

[111]  T. Yamada, J. Ishizuka, Y. Ōno, J. Phys. Soc. Jpn. 83(2014) 043704.

[112]  .S.Onari and H. Kontani, Phys. Rev. Lett 109(2012)137001.

[113]  .C.-C. Lee, W.-G. Yin, W. Ku,  Phys. Rev. Lett.103(2009)267001.

[114]  T. Shimojima, K. Ishizaka, Y. Ishida, N. Katayama, K. Ohgushi, T. Kiss, M. Okawa, T. Togashi, X. -Y. Wang, C. -T. Chen, S. Watanabe, R. Kadota, T. Oguchi, A. Chainani, S. Shin, Phys. Rev. Lett.104(2010) 057002.

[115]  M. Yi, D. H. Lu, J.-H. Chu, J. G. Analytis, A. P. Sorini, A. F. Kemper, B. Moritz, S.-K. Mo, R. G. Moore, M. Hashimoto, W. S. Lee, Z. Hussain, T. P. Devereaux, I. R. Fisher, Z.-X. Shen, PNAS 108(2011)6878.

[116]  A. E. Böhmer, T. Arai, F. Hardy, T. Hattori, T. Iye, T. Wolf, H. v. Löhneysen, K. Ishida, C. Meingast, arXiv:1407.5497.

[117]   Q. Zhang, R. M. Fernandes, J. Lamsal, J. Yan, S. Chi, G. S. Tucker, D. K. Pratt, J. W. Lynn, R. W. McCallum, P. C. Canfield, T. A. Lograsso, Alan I. Goldman, D. Vaknin, R. J. McQueeney, arXiv:1410.6855.

[118]   I.I. Mazin, D.J. Singh, M.D. Johannes, and M.H. Du, Phys. Rev. Lett. 101(2008) 057003.

[119]   H. Ikeda, J. Phys. Soc. Jpn. 77(2008) 123707.

[120]   R.Thomale, C. Platt, W. Hanke, B. A. Bernevig, Phys. Rev. Lett. 106(2011)187003.

[121]   F. Wang, H. Zhai, Y. Ran, A. Vishwanath, and D.-H. Lee, Phys. Rev. Lett. 102(2009) 47005.

[122]   F. Wang, H. Zhai, and D.-H. Lee, Phys. Rev. B 81(2010)184512.

[123]   T. Nomura, J. Phys. Soc. Jpn. 77(2008) Suppl. C 123.

[124]   F. Yang, H. Zhai, F. Wang, D.-H. Lee, Phys. Rev. B 83(2011)134502 .

[125]   H. Miyahara, R. Arita, H. Ikeda, Phys. Rev. B 87(2013) 045113.

[126]   A.G. Aronov and E.B. Sonin, Zh. Eksp. Teor. Fiz. 63(1972)1059 [Sov. Phys. -JETP 36(1973) 556.

[127]   A.I.Liechtenstein, I.I. Mazin, and O.K. Andersen, Phys. Rev. Lett. 74, (1995)2303.

[128]   K. Kuroki and R. Arita, Phys. Rev. B 64(2001) 024501.

[129]   N. Bulut, D.J. Scalapino, and R.T. Scalletar, Phys. Rev. B 45(1992)5577.

[130]   T. A. Maier and D. J. Scalapino, Phys. Rev. B 84(2013)180513(R).

[131]   H. Kontani and S. Onari, Phys. Rev. Lett. 104(2010) 157001.

[132]   Y. Yanagi, Y. Yamakawa, and Y. Ono, Phys. Rev. B 81(2010) 054518.

[133]   Y. Yanagi, Y. Yamakawa, N. Adachi, and Y. Ono, Phys. Rev. B 82(2011) 064518.

[134]   Y. Nomura, K. Nakamura, R. Arita, Phys. Rev. Lett. 112(2014) 027002.

[135]   H. Yamase and R. Zeyher , Phys.ReV.B 88(2013)180502(R).

[136]   Q. Si and E. Abrahams, Phys. Rev. Lett. 101(2008) 076401.

[137]   K. Seo, B. A. Bernevig, and J. Hu, Phys. Rev. Lett. 101(2008) 206404.

[138]   R. Yu, J.-X. Zhu, Q. Si, Phys. Rev. B 89(2014) 024509.

[139]   R. Yu and A. H. Nevidomskyy, arXiv:1404.1307.

[140]   J.C. Davis and D.-H. Lee, Proc.Nat.Acad.Sci. 110(2013) 17623.

[141]   Y. Suzumura, Y. Hasegawa, and H. Fukuyama, J. Phys. Soc. Jpn. 57(1988)2768.

[142]   G. Kotliar, Phys. Rev. B 37(1988) 3664.

[143]   H. Sakakibara, H. Usui, K. Kuroki, R. Arita, and H. Aoki, Phys. Rev. Lett. 105(2010) 057003.

[144]   H. Usui, K. Suzuki, and K. Kuroki, arXiv:1501.06303.

[145]   S. Miyasaka, A.Takemori, T. Kobayashi, S. Suzuki, S. Saijo, S. Tajima, J. Phys. Soc. Jpn. 82(2013) 124706.

[146]   K. T. Lai, A. Takemori, S. Miyasaka, F. Engetsu, H. Mukuda, S. Tajima, Phys. Rev. B 90(2014) 064504.

[147]   H. Mukuda, F. Engetsu, K. Yamamoto, K. T. Lai, M. Yashima, Y. Kitaoka, A. Takemori, S. Miyasaka, S. Tajima, Phys. Rev. B89(2014)064511.

[148]   S. Iimura, S. Matsuishi, H. Sato, T. Hanna, Y. Muraba, S.W. Kim, J. E. Kim, M. Takata and H. Hosono, Nat. Commun. 3(2012)943.

[149]   S. Iimura, S. Matsuishi, M. Miyakawa, T. Taniguchi, K. Suzuki, H. Usui, K. Kuroki, R. Kajimoto, M. Nakamura, Y. Inamura, K. Ikeuchi, S. Ji, and H. Hosono, Phys. Rev. B88(2013) 060501(R).

[150]   K. Suzuki, H. Usui, K. Kuroki, S. Iimura, Y. Sato, S.Matsuishi, and H. Hosono, J. Phys. Soc. Jpn. 82(2013) 083702.

[151]   K. Suzuki, H.Usui, S. Iimura, Y. Sato, S. Matsuishi, H. Hosono, and K. Kuroki   Phys. Rev. Lett. 113 (2014)027002.

[152]   T. Imai, K. Ahilan, F.L. Ning, T.M.McQueen and R.J. Cava, Phys. Rev. Lett. 102(2009) 177005.

[153]   F. L. Ning, K. Ahilan, T. Imai, A.S. Sefat, M.A. McGuire, B.C. Sales, D. Mandrus, P. Cheng, B. Shen, and H.-H.Wen, Phys. Rev. Lett. 104, (2010)037001.

[154]   Y. Nakai, T. Iye, S. Kitagawa, K. Ishida, H. Ikeda, S. Kasahara, H. Shishido, T. Shibauchi, Y. Matsuda, and T. Terashima, Phys. Rev. Lett. 105(2010)107003.

[155]   D. Guterding, H.O. Jeschke, P. J. Hirschfeld, and Roser Valenti, Phys. Rev. B 91()2015) 041112 (R).

[156]   Y. Yamakawa, S. Onari, H. Kontani, N. Fujiwara, S. Iimura, and H. Hosono, Phys. Rev. B 88 (2013)041106.

[157]   S. Onari, Y. Yamakawa, and H. Kontani, Phys. Rev. Lett. 112 (2014)187001.

[158]   Y. Bang, New J. Phys. 16 (2014)023029.

[159]   K. Kuroki and K. Suzuki, unpublished.

[160]   It should be noted that the vacancy ordering and the mixture of

various structural phases in this material makes the analysis of superconductivity difficult, see, e.g., E.Dagotto, Rev. Mod. Phys. 85(2013)849, or X. Ding, D. Fang, Z. Wang, H. Yang, J. Liu, Q. Deng, G. Ma, C. Meng, Y. Hu, H.-H.Wen, Nat. Comm. 4, (2013) 1897 and references therein.

[161]   T. Qian, X.-P. Wang, W.-C. Jin, P. Zhang, P. Richard, G. Xu, X. Dai, Z. Fang, J.-G. Guo, X.-L. Chen, H. Ding, Phys. Rev. Lett. 106(2011)187001.

[162]   Y. Zhang, L. X. Yang, M. Xu, Z. R. Ye, F. Chen, C. He, J. Jiang, B. P. Xie, J. J. Ying, X. F. Wang, X. H. Chen, J. P. Hu, D. L. Feng, Nature Mat. 10(2011)273.

[163]   D. Mou, S.Liu, X. Jia, J. He, Y Peng, L. Zhao, L. Yu, G.Liu, S.He, X. Dong, J.Zhang, H.Wang, C.Dong, M. Fang, X.Wang, Q.Peng, Z.Wang, S.Zhang, F.Yang, Z.Xu, C.Chen, and X. J. Zhou, Phys. Rev. Lett. 106(2011)107001.

[164]   M. Yi, D. H. Lu, R. Yu, S. C. Riggs, J.-H. Chu, B. Lv, Z. K. Liu, M. Lu, Y.-T. Cui, M. Hashimoto, S.-K. Mo, Z. Hussain, C. W. Chu, I. R. Fisher, Q. Si, and Z.-X. Shen, Phys. Rev. Lett. 110(2013) 067003.

[165]   T.A. Maier, S. Graser, P.J. Hirschfeld, and D.J. Scalapino, Phys. Rev. B 83(2011) 100515.

[166]   F. Wang, F. Yang, M. Gao, Z.-Yi Lu, T. Xiang, and D.-H. Lee, Europhys. Lett. 93(2011) 57003..

[167]   I.I. Mazin, Phys. Rev. B 84(2011) 024529.

[168]   M. Khodas and  A. V. Chubukov,   Phys. Rev. Lett. 108(2012) 247003.

[169]   A. Kreisel, Y. Wang, T.A. Maier, P.J. Hirschfeld, and D.J. Scalapino, Phys. Rev. B 88(2013) 094522.

[170]   C. Fang, Y.-L. Wu, R. Thomale, B.A. Bernevig, and J. Hu,   Phys. Rev. X 1(2011) 011009.

[171]   F. Yang, F. Wang, & D. -H. Lee, Phys. Rev. B 88(2013) 100504 (2013).

[172]   R. Yu, P. Goswami, Q.Si, P. Nikolic and J. -X. Zhu,   Nat. Comm. 4, (2013)2783.

[173]   T. Saito, S. Onari and H. Kontani, Phys. Rev. B 83(2011)140512.

[174]   M. Sunagawa, K. Terashima, T. Hamada, H. Fujiwara, M. Tanaka, H. Takeya, Y. Takano, M. Arita, K. Shimada, H. Namatame, M. Taniguchi, K. Suzuki, H. Usui, K. Kuroki, T. Wakita, Y. Muraoka, T. Yokoya, unpublished.

[175]   K. Suzuki, H. Usui, and K. Kuroki, Phys. Rev. B 84(2011)144514 .

[176]   H. Ding, P. Richard, K. Nakayama, T. Sugawara, T. Arakane, Y. Sekiba, A. Takayama,S. Souma, T. Sato, T. Takahashi, Z. Wang, X. Dai, Z. Fang, G. F. Chen, J. L. Luo, N. L. Wang, Europhys. Lett. 83(2008)47001.

[177]   K. Nakayama, T. Sato, P. Richard, Y.-M. Xu, Y. Sekiba, S. Souma, G.F. Chen, J.L. Luo,N.L.Wang, H. Ding, and T. Takahashi, Europhys. Lett. 85 (2009)67002.

[178]   K. Hashimoto, T. Shibauchi, T. Kato, K. Ikada, R. Okazaki, H. Shishido, M. Ishikado, H. Kito, A. Iyo, H. Eisaki, S. Shamoto, and Y. Matsuda, Phys. Rev. Lett. 102 (2009)017002.

[179]   H. Luetkens, H.-H. Klauss, R. Khasanov, A. Amato, R. Klingeler, K. Hellmann, N.Leps, A. Kondrat, C. Hess, A. Kohler, G. Behr, J. Werner, and B. Buchner, Phys. Rev.Lett. 101(2008) 097009.

[180]   A. A. Aczel, E. Baggio-Saitovitch, S. L. Budko, P.C. Canfield, J. P. Carlo, G.F. Chen, P. Dai, T. Goko, W. Z. Hu, G. M. Luke, J. L. Luo, N. Ni, D. R. Sanchez-Candela, F. F. Tafti, N. L.Wang, T. J.Williams,W. Yu, and Y. J. Uemura, Phys. Rev. B 78(2008)214503.

[181]   T. Goko, A. A. Aczel, E. Baggio-Saitovitch, S. L. Bud'ko, P. C. Canfield, J. P. Carlo,G. F. Chen, P. Dai, A. C. Hamann, W. Z. Hu, H. Kageyama, G. M. Luke, J. L. Luo, B. Nachumi, N. Ni, D. Reznik, D. R. Sanchez-Candela, A. T. Savici, K. J. Sikes, N. L.Wang, C. R.Wiebe, T. J.Williams, T. Yamamoto, W. Yu, and Y. J. Uemura, Phys. Rev. B 80(2009) 024508.

[182]   M. Hiraishi, R. Kadono, S. Takeshita, M. Miyazaki, A. Koda, H. Okabe, and J. Akimitsu, J. Phys. Soc. Jpn. 78(2009) 023710.

[183]   J.D. Fletcher, A. Serafin, L. Malone, J. Analytis, J-H Chu, A.S. Erickson, I.R. Fisher, and A. Carrington, Phys. Rev. Lett. 102(2009)147001.

[184]   C.W.Hicks, T.M. Lippman, M.E. Huber, J.G. Analytis, J.-H. Chu, A.S. Erickson, I.R.Fisher, and K.A.Moler, Phys. Rev. Lett.103(2009)127003.

[185]   K. Hashimoto, M. Yamashita, S. Kasahara, Y. Senshu, N. Nakata, S. Tonegawa, K.Ikada, A. Serafin, A. Carrington, T. Terashima, H. Ikeda, T. Shibauchi and Y. Matsuda, Phys. Rev. B 81(2010)220501(R).

[186]   Y. Nakai, T. Iye, S. Kitagawa, K. Ishida, H. Ikeda, S. Kasahara, H. Shishido, T. Shibauchi, Y. Matsuda and T. Terashima, Phys. Rev. B81,

(2010)020503(R).

[187]  Y. Zhang, Z. R. Ye, Q. Q. Ge, F. Chen, Juan Jiang, M. Xu, B. P. Xie and D. L. Feng, Nature Physics 8(2012)371.

[188]  K. Suzuki, H. Usui, and K. Kuroki, J. Phys. Soc. Jpn. 80 (2011) 013710.

[189]  S. Graser, A. F. Kemper, T. A. Maier, H.-P. Cheng, P. J. Hirschfeld, and D. J. Scalapino, Phys. Rev. B 81(2010) 214503.

[190]  A.V. Chubukov, M.G.Vavilov, and A.B. Vorontsov, Phys. Rev. B80(2009)140515(R).

[191]  T. Saito, S. Onari, H. Kontani, Phys. Rev. B 88(2013)045115.

[192]  G. Friemel, J. T. Park, T. A. Maier, V. Tsurkan, Yuan Li, J. Deisenhofer, H.-A. Krug von Nidda, A. Loidl, A. Ivanov, B. Keimer, and D. S. Inosov, Phys. Rev. B 85(2012)140511.

[193]  Y. Wang, A. Kreisel, V. B. Zabolotnyy, S. V. Borisenko, B. B̈uchner, T. A. Maier, P. J. Hirschfeld, and D. J.Scalapino, Phys. Rev. B 88(2013)174516.

[194]  S.V. Borisenko, V.B. Zabolotnyy, A.A. Kordyuk, D.V. Evtushinsky, T.K. Kim, I.V. Morozov, R. Follath and B.B̈uchner, Symmetry 4(2012)251.

[195]  K. Umezawa, Y. Li, H. Miao, K. Nakayama, Z.-H. Liu, P. Richard, T. Sato, J. B. He, D.-M. Wang, G. F. Chen, H. Ding, T. Takahashi, and S.-C. Wang, Phys. Rev. Lett.108(2012) 037002.

[196]  M. P. Allan, A. W. Rost, A. P. Mackenzie, Yang Xie, J.C. Davis, K. Kihou, C. H. Lee, A. Iyo, H. Eisaki, and T.-M. Chuang, Science 336(2012) 563.

[197]  F. Ahn, I. Eremin, J. Knolle, V.B. Zabolotnyy, S.V. Borisenko, B. B̈uchner, A.V. Chubukov, Phys. Rev. B 89,(2014)144513.

[198]  Z. P. Yin, K. Haule, G. Kotliar, Nat. Phys. 10(2014) 845.

[199]  T. Saito, S. Onari, Y. Yamakawa, H. Kontani, Sergey V. Borisenko, and Volodymyr B. Zabolotnyy, Phys. Rev. B 90(2014) 035102.

[200]  K. Suzuki and K. Kuroki, unpublished.

[201]  R. Thomale, C. Platt, W. Hanke, J. Hu, and B. A. Bernevig, Phys. Rev. Lett. 107(2011)117001.

[202]  S. Maiti, M. M. Korshunov, T. A. Maier, P. J. Hirschfeld, A. V. Chubukov,  Phys. Rev. Lett. 107(2011) 147002.

[203]  S. Maiti, M.M. Korshunov, and A.V. Chubukov, Phys. Rev. B 85(2012) 014511.

[204]  K. Okazaki, Y. Ota, Y. Kotani, W. Malaeb, Y. Ishida, T. Shimojima, T. Kiss, S. Watanabe, C.-T. Chen, K. Kihou, C. H. Lee, A. Iyo, H. Eisaki, T. Saito, H. Fukazawa, Y. Kohori, K. Hashimoto, T. Shibauchi, Y. Matsuda, H. Ikeda, H. Miyahara, R. Arita, A. Chainani, S. Shin  Science 337(2012)1314.

[205]  F. F. Tafti, A. Juneau-Fecteau, M.-` E. Delage, S. R. de Cotret, J.-P. Reid, A. F. Wang, X.-G. Luo, X. H. Chen, N. Doiron-Leyraud, and L. Taillefer, Nature Phys. 9(2013)349.

[206]  T. Terashima, K. Kihou, K. Sugii, N. Kikugawa, T. Matsumoto, S. Ishida, C.H. Lee, A. Iyo, H. Eisaki, and S. Uji,  Phys. Rev. B 89(2014) 134520.

[207]  M. Sato, Y. Kobayashi, S. C. Lee, H. Takahashi, E. Satomi, and Y. Miura, J. Phys. Soc. Jpn. 79 (2010) 014710.

[208]  C. Tarantini, M. Putti, A. Gurevich, Y. Shen, R.K. Singh, J. M. Rowell, N. Newman, D.C. Larbalestier, P. Cheng, Y. Jia, and H.-H.Wen, Phys. Rev. Lett. 104 (2010)087002.

[209]  Y. Nakajima, T. Taen, Y. Tsuchiya, T. Tamegai, H. Kitamura, and T. Murakami, Phys. Rev. B 82(2010)220504.

[210]  S. Onari and H. Kontani, Phys. Rev. Lett. 103 (2009) 177001.

[211]  Y. Mizukami, M. Konczykowski, Y. Kawamoto, S. Kurata, S. Kasahara, K. Hashimoto, V. Mishra, A. Kreisel, Y. Wang, P. J. Hirschfeld, Y. Matsuda, and T. Shibauchi, Nat. Comm. 5(2014)5657.

[212]  M. M. Korshunov and I. Eremin, Phys. Rev. B 78(2008)140509(R).

[213]  T.A. Maier and D. J. Scalapino, Phys. Rev. B 78(2008) 020514(R).

[214]  A. D. Christianson, E. A. Goremychkin, R. Osborn, S. Rosenkranz, M. D. Lumsden, C.D. Malliakas, I. S. Todorov, H. Claus, D. Y. Chung, M. G. Kanatzidis, R. I. Bewley and T.Guidi, Nature 456(2008)930.

[215]  D. S. Inosov, J. T. Park, P. Bourges, D. L. Sun, Y. Sidis, A. Schneidewind, K. Hradil, D.Haug, C. T. Lin, B. Keimer and V. Hinkov, Nature Phys. 6(2010) 178.

[216]  C.-T. Chen, C.C. Tsuei, M.B. Ketchen, Z.-A. Ren and Z.X. Zhao, Nature Physics 6(2010)260.

[217]  T. Hanaguri, S. Niitaka, K. Kuroki, and H. Takagi, Science 328(2010)474.

[218]  S. Onari, H. Kontani, and M. Sato, Phys. Rev. B 81(2010) 060504(R).

[219]  Y. Nagai and K. Kuroki, Phys. Rev. B 83(2011)220516(R).

[220]   Y. Nagai and K. Kuroki, Phys. Rev. B 85(2012) 134521.

[221]   T.A. Maier, S. Graser, P.J. Hirschfeld, and D.J. Scalapino, Phys. Rev. B 83(2011) 220505(R).

[222]   T. Kariyado and M. Ogata, J. Phys. Soc. Jpn. 79(2010) 083704 .

**Figure Captions**

Fig.1 Summary of properties of layered $LaT_MOPn$ compounds.
($T_M$: 3d transition metal, Pn= P or As).

Fig.2. Parent compounds.
(A)Crystal structure. The square lattice of $Fe^{2+}$ which is tetrahedrally coordinated with pnictogen or chalcogenide anions is commonly contained as a building block. (a)122, (b)245, (c)intercalated 11 ,(d)11, (e)111, (f)1111, (g)11-derived 1111, (h)112, (i)1048, and (j) perovskite blocking layer-type.

(B) Classification of crystal structures of the representative iron-based superconductors The crystal structures of the representative iron-based superconductors are classified as 6-types by the varieties of ions and inserted structures sandwiched by the conduction layers. The A, Ae, and Ln denote alkali metal, alkaline earth, and lanthanides, respectively. The ab-planes of the 245 and 1048-systems form superstructure with a unit cell of $\sqrt{5} \times \sqrt{5}$ due to the ordered iron-vacancy or size-mismatch between Pt- and $Fe2As2$-square lattices, respectively.

Fig.3. Schematic phase diagrams of representative IBSCs.

Fig.4. Tc-domes in electron-doped LnFeAsO. Note that electron-doping using $H^-$ realizes much wider superconducting ranges using $F^-$ and two dome structure is seen in the $LaFeAsO_{1-x}H_x$ [15].

Fig.5. Tc vs. electron doping level in $Ba(Fe_{1-x}TMx)_2As_2$ (TM: transition metal) [21]. Note that the Tc may be scaled by doped electron count per Fe ions.

Fig.6. Tc vs. inter layer spacing.

Fig.7. Correlation between Tc and An-Fe-An-angle (An = pnicogen or chalcogen ion). Data of non-optimally doped samples are plotted as well.
[F7_1] Phys. Rev. B 81 (2010) 094115. [F7_2] Solid State Commun. 148 (2008) 538. [F7_3] Phys. Rev. Lett. 104 (2010) 057007. [F7_4] Phys. Rev. B 85 (2012) 224521. [F7_5] Phys. Rev. B 84 (2011) 184534. [F7_6] Phys. Rev. B 82

(2010) 024510. [F7_7] Phys. Rev. Lett. 101 (2008) 107007. [F7_8] ANGEW. CHEM. INT. ED. 47 (2008) 7949. [F7_9] J. Phys.: Condens. Matter 23 (2011) 455702. [F7_10] Y. Muraba, Private communication. [F7_11] J. Phys. Soc. Jpn. 77 (2008) 113709. [F7_12] Phys. Rev. B 89 (2014) 094501. [F7_13] J. Phys. Soc. Jpn. 83 (2014) 033705. [F7_14] Nat. Commun. 3 (2012) 943. [F7_15] Phys. Rev. B 85 (2012) 014514. [F7_16] Phys. Rev. B 84 (2011) 024521. [F7_17] J. Phys. Soc. Jpn. 82 (2013) 123702. [F7_18] J. Phys. Soc. Jpn. 80 (2011) 093704. [F7_19] Phys. Rev. B 79 (2009) 220512. [F7_20] Chin. Phys. Lett. 28 (2011) 086104. [F7_21] Nat. Mater. 12 (2013) 15. [F7_22] arXiv: 1408.2006 (2014).

Fig.8. Phase diagram of $LaFeAsO_{1-x}H_x$ system. There exit two Tc domes and two parent compounds with different nature [39].

Fig.9. Pressure dependence of Tc in $LaFeAsO_{1-x}H_x$ [42].

Fig.10. Progress in thin film fabrication and then film performance. (a) 122 system, (b) 11 system, and (c) 1111 system.

Fig.11. Critical current density (Jc) versus misorientation angle. $BaFe_2As_2$:Co (55) BGB junctions grown on [001]-tilt bicrystal substrates of MgO (indicated by open symbols) and LSAT (closed symbols) with $\theta_{GB}$ = 3°–45° [53].

Fig.12. Critical current density of various superconducting wires as a function of applied magnetic field [55].

Fig.13 Left: a typical five orbital model of the iron-based superconductor and its orbital component. Right: the corresponding Fermi surface.

Fig.14 Left: dxy portion of the band structure extracted. Right : schematic of how the band structure evolves with decreasing the Pn-Fe-Pn bond angle.

Fig.15  Debated issues regarding the pairing mechanism

Fig.16   Left:   stripe-type   antiferromagnetic   state.   Right   :
ferro-orbital-ordering state.

Fig.17. The relation among pairing interactions mediated by spin or orbital
fluctuations.

Fig.18 Left : orbital-polarization interaction that induces ferro-orbital
fluctuation. Middle : (a) the spin ($\alpha_s$) and charge-orbital ($\alpha_c$) Stoner factor
and (b) eigenvalue of the s$\pm$ pairing superconductivity plotted against the
orbital-polarization interaction. Right : the corresponding superconducting
gap function (taken from [104]).

Fig.19.   $J_1$-$J_2$ model.

Fig.20 Correspondence between real space and momentum space gap
functions in the iron-based superconductors (left) and the cuprates (right).

Fig.21 Upper panel : schematic figure of the mechanism of the $t_1$ reduction.
Lower panel : the hopping integrals $t_1$ and $t_2$ within the $d_{xy}$ orbital as
functions of the electron doping in the model of $Ln$FeAs(O,H). (taken from
[142])

Fig.22 (a) Eigenvalue of the linearized Eliashberg equation $\lambda_E$ plotted
against the hypothetically varied Pn-Fe-Pn bond angle for the model of
$Ln$Fe(As,P)(O,F). $n$ is the electron density and $n$-6 corresponds to the
electron doping rate (taken from [135]). (b) $\lambda_E$ plotted against the electron
doping rate for the model of $Ln$Fe(As,P)(O,H). $\Delta\alpha$ is the Pn-Fe-Pn bond angle
deviation with respect to that of LaFeAsO (taken from [142]). In both (a) and
(b),  $\lambda_E$ can be considered as a qualitative measure for $T_c$ ,and smaller bond
angle corresponds to smaller P/As content ratio or replacing the rare earth as
La→Nd or Sm. (c) A theoretical interpretation of the superconducting phase
diagram of the 1111 family (taken from [135]).

Fig.23 (a) Various spin-fluctuation-mediated gap functions depending on the
strength of the pairing interaction (taken from [84]). (b) A schematic figure of
s-wave   vs.   d-wave   phase   diagram   in   $K_x$Fe$_{2-y}$Se$_2$   in   the hybridization-temperature space (taken from [155]).

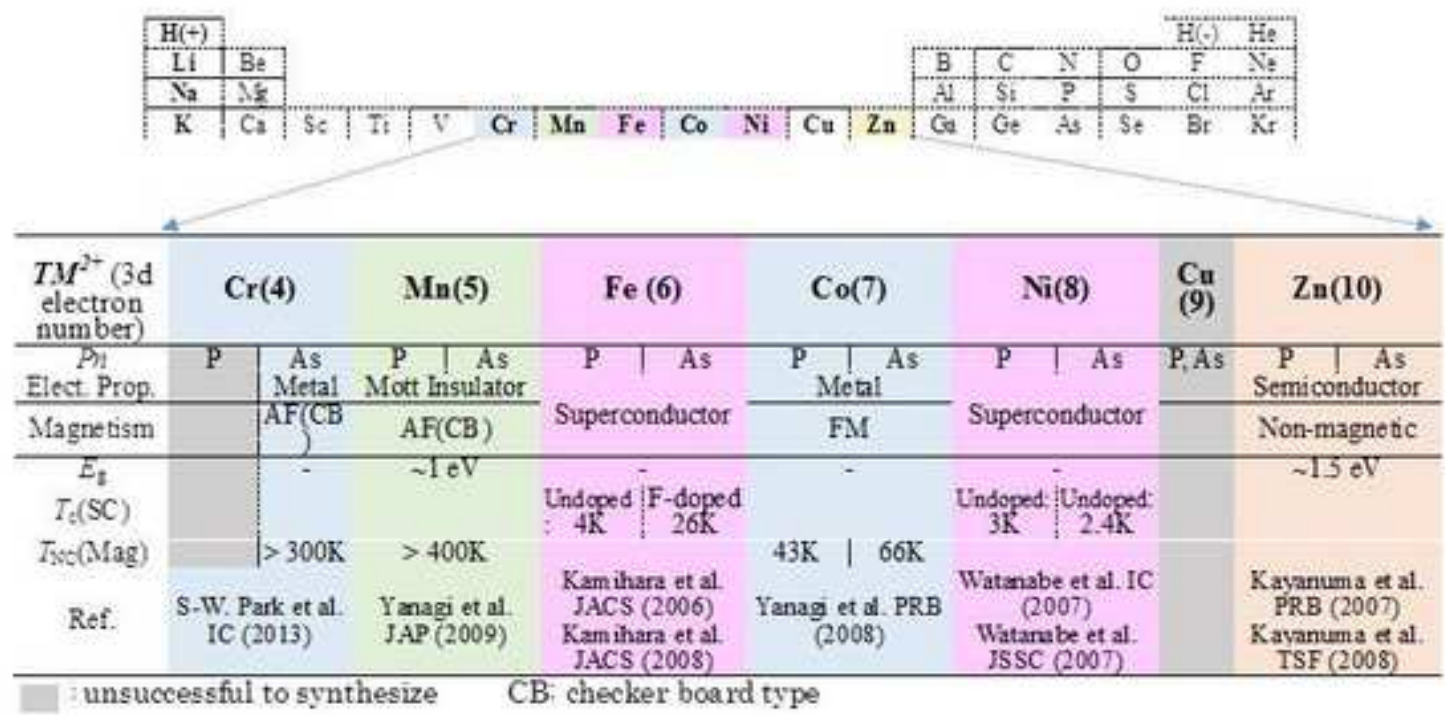

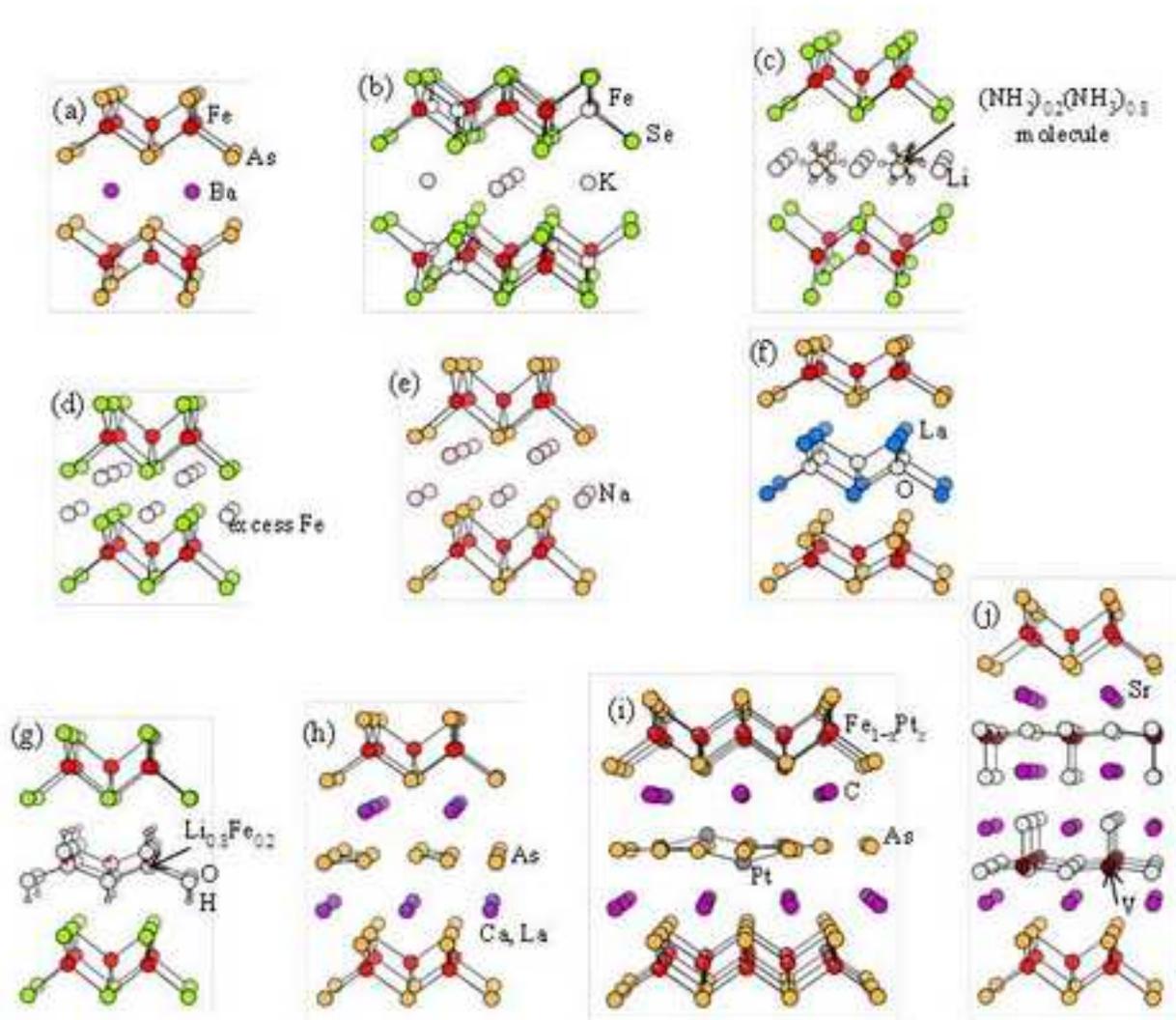

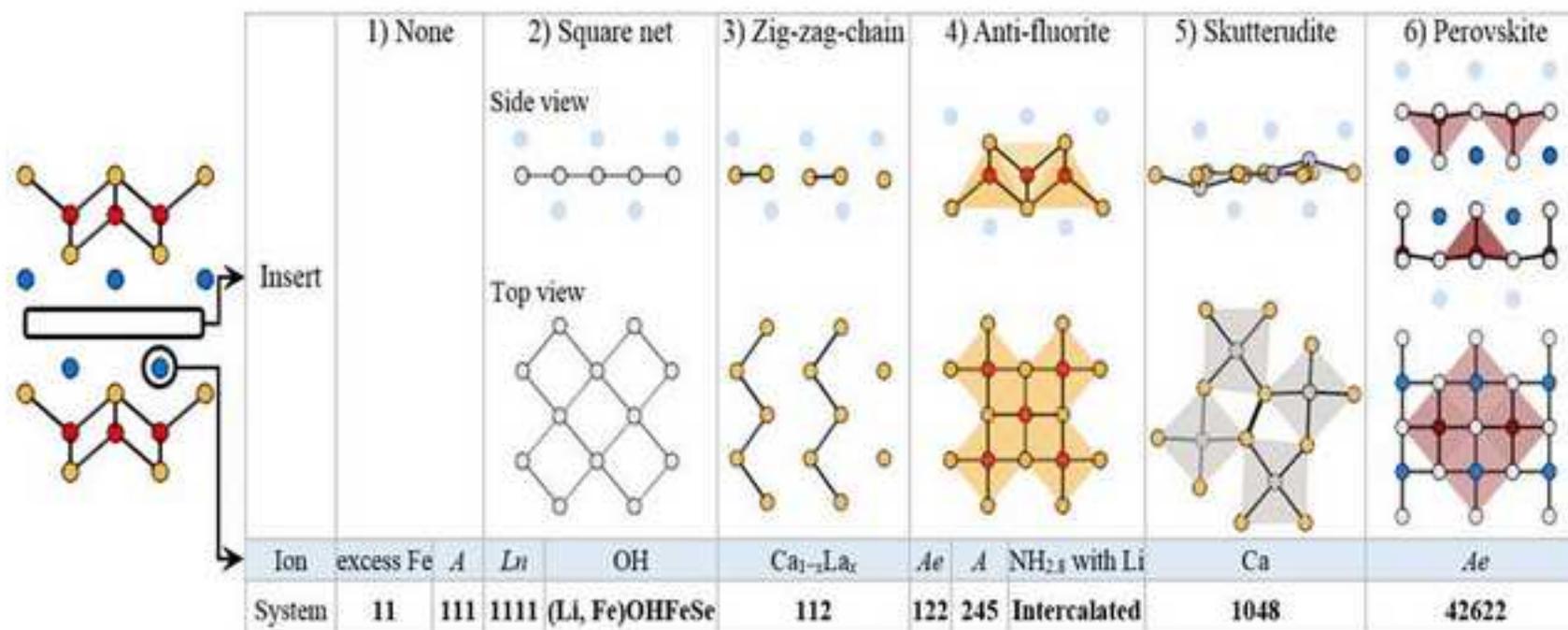

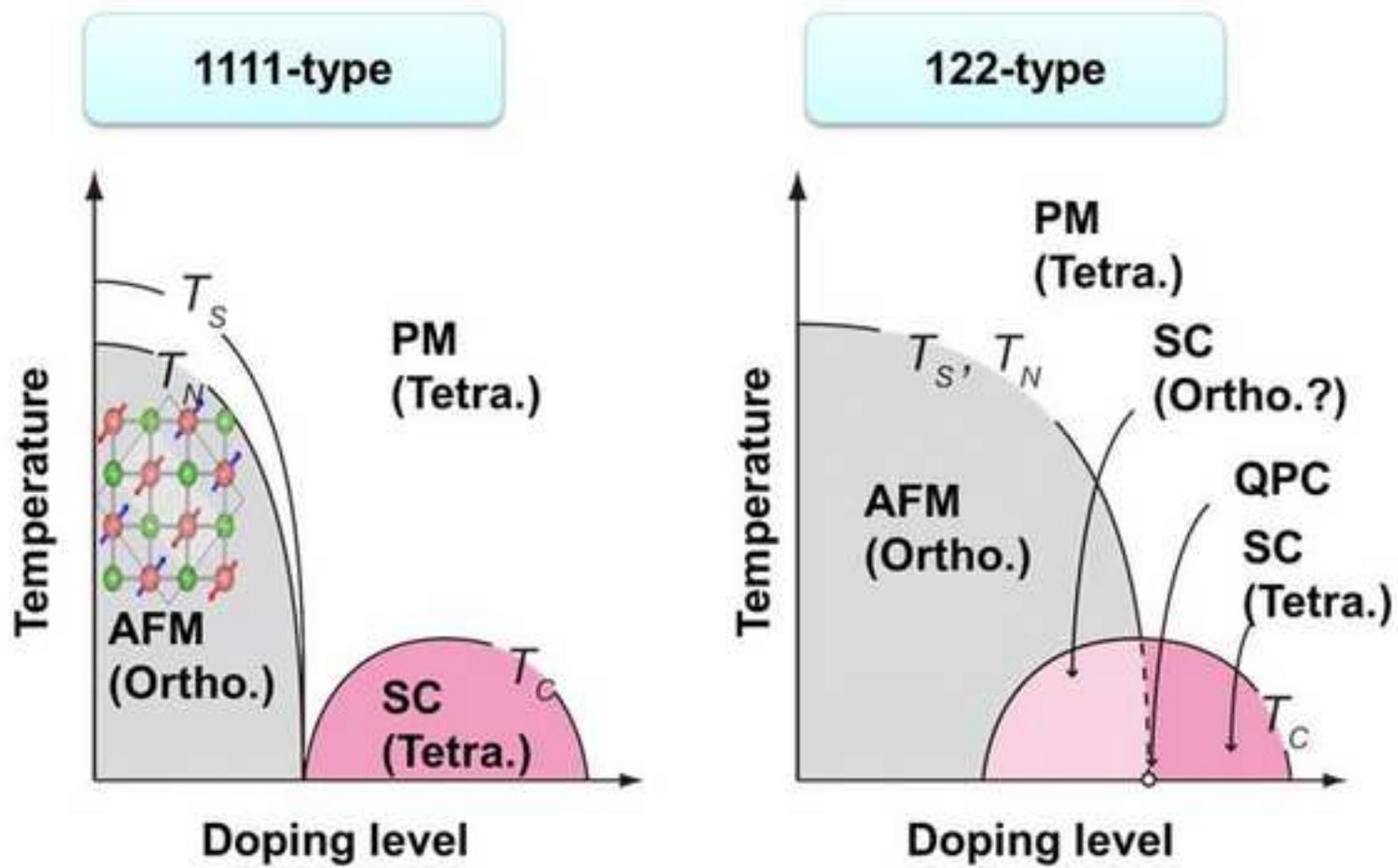

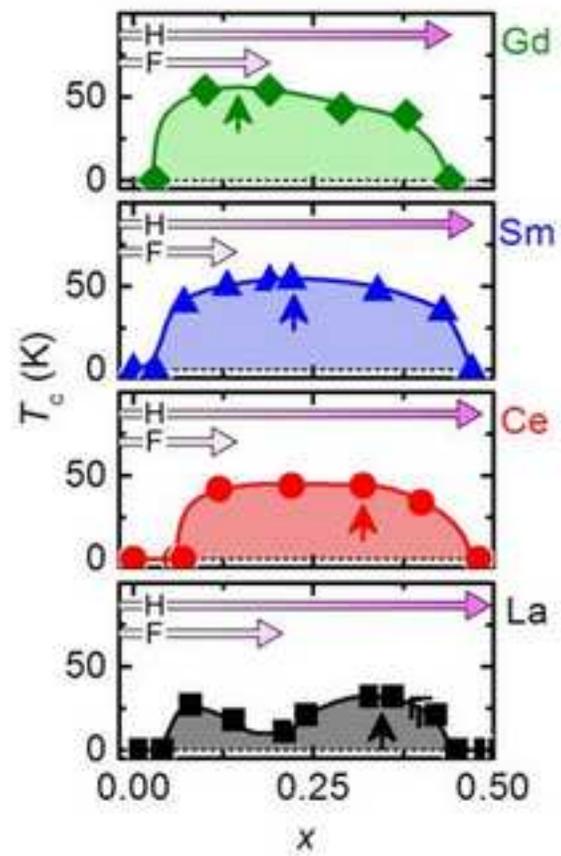

Figure with legend:
Ba(Fe$_{1-x}$Co$_x$)$_2$As$_2$
Ba(Fe$_{1-x}$Ni$_x$)$_2$As$_2$
Ba(Fe$_{1-x}$Rh$_x$)$_2$As$_2$
Ba(Fe$_{1-x}$Pd$_x$)$_2$As$_2$
Ba(Fe$_{1-x}$Cu$_x$)$_2$As$_2$
Ba(Fe$_{1-x-y}$Co$_x$Cu$_y$)$_2$As$_2$

Structural/Magnetic
R

Superconductivity
onset of R
offset of R

$T$ (K)

Excess electron number /Fe

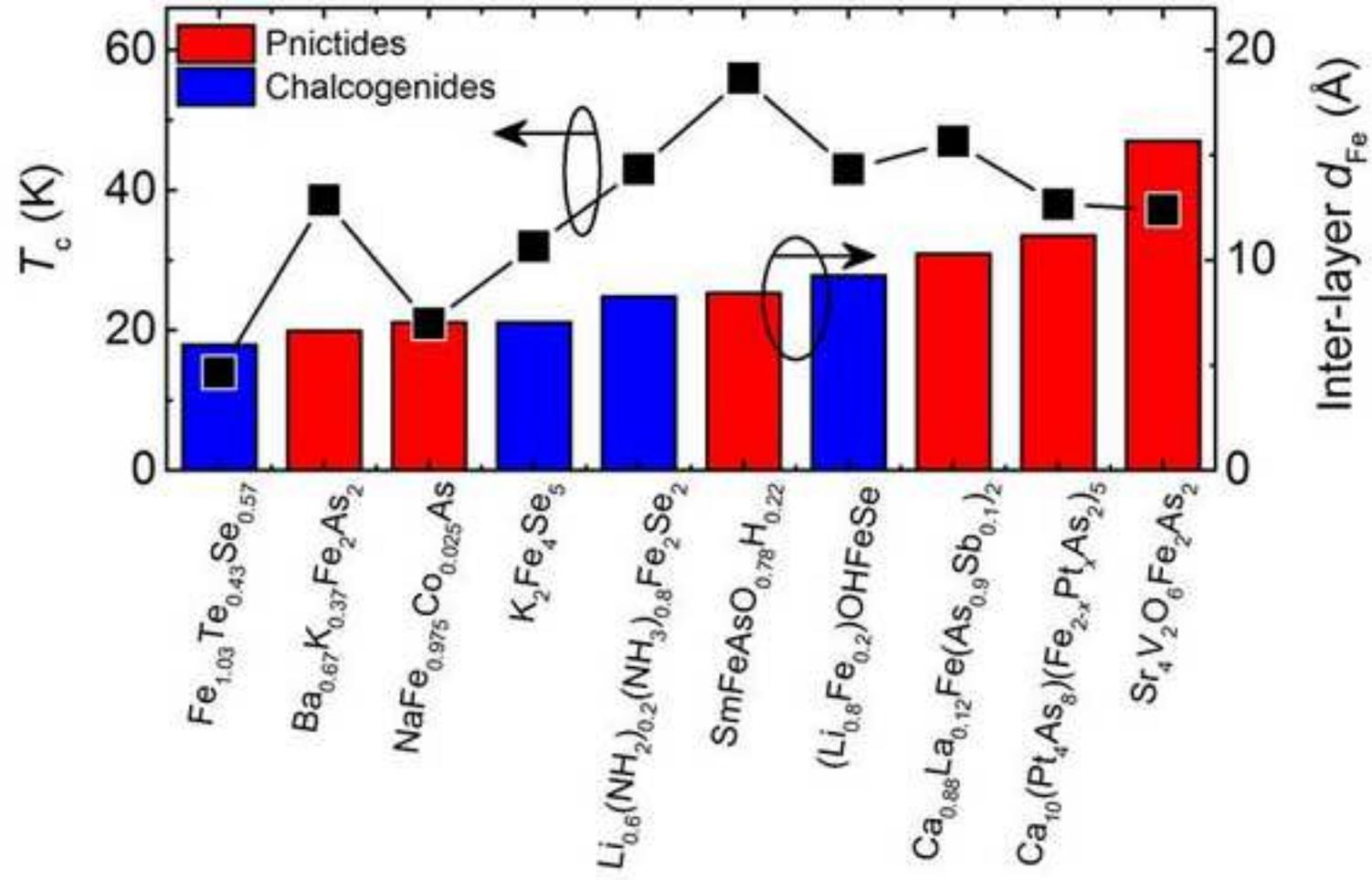

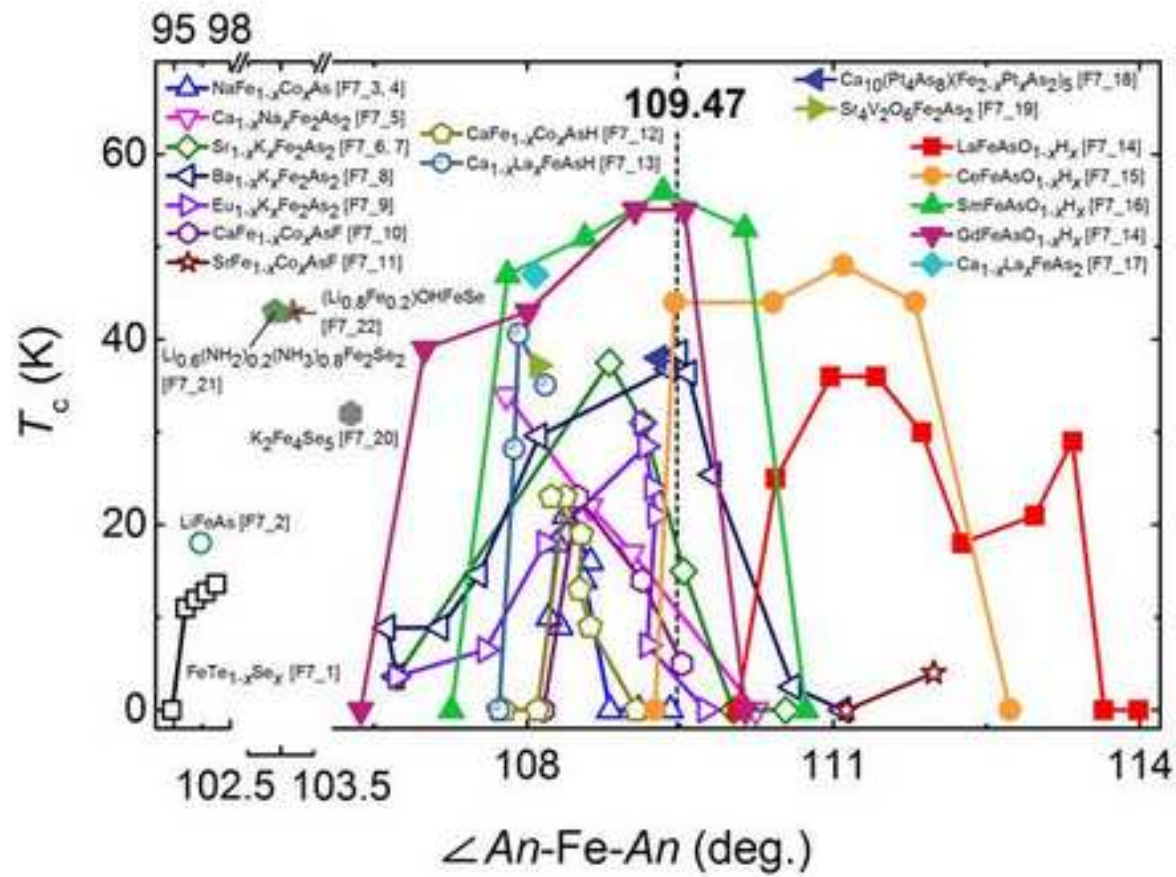

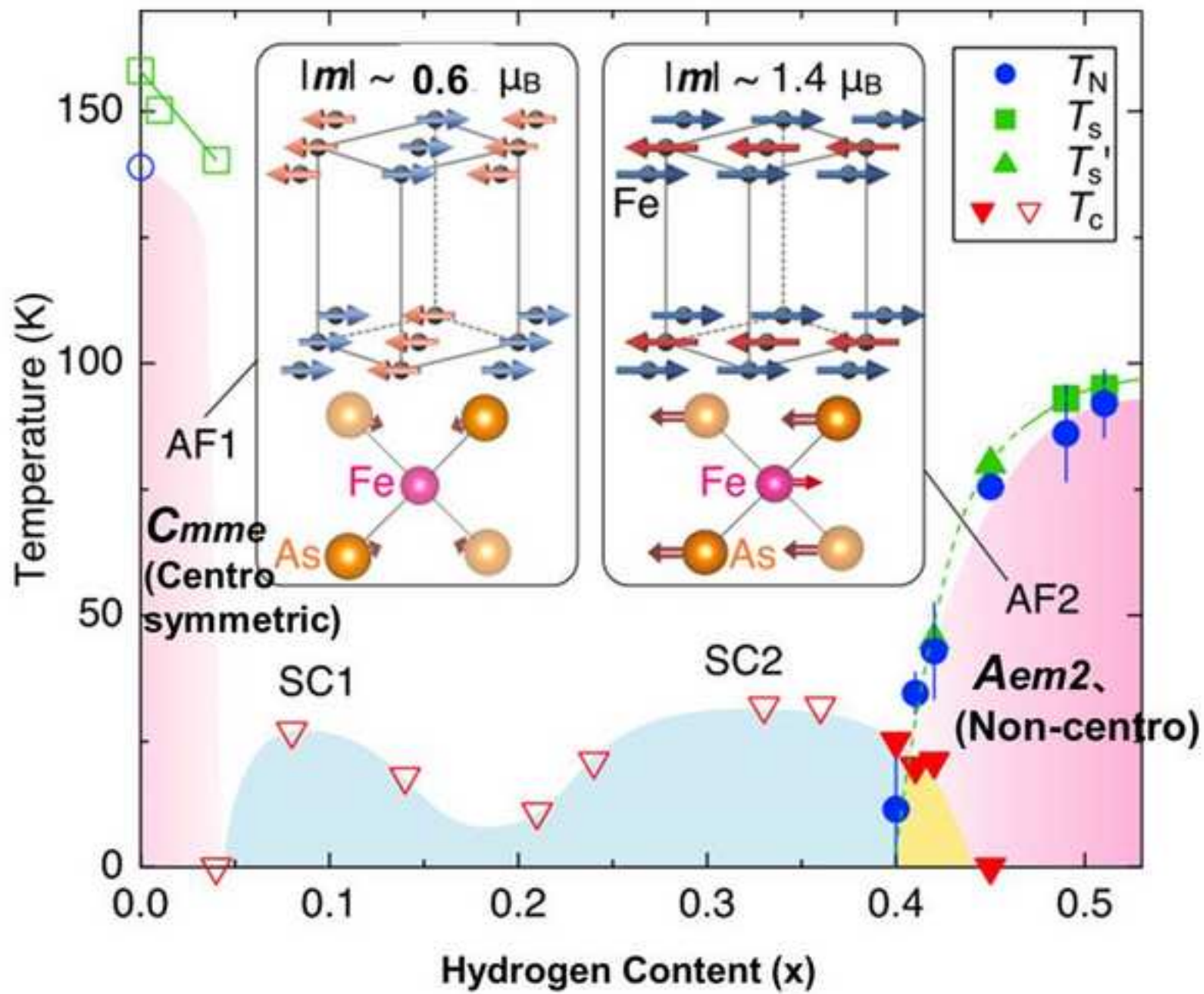

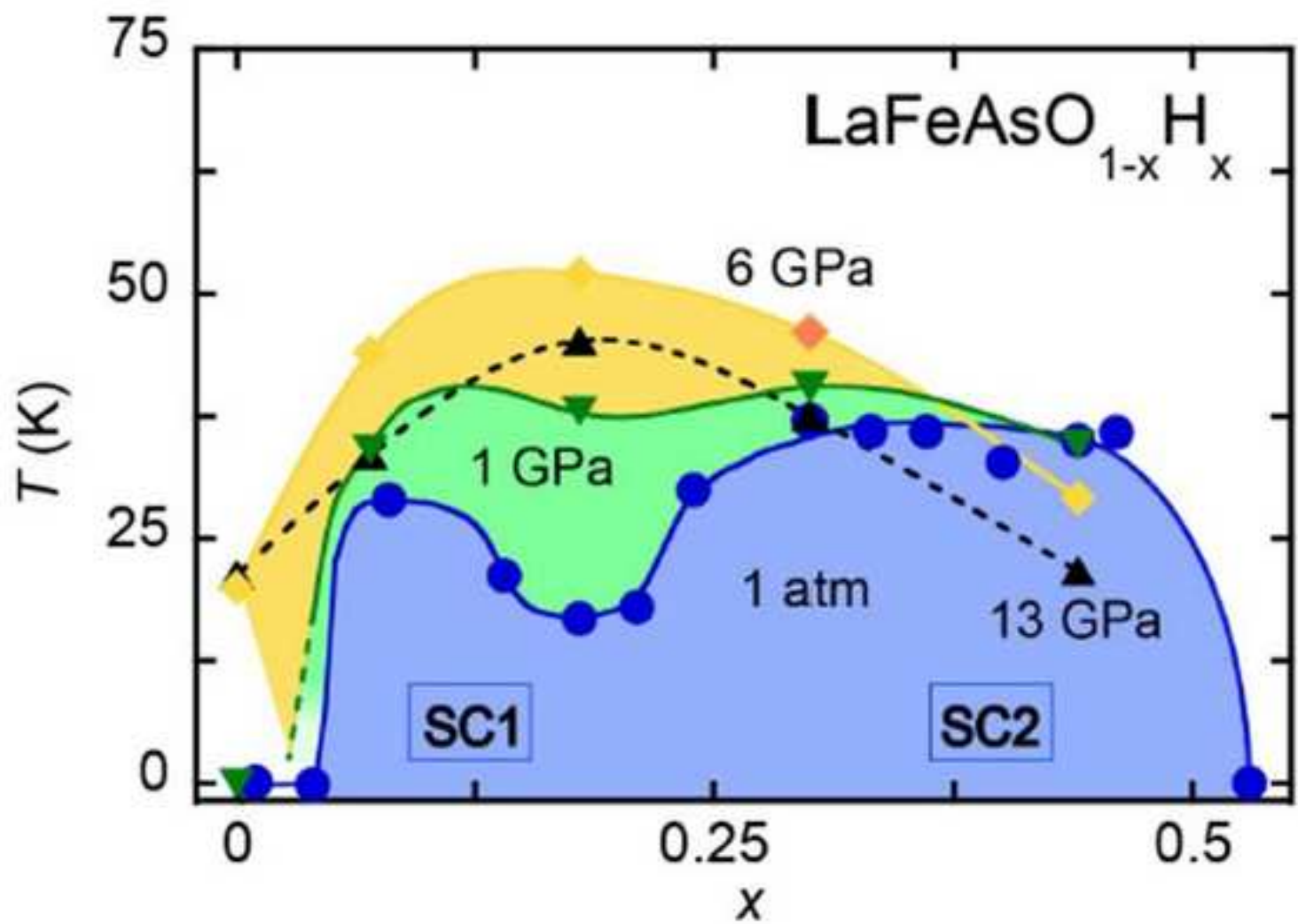

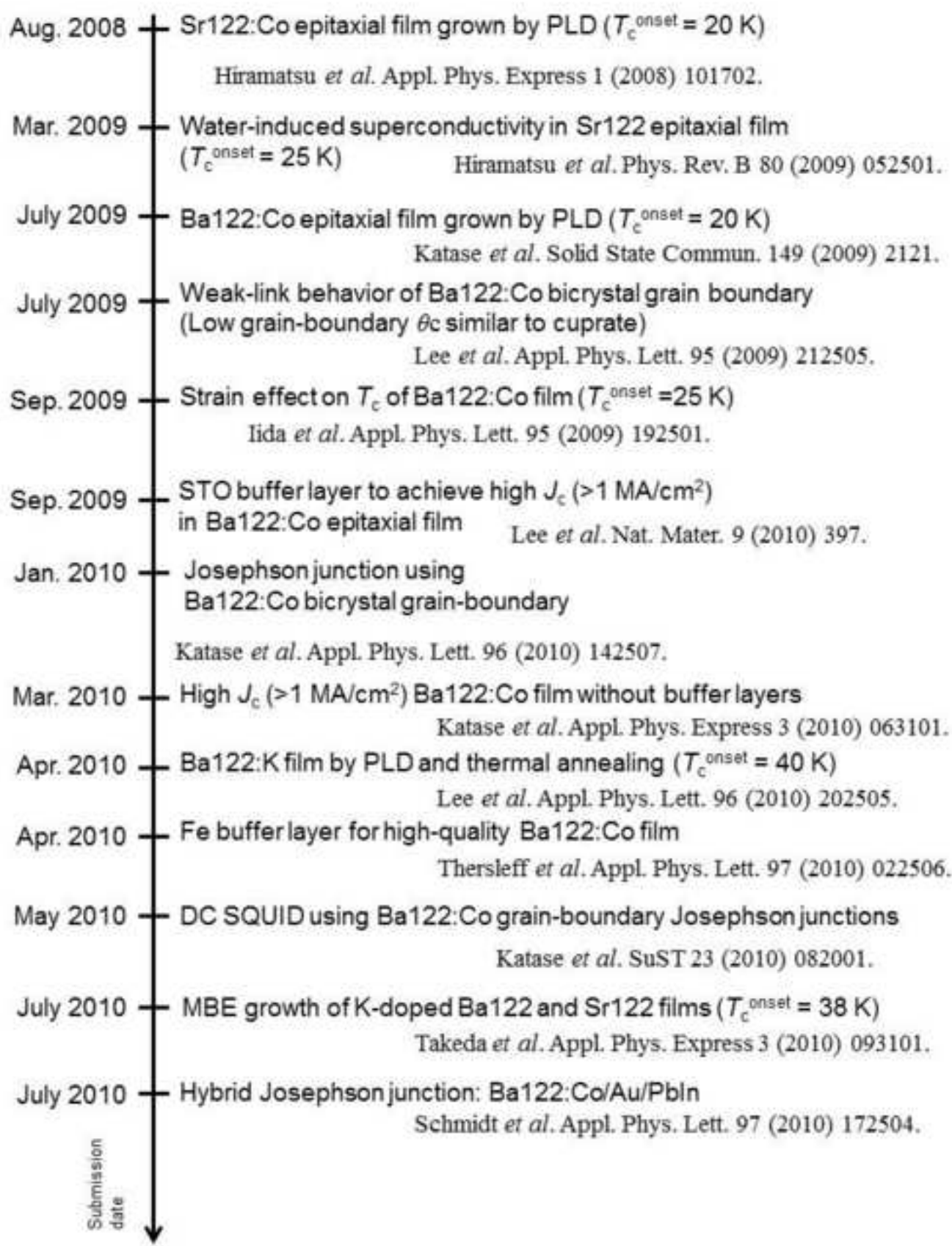

Fig 10(a-1)

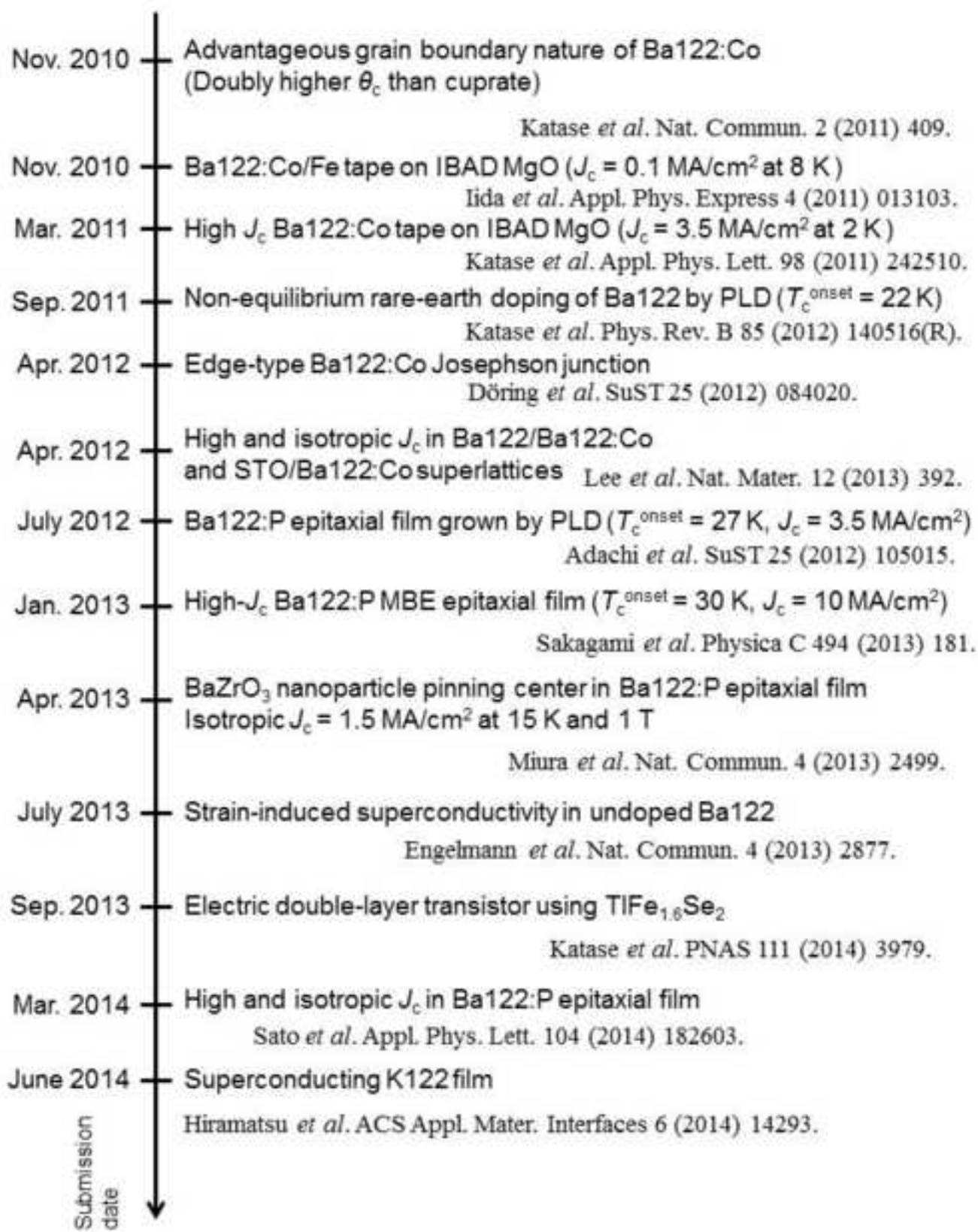

Nov. 2010 — Advantageous grain boundary nature of Ba122:Co
(Doubly higher $\theta_c$ than cuprate)

                   Katase *et al.* Nat. Commun. 2 (2011) 409.

Nov. 2010 — Ba122:Co/Fe tape on IBAD MgO ($J_c$ = 0.1 MA/cm$^2$ at 8 K )
               Iida *et al.* Appl. Phys. Express 4 (2011) 013103.

Mar. 2011 — High $J_c$ Ba122:Co tape on IBAD MgO ($J_c$ = 3.5 MA/cm$^2$ at 2 K)
               Katase *et al.* Appl. Phys. Lett. 98 (2011) 242510.

Sep. 2011 — Non-equilibrium rare-earth doping of Ba122 by PLD ($T_c^{onset}$ = 22 K)
              Katase *et al.* Phys. Rev. B 85 (2012) 140516(R).

Apr. 2012 — Edge-type Ba122:Co Josephson junction
            Döring *et al.* SuST 25 (2012) 084020.

Apr. 2012 — High and isotropic $J_c$ in Ba122/Ba122:Co
and STO/Ba122:Co superlattices   Lee *et al.* Nat. Mater. 12 (2013) 392.

July 2012 — Ba122:P epitaxial film grown by PLD ($T_c^{onset}$ = 27 K, $J_c$ = 3.5 MA/cm$^2$)
              Adachi *et al.* SuST 25 (2012) 105015.

Jan. 2013 — High-$J_c$ Ba122:P MBE epitaxial film ($T_c^{onset}$ = 30 K, $J_c$ = 10 MA/cm$^2$)
             Sakagami *et al.* Physica C 494 (2013) 181.

Apr. 2013 — BaZrO$_3$ nanoparticle pinning center in Ba122:P epitaxial film
Isotropic $J_c$ = 1.5 MA/cm$^2$ at 15 K and 1 T

             Miura *et al.* Nat. Commun. 4 (2013) 2499.

July 2013 — Strain-induced superconductivity in undoped Ba122
           Engelmann *et al.* Nat. Commun. 4 (2013) 2877.

Sep. 2013 — Electric double-layer transistor using TlFe$_{1.6}$Se$_2$

             Katase *et al.* PNAS 111 (2014) 3979.

Mar. 2014 — High and isotropic $J_c$ in Ba122:P epitaxial film
        Sato *et al.* Appl. Phys. Lett. 104 (2014) 182603.

June 2014 — Superconducting K122 film

Hiramatsu *et al.* ACS Appl. Mater. Interfaces 6 (2014) 14293.

Submission date

Fig 10(a-2)

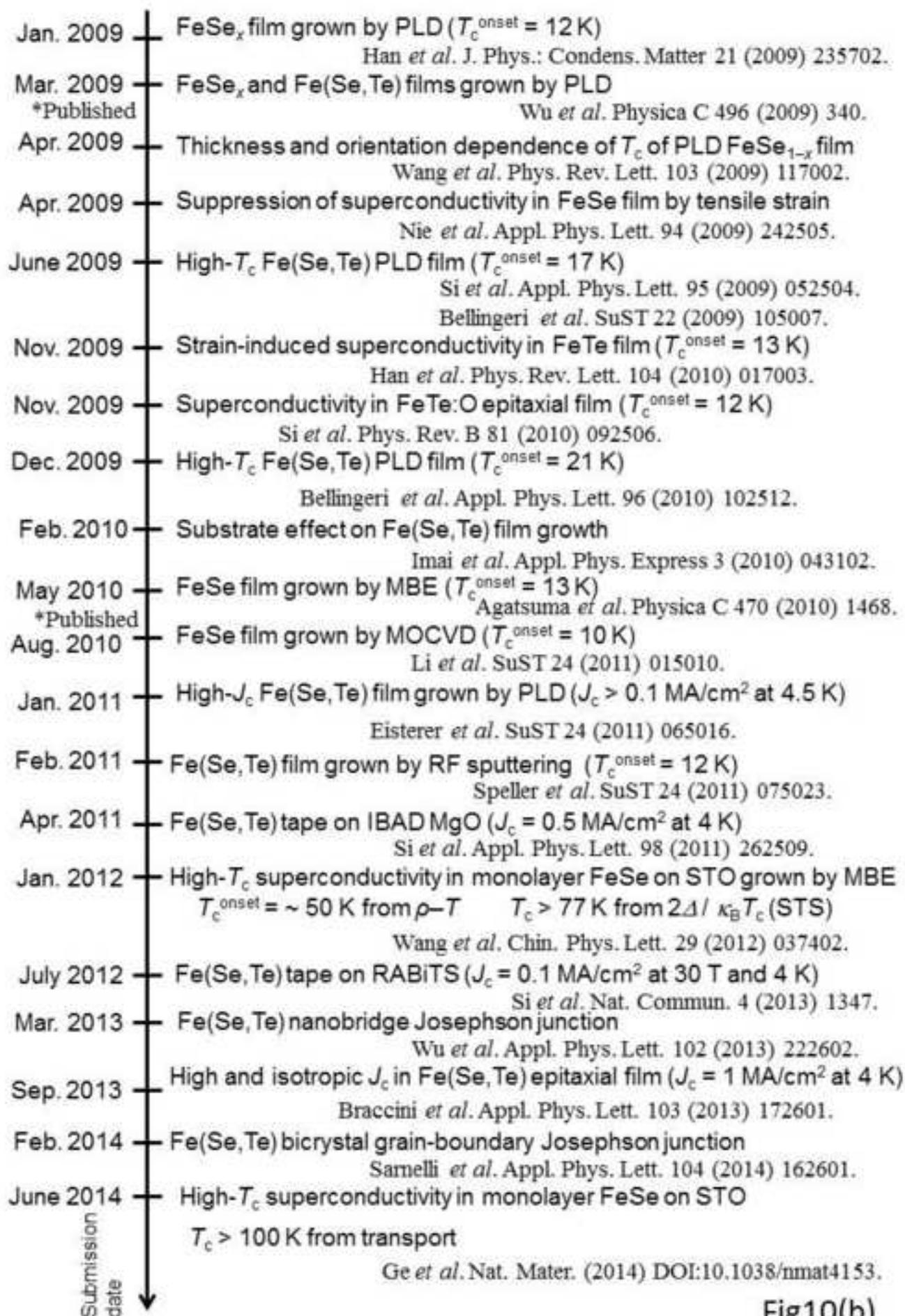

| Jan. 2009 | FeSe$_x$ film grown by PLD ($T_c^{onset}$ = 12 K) |
| | Han *et al.* J. Phys.: Condens. Matter 21 (2009) 235702. |
| Mar. 2009 | FeSe$_x$ and Fe(Se,Te) films grown by PLD |
| *Published | Wu *et al.* Physica C 496 (2009) 340. |
| Apr. 2009 | Thickness and orientation dependence of $T_c$ of PLD FeSe$_{1-x}$ film |
| | Wang *et al.* Phys. Rev. Lett. 103 (2009) 117002. |
| Apr. 2009 | Suppression of superconductivity in FeSe film by tensile strain |
| | Nie *et al.* Appl. Phys. Lett. 94 (2009) 242505. |
| June 2009 | High-$T_c$ Fe(Se,Te) PLD film ($T_c^{onset}$ = 17 K) |
| | Si *et al.* Appl. Phys. Lett. 95 (2009) 052504. |
| | Bellingeri *et al.* SuST 22 (2009) 105007. |
| Nov. 2009 | Strain-induced superconductivity in FeTe film ($T_c^{onset}$ = 13 K) |
| | Han *et al.* Phys. Rev. Lett. 104 (2010) 017003. |
| Nov. 2009 | Superconductivity in FeTe:O epitaxial film ($T_c^{onset}$ = 12 K) |
| | Si *et al.* Phys. Rev. B 81 (2010) 092506. |
| Dec. 2009 | High-$T_c$ Fe(Se,Te) PLD film ($T_c^{onset}$ = 21 K) |
| | Bellingeri *et al.* Appl. Phys. Lett. 96 (2010) 102512. |
| Feb. 2010 | Substrate effect on Fe(Se,Te) film growth |
| | Imai *et al.* Appl. Phys. Express 3 (2010) 043102. |
| May 2010 | FeSe film grown by MBE ($T_c^{onset}$ = 13 K) |
| *Published | Agatsuma *et al.* Physica C 470 (2010) 1468. |
| Aug. 2010 | FeSe film grown by MOCVD ($T_c^{onset}$ = 10 K) |
| | Li *et al.* SuST 24 (2011) 015010. |
| Jan. 2011 | High-$J_c$ Fe(Se,Te) film grown by PLD ($J_c$ > 0.1 MA/cm$^2$ at 4.5 K) |
| | Eisterer *et al.* SuST 24 (2011) 065016. |
| Feb. 2011 | Fe(Se,Te) film grown by RF sputtering ($T_c^{onset}$ = 12 K) |
| | Speller *et al.* SuST 24 (2011) 075023. |
| Apr. 2011 | Fe(Se,Te) tape on IBAD MgO ($J_c$ = 0.5 MA/cm$^2$ at 4 K) |
| | Si *et al.* Appl. Phys. Lett. 98 (2011) 262509. |
| Jan. 2012 | High-$T_c$ superconductivity in monolayer FeSe on STO grown by MBE |
| | $T_c^{onset}$ = ~ 50 K from $\rho-T$     $T_c$ > 77 K from $2\Delta / \kappa_B T_c$ (STS) |
| | Wang *et al.* Chin. Phys. Lett. 29 (2012) 037402. |
| July 2012 | Fe(Se,Te) tape on RABiTS ($J_c$ = 0.1 MA/cm$^2$ at 30 T and 4 K) |
| | Si *et al.* Nat. Commun. 4 (2013) 1347. |
| Mar. 2013 | Fe(Se,Te) nanobridge Josephson junction |
| | Wu *et al.* Appl. Phys. Lett. 102 (2013) 222602. |
| Sep. 2013 | High and isotropic $J_c$ in Fe(Se,Te) epitaxial film ($J_c$ = 1 MA/cm$^2$ at 4 K) |
| | Braccini *et al.* Appl. Phys. Lett. 103 (2013) 172601. |
| Feb. 2014 | Fe(Se,Te) bicrystal grain-boundary Josephson junction |
| | Sarnelli *et al.* Appl. Phys. Lett. 104 (2014) 162601. |
| June 2014 | High-$T_c$ superconductivity in monolayer FeSe on STO |
| | $T_c$ > 100 K from transport |
| | Ge *et al.* Nat. Mater. (2014) DOI:10.1038/nmat4153. |

Submission date

Fig10(b)

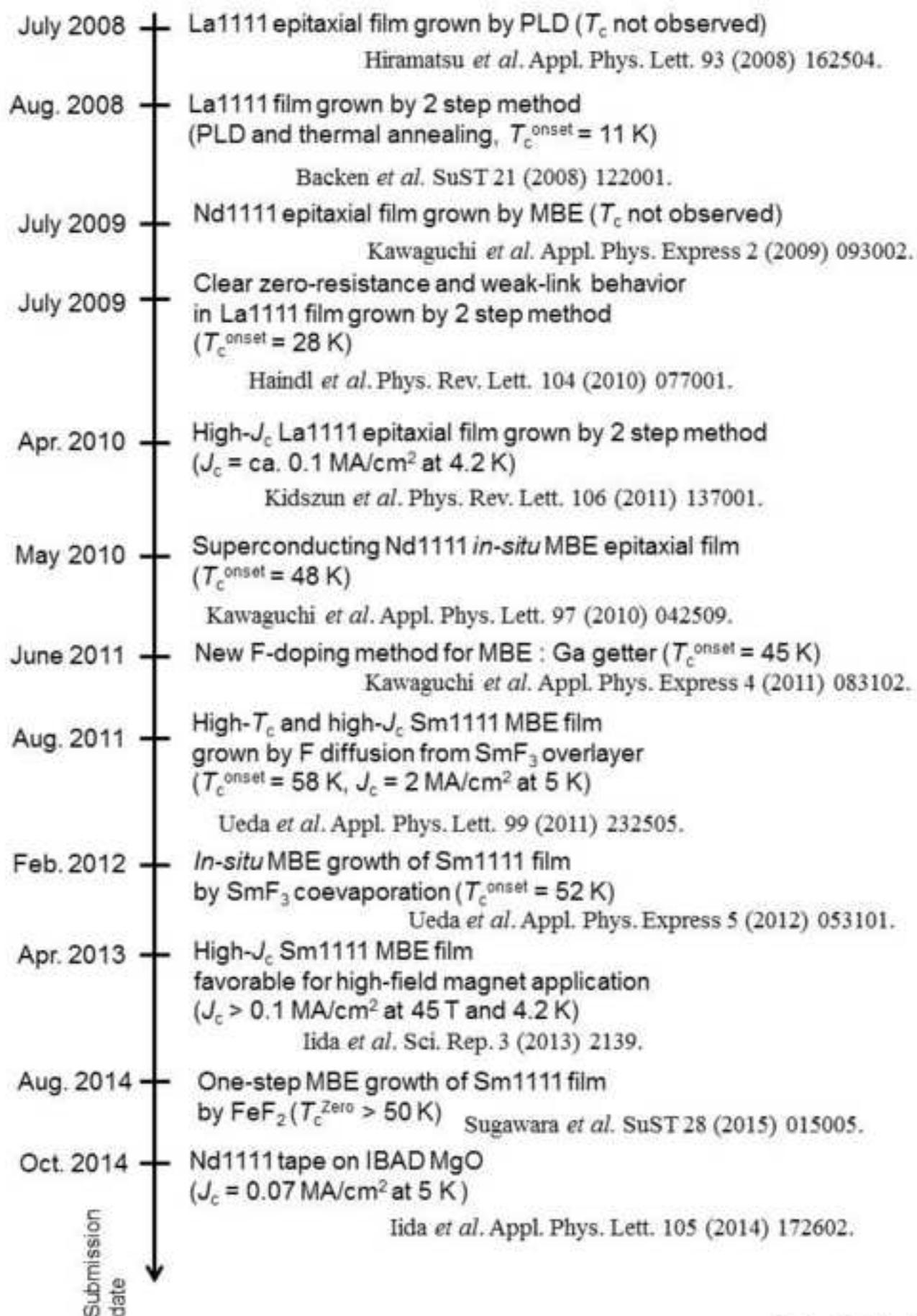

Fig.10(c)

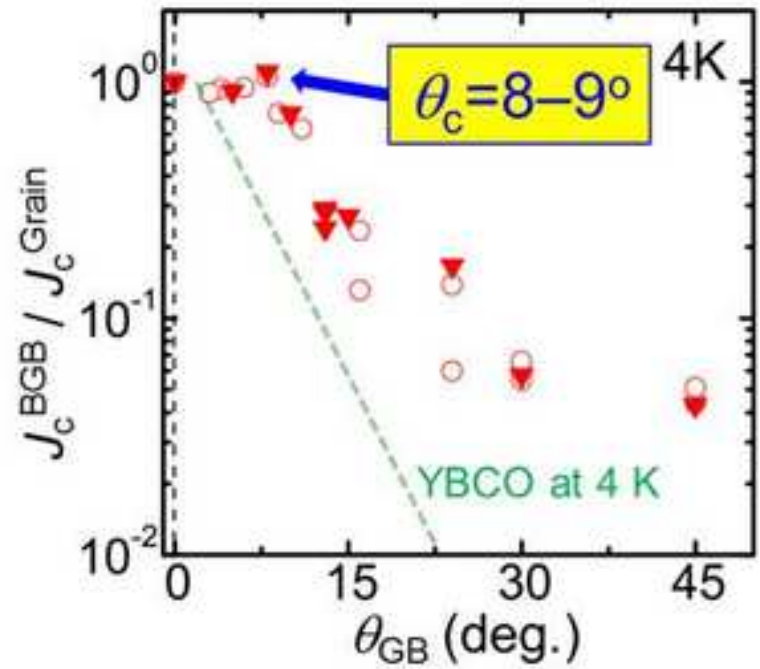

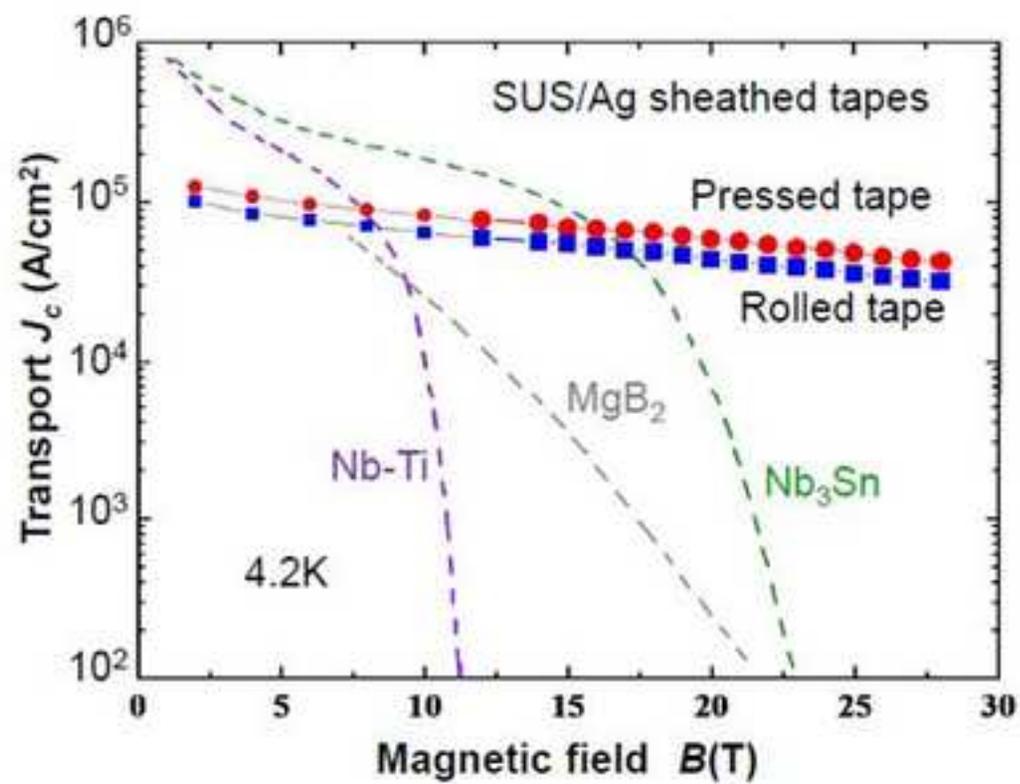

Fig.13

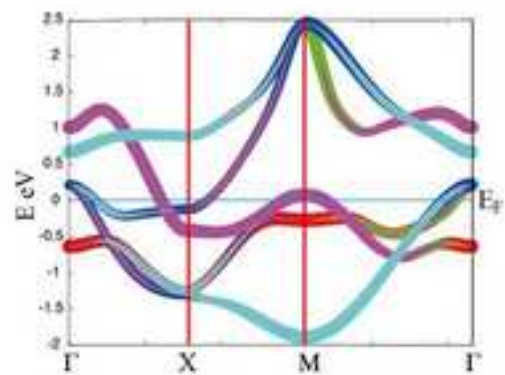

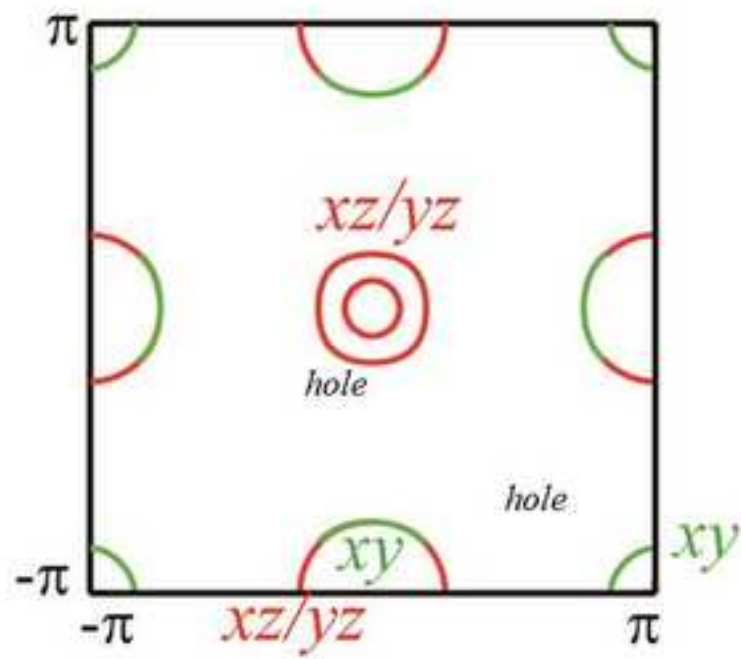

Fig.14

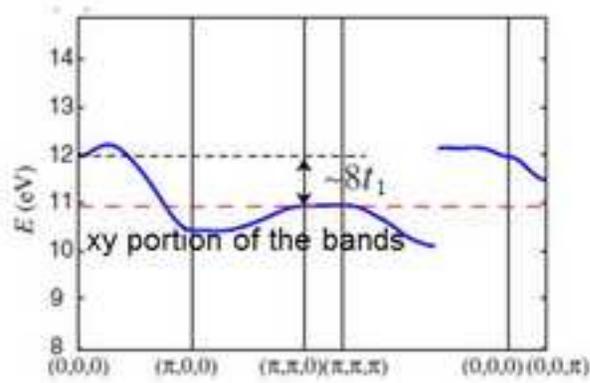

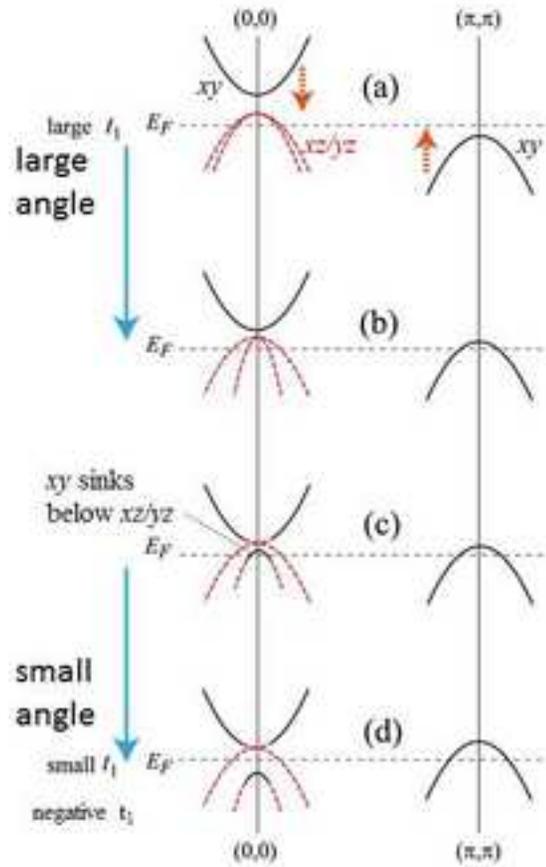

Fig.15

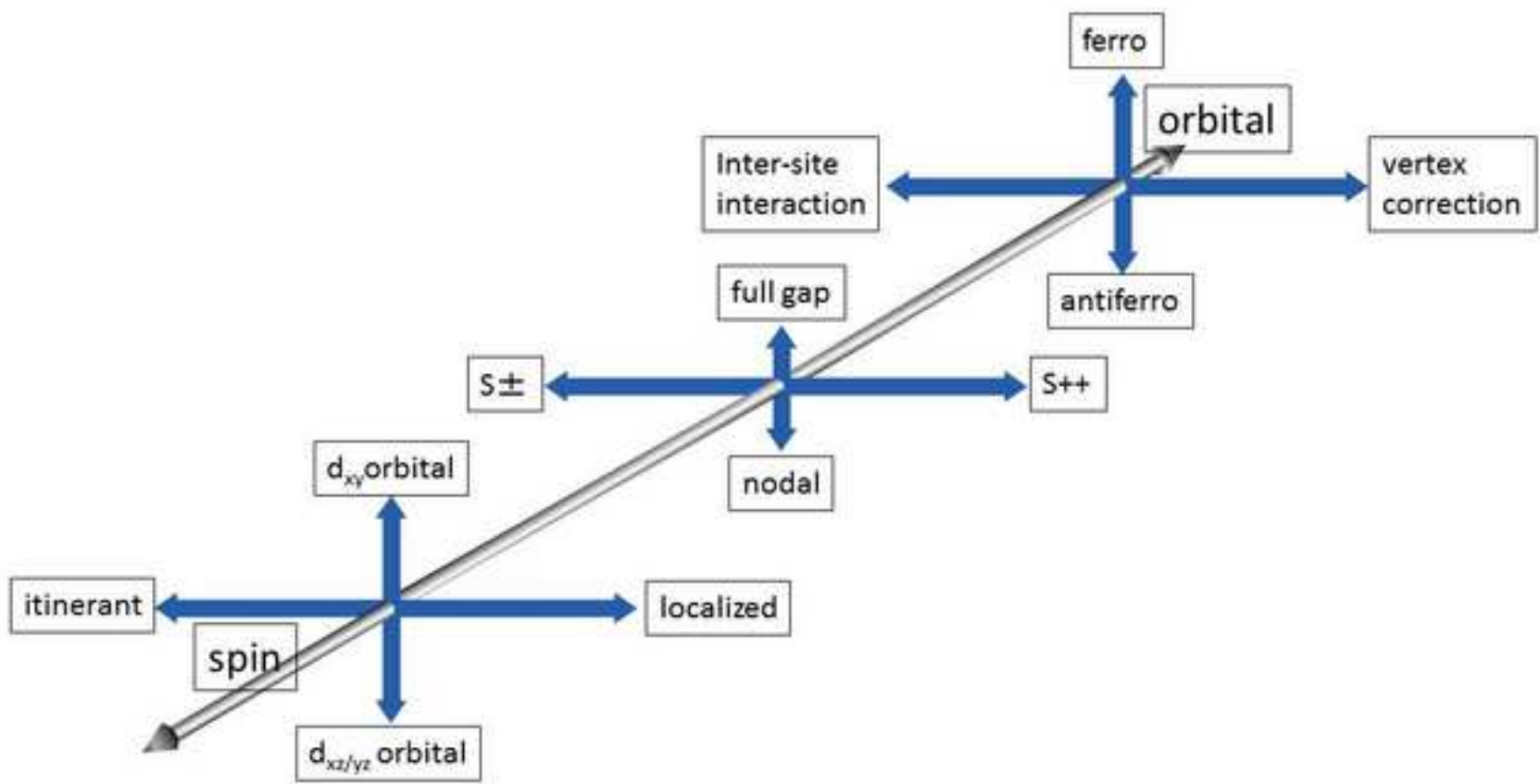

Fig.16

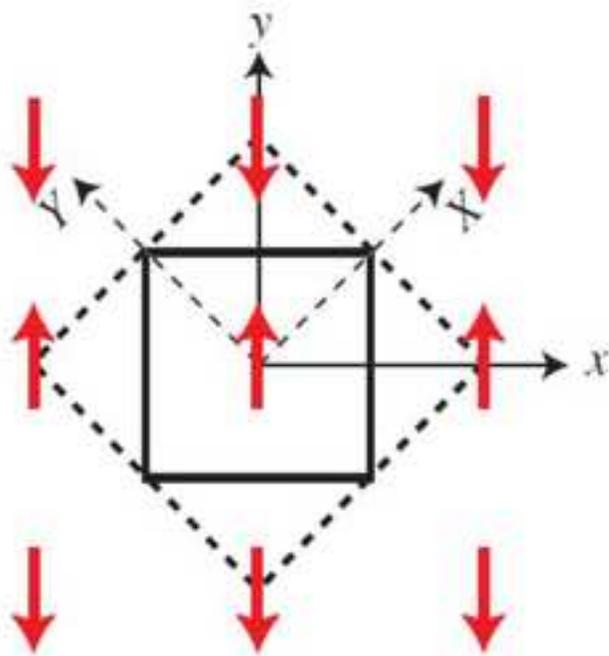

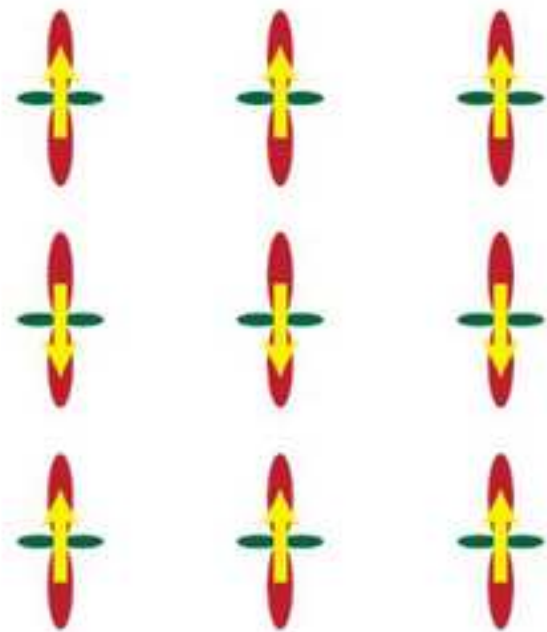

Fig.17

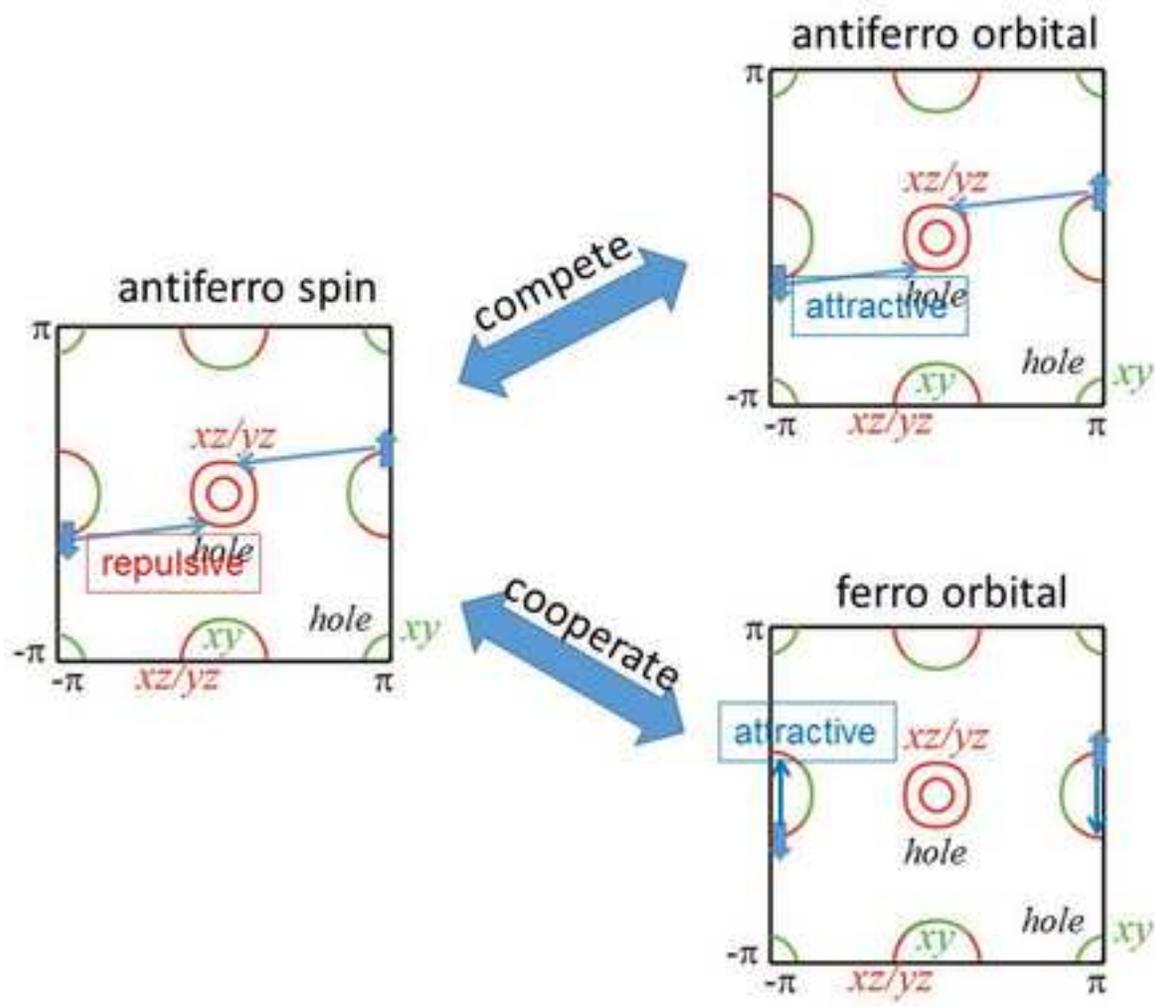

Fig.18

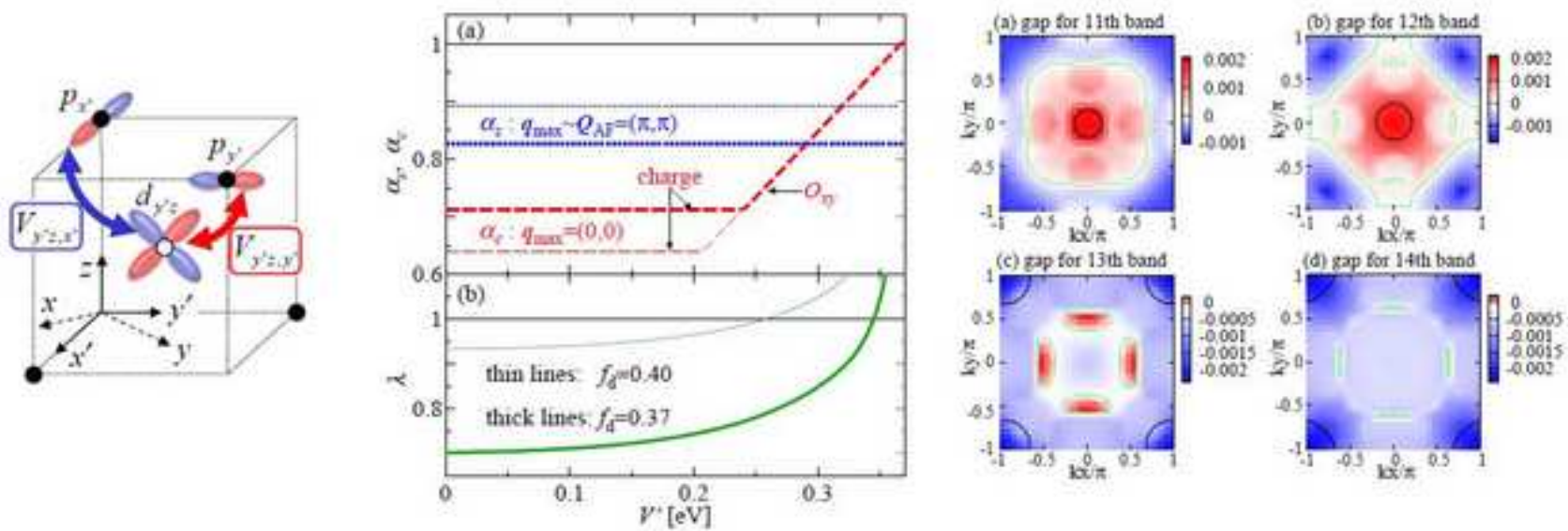

Fig.19

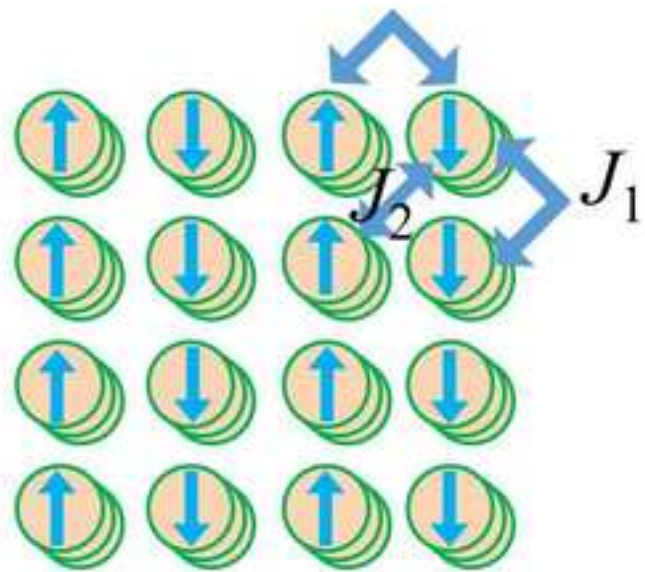

Fig.20

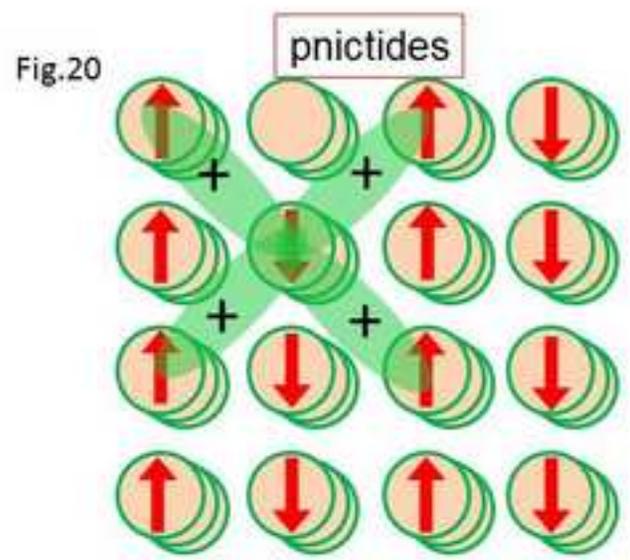

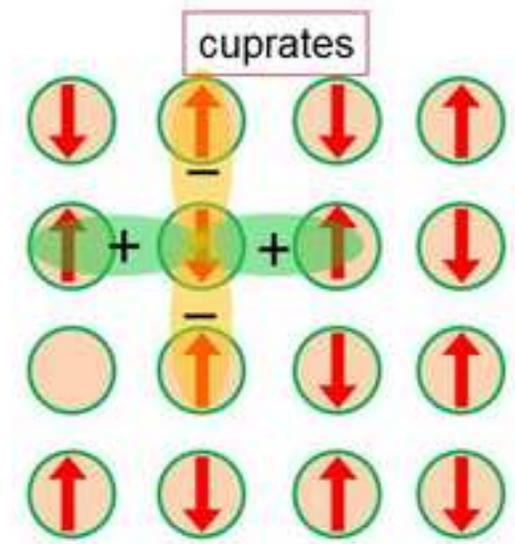

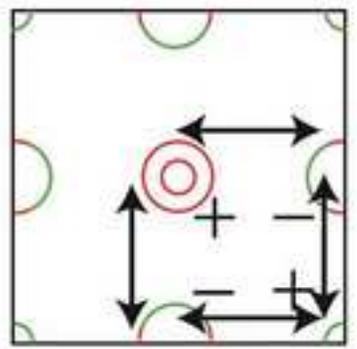

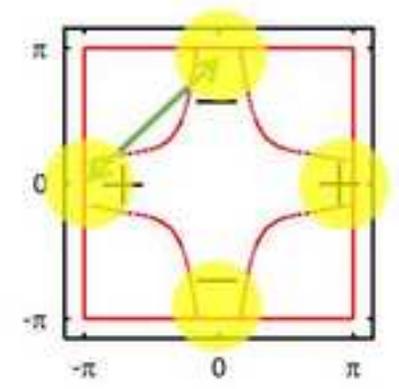

Fig.21

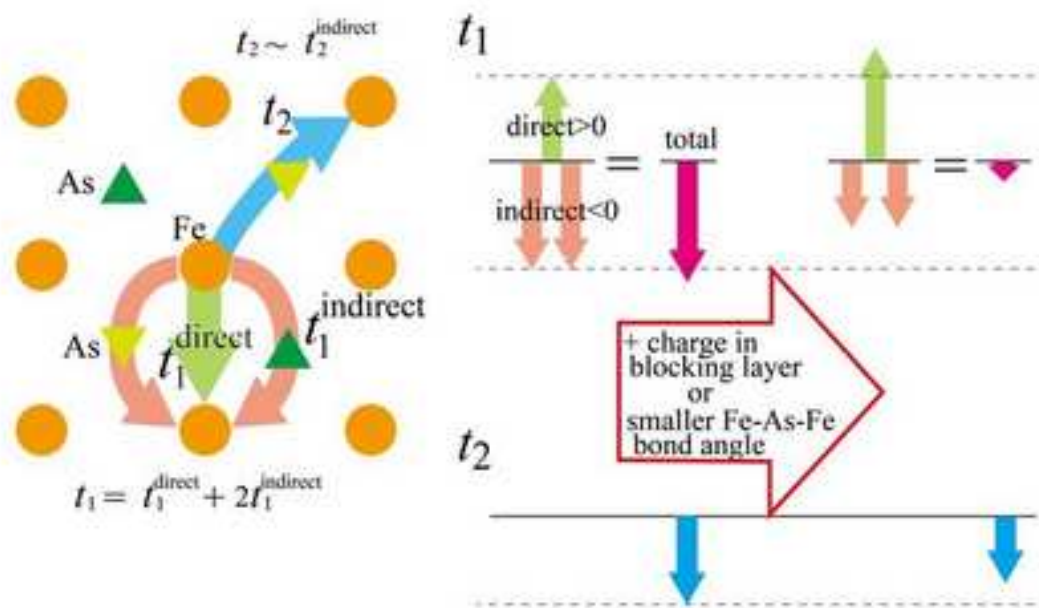

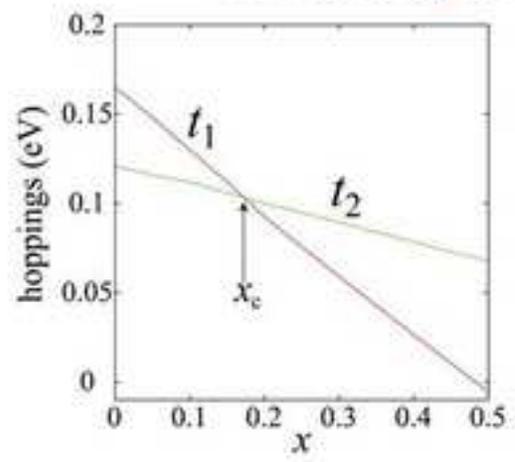

Fig.22

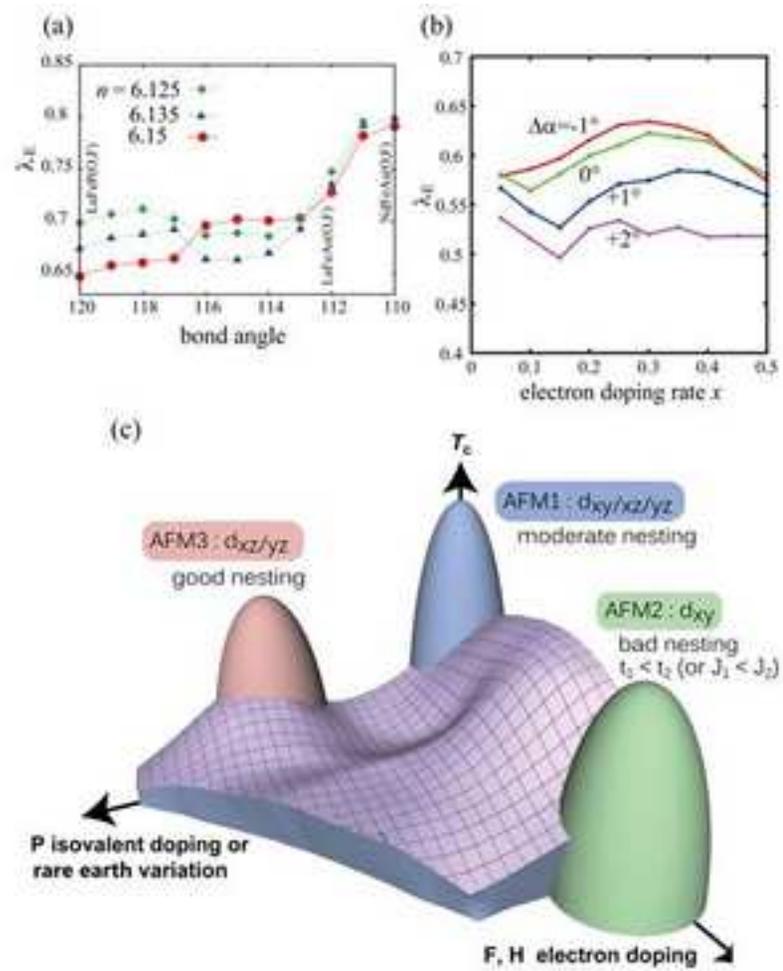

**Figure 23**
Click here to download high resolution image

Fig.23

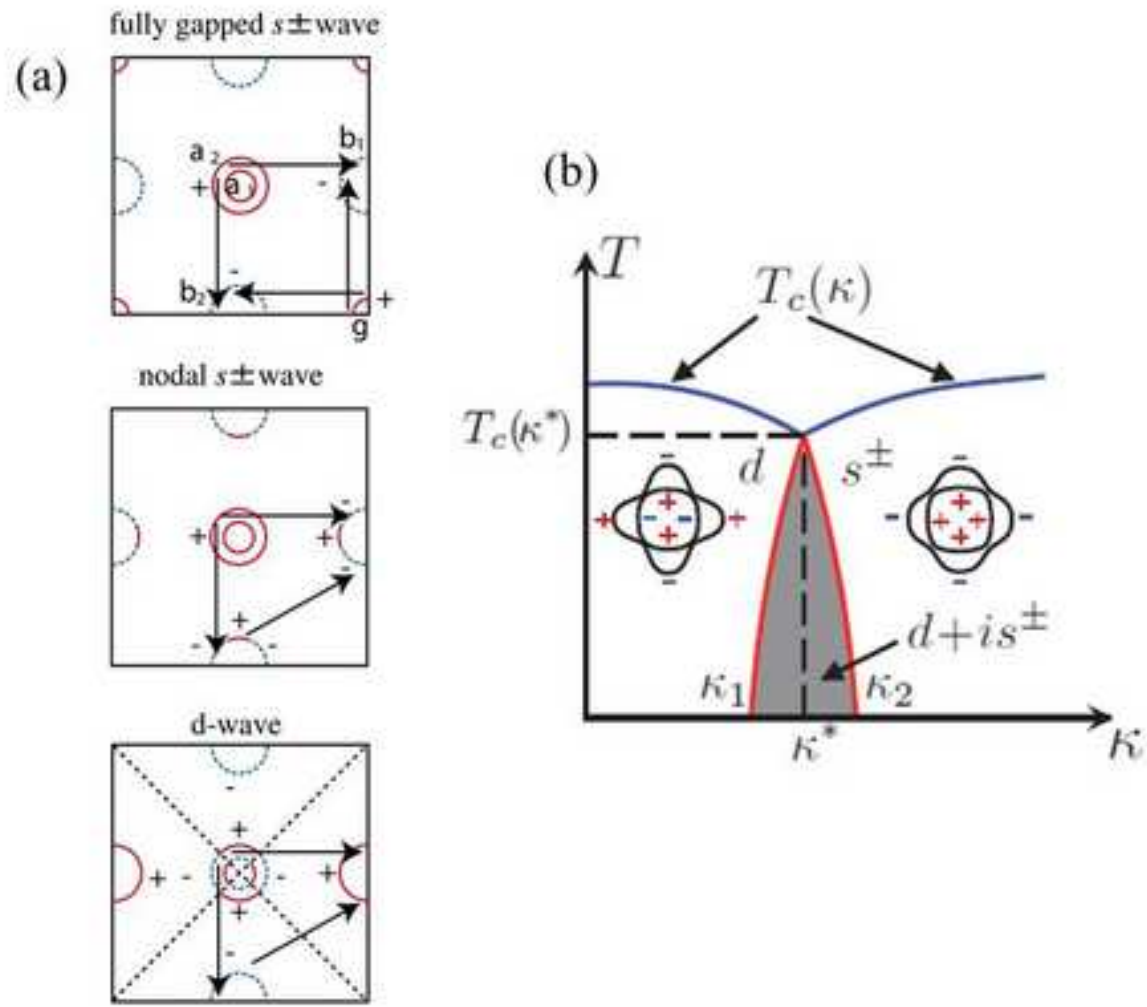

Table 1: Physical properties of parent phases of iron based superconductors

| System | Parent Compound | $d_{Fe}^{inter}$ (Å) | $d_{Fe}^{intra}$ (Å) | $An$-Fe-$An$ (deg.) | $T_s$ (K) | $T_N$ (K) | $(a_L-b_L)/(a_L+b_L)$ $\times 10^3$ | $p$ ($\mu_B$/Fe) | $\Theta$ (K) | Ref. |
|---|---|---|---|---|---|---|---|---|---|---|
| | | @ $T$ = 293-300 K, RT | | | | | | | | |
| 11 or 111 | $Fe^{II}_{0.11}Fe^{I}_1Te$ | 6.28 | 2.70 | 94.39 | 64 | 64 | 6.44 | 1.9 (2 K) | 189 | [T1_1] |
| | $Fe^{II}_{1.00}Fe^{I}_1As$ | 5.98 | 2.57 | 97.8 | - | 353 | - | 1.28$^I$ (RT), 2.05$^{II}$ (RT) | 296 | [T1_2-3] |
| | NaFeAs | 7.04 | 2.79 | 108.28 | 50 | 39 | 1.8 | 0.09 (5 K) | 254*1 | [T1_4-6] |
| 122 | $CaFe_2As_2$ | 5.80 | 2.76 | 110.5 | 173 | 173 | 5.1 | 0.80 (10 K) | 258 | [T1_7-9] |
| | $SrFe_2As_2$ | 6.18 | 2.77 | 110.54 | 205 | 205 | 5.5 | 1.01 (1.5 K) | 245 | [T1_10-13] |
| | $BaFe_2As_2$ | 6.51 | 2.80 | 111.06 | 142 | 142 | 4 | 0.87 (5 K) | 260 | [T1_14-16] |
| | $EuFe_2As_2$ | 6.06 | 2.76 | 110.14 | 190 | 190 | 2.9 | 0.98 (2.5 K) | 290*2 | [T1_17, 18] |
| 1111 | CaFeAsF | 8.58 | 2.74 | 108.11 | 134 | 114 | 3.4 | 0.49 (2 K) | ND | [T1_19, 20] |
| | SrFeAsF | 8.97 | 2.83 | 112.09 | 180 | 133 | 3.8 | 0.58 (2 K) | 339 | [T1_21, 22] |
| | CaFeAsH | 8.26 | 2.74 | 107.63 | ND | ND | 3.25 | ND | ND | [T1_23] |
| | LaFeAsO | 8.72 | 2.85 | 113.62 | 150 | 137 | 2.4 | 0.63 (2 K) | 282, 365 | [T1_24-27] |
| | $LaFeAsO_{0.5}H_{0.5}$ | 8.65 | 2.81 | 109.19 | 95 | 89 | 2.1 | 1.21 (10 K) | ND | [T1_28] |
| | CeFeAsO | 8.63 | 2.83 | 112.73 | 158, 149 | 138 | 2.6 | 0.94 (1.7 K) | 377 | [T1_26, 27, 29, 30] |
| | PrFeAsO | 8.60 | 2.82 | 112.14 | 153 | 127 | 2.8 | 0.48 (5 K) | 376, 355 | [T1_26, 27, 31, 32] |
| | NdFeAsO | 8.57 | 2.81 | 111.7 | 142 | 137 | 2.6 | 0.54 | 381 | [T1_26, 27, 33] |
| | SmFeAsO | 8.50 | 2.79 | 110.85 | 143.7, 144 | 135.3 | 2.7 | ND | 382 | [T1_26, 27, 29] |
| | GdFeAsO | 8.45 | 2.77 | 110.07 | 135 | 128 | 2.5 | ND | ND | [T1_26, 27, 34, 35] |
| | TbFeAsO | 8.41 | 2.76 | 109.56 | 126 | 122*3 | 2.4 | ND | ND | [T1_26, 34] |
| | NpFeAsO | 8.37 | 2.73 | 106.04 | - | - | - | - | 287 | [T1_36] |
| | PuFeAsO | 8.50 | 2.77 | 110.29 | - | - | - | - | 194 | [T1_37] |

*1: 1.9% of Co are substituted. *2: 2% of P are substituted. *3: From resistivity measurement

[T1_1] Eur. Phys. J. B 82 (2011) 113. [T1_2] J. Phys. Soc. Jpn. 21 (1966) 2238. [T1_3] Supercond. Sci. Technol. 25 (2012) 084018. [T1_4] Phys. Rev. B 80 (2009) 020504. [T1_5] Phys. Rev. B 85 (2012) 224521. [T1_6] Chem. Commun. (2009) 2189. [T1_7] Phys. Rev. B 78 (2008) 100506. [T1_8] Phys. Rev. B 78 (2008) 014523. [T1_9] Sci. Rep. 4, (2014) 4120. [T1_10] J. Phys.: Condens. Matter 20 (2008) 452201. [T1_11] Phys. Rev. B 78 (2008) 212502. [T1_12] Chem. Mater. 23 (2011) 1009. [T1_13] Phys. Rev. B 78 (2008) 224512. [T1_14] Phys. Rev. B 78 (2008) 020503. [T1_15] Phys. Rev. Lett. 101 (2008) 257003. [T1_16] Phys. Rev. B 79 (2009) 094508. [T1_17] Phys. Rev. B 80 (2009) 174424. [T1_18] arXiv:1406.7715 (2014). [T1_19] Phys. Rev. B 79 (2009) 060504. [T1_20] Supercond. Sci. Technol. 22 (2009) 055016. [T1_21] Phys. Rev. B 81 (2010) 094523. [T1_22] EPL 84 (2008) 67007. [T1_23] Phys. Rev. B 84 (2011) 024521. [T1_24] Phys. Rev. B 82 (2010) 184521. [T1_25] EPL 83 (2008) 27006. [T1_26] Phys. Rev. B 87 (2013) 064302. [T1_27] Phys. Rev. B 82 (2010) 134514. [T1_28] Nat. Phys. 10 (2014) 300. [T1_29] arXiv:1210.6959v1 (2012). [T1_30] Nat. Mater. 7 (2008) 953. [T1_31] Phys. Rev. B 78 (2008) 132504. [T1_32] Phys. Rev. B 78 (2008) 140503. [T1_33] Phys. Rev. B 82 (2010) 060514. [T1_34] Phys. Rev. B 80 (2009) 224511. [T1_35] Phys. Rev. B 83 (2011) 094526. [T1_36] Phys. Rev. B 85 (2012) 174506. [T1_37] Phys. Rev. B 86 (2012) 174510.

Table 2: Pressure induced superconductivity in parent phases

| System | Parent Compound | $T_c^{\text{opt.}}$ (K) | $p^{\text{opt.}}$ (GPa) | Ref. |
|---|---|---|---|---|
| 11 or 111 | $Fe^{II}_{0.03}Fe^{I}_{1}Te$ | Ferromagnetism in $p > 2$ GPa | | [T2_1] |
| | $Fe^{II}_{1.00}Fe^{I}_{1}As$ | Metallic in $p < 32.4$ GPa | | [T2_2] |
| | NaFeAs | 33 | 4 | [T2_3] |
| 122 | $CaFe_2As_2$ | 12 | 0.5 | [T2_4] |
| | $SrFe_2As_2$ | 40 | 2.5 | [T2_5] |
| | $BaFe_2As_2$ | 35 | 1.5 | [T2_6] |
| | $EuFe_2As_2$ | 41 | 10 | [T2_7] |
| 1111 | CaFeAsF | 29 | 5 | [T2_8] |
| | SrFeAsF | 25 | 16.5 | [T2_9] |
| | CaFeAsH | 28 | 3.3 | [T2_10] |
| | LaFeAsO | 21 | 12 | [T2_11] |
| | SmFeAsO | 11 | 9 | [T2_11] |

[T2_1] Phys. Rev. B 87 (2013) 060409. [T2_2] Supercond. Sci. Technol. 25 (2012) 084018. [T2_3] JPS Conf. Proc. 3 (2014) 015031. [T2_4] Phys. Rev. B 80 (2009) 024519. [T2_5] Phys. Rev. B 78 (2008) 184516. [T2_6] EPL 87 (2009) 17004. [T2_7] J. Phys.: Condens. Matter 22 (2010) 292202. [T2_8] Phys. Rev. B 81 (2010) 054507. [T2_9] J. Phys.: Condens. Matter 26 (2014) 155702. [T2_10] J Supercond. Nov. Magn. 25 (2012) 1293. [T2_11] J. Supercond. Nov. Magn. 22 (2009) 595.

Table 3: Pressure effects on $T_c$ of superconductors

| System | Superconductor | $T_c^0$ (K) | $T_c^{\mathrm{opt.}}$ (K) | $p^{\mathrm{opt.}}$ (GPa) | $dT_c/dp$ (K/GPa) | Ref. |
|---|---|---|---|---|---|---|
| 11 or 111 | $Fe_{1.01}Se$ | 8 | 36.7 | 8.9 | 3.2 | [T3_1] |
| | LiFeAs | 18 | 7 | 8 | -1.4 | [T3_2] |
| | $NaFe_{0.972}Co_{0.028}As$ | 20 | 31 | 2.28 | 4.8 | [T3_3] |
| 122 | $Ba_{0.55}K_{0.45}Fe_2As_2$ | 30 | 27 | 20 | -0.2 | [T3_4] |
| | $Ba(Fe_{0.926}Co_{0.074})_2As_2$ | 22 | 10 | 5.5 | -2.2 | [T3_5] |
| | $BaFe(As_{0.65}P_{0.35})_2$ | 31 | 19 | 38 | -0.3 | [T3_6] |
| | $Ba_{0.87}La_{0.13}Fe_2As_2$ | 22.5 | 30 | 2.8 | 2.7 | [T3_7] |
| 1111 | $Ca(Fe_{1-x}Co_x)AsF$ | 23.8 | 24.7 | 1 | 0.9 | [T3_8] |
| | $LaFeAsO_{0.89}F_{0.11}$ | 26 | 43 | 3 | 5.7 | [T3_9] |
| | $LaFeAsO_{0.65}H_{0.35}$ | 36 | 46 | 3 | 3.3 | [T3_10] |
| | $CeFeAsO_{0.88}F_{0.12}$ | 44 | 1.1 | 26.5 | -1.6 | [T3_11] |
| | $NdFeAsO_{0.85}$ | 53 | 35 | 7 | -2.6 | [T3_12] |
| | $SmFeAsO_{0.85}$ | 55 | 41 | 7 | -2.0 | [T3_12] |
| 42622 | $Sr_4V_2O_6Fe_2As_2$ | 36 | 46 | 4 | 2.5 | [T3_13] |
| | $Sr_4Sc_2O_6Fe_2P_2$ | 15.6 | 6 | 4 | -2.4 | [T3_13] |
| 245 | $K_{0.8}Fe_{1.78}Se_2$ | 32 | 48.7 | 12.5 | 1.3 | [T3_14] |
| | $Rb_{0.8}Fe_{1.6}Se_2$ | 32.4 | 0 | 6 | -5.4 | [T3_15] |
| PtAs-blocking layers | $Ca_{10}(Pt_4As_8)(Fe_{2-x}Pt_xAs_2)_5$ | 25 | 7.5 | 10 | -1.8 | [T3_16] |

[T3_1] Nat. Mater. 8 (2009) 630. [T3_2] Phys. Rev. B 80 (2009) 014506. [T3_3] New J. Phys. 14 (2012) 113043. [T3_4] Phys. Rev. B 78 (2008) 104527. [T3_5] Supercond. Sci. Technol. 23 (2010) 054003. [T3_6] J. Phys. Soc. Jpn. 79 (2010) 123706. [T3_7] Phys. Rev. B 88 (2013) 140503. [T3_8] Phys. Rev. B 81 (2010) 054507. [T3_9] Nature 453 (2008) 376. [T3_10] Nat. Commun. 3 (2012) 943. [T3_11] Physica C: Superconductivity 468 (2008) 2229. [T3_12] EPL 83 (2008) 57002. [T3_13] J. Phys. Soc. Jpn. 78 (2009) 123707. [T3_14] Nature 483 (2012) 67. [T3_15] Phys. Rev. B 85 (2012) 214519. [T3_16] Adv. Mater. 26 (2014) 2346.

Table 4: Doping variation in optimum $T_c$ of 11, 111, and 122-systems

| System | "Parent" Compound | $T_c^{opt.}$ (K) | | | | | | |
| | | *dopant/sites* | | | | | | |
| | | Li$^+$/Fe$^{II\,2+}$ | Co($d^7$)/Fe$^I$ ($d^6$) | $A^{1+}$/$Ae^{2+}$ | $Ln^{3+}$/$Ae^{2+}$ | P/As | La$^{3+}$/Ca$^{2+}$ & P/As | Se/Te |
|---|---|---|---|---|---|---|---|---|
| 11 or 111 | Fe$^{II}$Fe$^I$Te | | | | | | | 14 [T4_1] |
| | Fe$^{II}_{1.00}$Fe$^I_1$As | 18 [T4_2] | | | | | | |
| | NaFeAs | | 21 [T4_3] | | | | | |
| 122 | CaFe$_2$As$_2$ | | 20 [T4_4] | 34 [T4_5] | | 15 [T4_6] | 45 [T4_7] | |
| | SrFe$_2$As$_2$ | | 19.2 [T4_8] | 37 [T4_9] | 21 [T4_10] | 33 [T4_11] | | |
| | BaFe$_2$As$_2$ | | 22 [T4_12] | 37 [T4_13] | 22 [T4_14] | 30 [T4_15] | | |
| | EuFe$_2$As$_2$ | | 20.5 [T4_16] | 33 [T4_17] | | 29 [T4_18] | | |

[T4_1] Phys. Rev. B 78 (2008) 224503. [T4_2] Phys. Rev. B 78 (2008) 060505. [T4_3] Phys. Rev. B 89 (2014) 054502. [T4_4] Phys. Rev. B 83 (2011) 094523.   [T4_5] Phys. Rev. B 84 (2011) 184534. [T4_6] Phys. Rev. B 83 (2011) 060505. [T4_7] Sci. Rep. 3 (2013) 1478. [T4_8] Phys. Rev. Lett. 101 (2008) 207004. [T4_9] Phys. Rev. Lett. 101 (2008) 107007. [T4_10] IEEE Transactions on Applied Superconductivity 23 (2013) 7300405. [T4_11] J. Phys. Soc. Jpn. 83 (2014) 104702.
[T4_12] Phys. Rev. Lett. 101 (2008) 117004. [T4_13] Phys. Rev. Lett. 101 (2008) 107006. [T4_14] Phys. Rev. B 85 (2012) 140516. [T4_15] Phys. Rev. B 81 (2010) 184519. [T4_16] Phys. Rev. B 83 (2011) 224505. [T4_17] J. Phys.: Condens. Matter 23 (2011) 455702. [T4_18] J. Phys.: Condens. Matter 23 (2011) 464204.

Table 5: Doping variation in optimum $T_c$ of 1111-system

| System | | "Parent" Compound | $T_c^{opt.}$ (K) dopant/sites | | | | | | | | | |
|---|---|---|---|---|---|---|---|---|---|---|---|---|
| | | | Co($d^7$)/Fe($d^6$) | F$^-$/O$^-$ | Va$^{\pm0}$/O$^{2-}$ | Va$^{\pm0}$/A$^{1-}$ | H$^-$/O$^{2-}$ | La$^{3+}$/Ae$^{2+}$ | Th$^{4+}$/Ln$^{3+}$ | Sr$^{2+}$/Ln$^{3+}$ | O$^{2-}$/H$^-$ | P/As |
| 1111 | $Ae$FeAs(F, H) | CaFeAsF | 22 [T5_1] | | | 29 [T5_2] | | | | | | |
| | | SrFeAsF | 4 [T5_3] | | | | | | | | | |
| | | CaFeAsH | 22 [T5_4] | | | 29 [T5_2] | 47 [T5_5] | | | | | |
| | $Ln$FeAsO | LaFeAsO | 13 [T5_6] | 26 [T5_7] | 29 [T5_8] | | 36 [T5_9] | | 30.3 [T5_10] | 25 [T5_11] | | 11 [T5_12] |
| | | LaFeAsO$_{0.5}$H$_{0.5}$ | | | | | | | | | 36 [T5_9] | |
| | | CeFeAsO | 12.5 [T5_13] | 41 [T5_14] | 41.2 [T5_8] | | 47 [T5_15] | | | | | no SC [T5_16] |
| | | PrFeAsO | 16 [T5_17] | 47 [T5_18] | 49.2 [T5_8] | | | | | 16.3 [T5_19] | | |
| | | NdFeAsO | 16.5 [T5_20] | 45 [21] | 54.3 [T5_8] | | | | 38 [T5_22] | 13.5 [T5_23] | | |
| | | SmFeAsO | 17.2 [T5_6] | 58.1 [24] | 53.3 [T5_8] | | 55 [T5_25] | | 51.5 [T5_26] | | | no SC [T5_27] |
| | | GdFeAsO | 20 [T5_28] | 53 [29] | 54 [T5_8] | | 55 [T5_9] | | 56 [T5_30] | | | |
| | | TbFeAsO | | 45.5 [31] | 52.6 [T5_8] | | | | 52 [T5_32] | | | |
| | | DyFeAsO | | 45.3 [31] | 51.6 [T5_8] | | | | 49.3 [T5_33] | | | |
| | | HoFeAsO | | | 50.3 [T5_34] | | | | | | | |
| | | ErFeAsO | | | 44.5 [T5_35] | | | | | | | |
| | | YFeAsO | | 10.2 [36] | 46.5 [T5_34] | | | | | | | |

[T5_1] J. Am. Chem. Soc. 130 (2008) 14428. [T5_2] Appl. Phys. Lett. 103 (2013) 142601. [T5_3] J. Phys. Soc. Jpn. 77 (2008) 113709. [T5_4] Phys. Rev. B 89 (2014) 094501. [T5_5] J. Phys. Soc. Jpn. 83 (2014) 033705. [T5_6] Phys. Rev. B 79 (2009) 054521. [T5_7] J. Am. Chem. Soc. 130 (2008) 3296. [T5_8] J. Phys. Soc. Jpn. 78 (2009) 034712. [T5_9] Nat Commun 3 (2012) 943. [T5_10] J. Phys.: Condens. Matter 21 (2009) 175705. [T5_11] J. Phys. Soc. Jpn. 77 (2009) 15. [T5_12] EPL 86 (2009) 47002. [T5_13] J. Phys.: Condens. Matter 22 (2010) 115701. [T5_14] Phys. Rev. Lett. 100 (2008) 247002. [T5_15] Phys. Rev. B 85 (2012) 014514. [T5_16] Phys. Rev. B 81 (2010) 134422. [T5_17] Phys. Rev. B 83 (2011) 014503. [T5_18] Phys. Rev. B 80 (2009) 144517. [T5_19] Phys. Rev. B 79 (2009) 104501. [T5_20] Phys. Rev. B 81 (2010) 064511. [T5_21] J. Am. Chem. Soc. 132 (2010) 2417. [T5_22] Chem. Mater. 20 (2008) 7201. [T5_23] Chem. Commun. (2009) 707. [T5_24] Supercond. Sci. Technol. 26 (2013) 085023. [T5_25] Phys. Rev. B 84 (2011) 024521. [T5_26] Phys. Rev. B 82 (2010) 064517. [T5_27] Phys. Rev. B 84 (2011) 134526. [T5_28] Phys. Rev. B 87 (2013) 075148. [T5_29] Jetp Lett. 90 (2009) 387. [T5_30] EPL 83 (2008) 67006. [T5_31] Chem. Commun. (2008) 3634. [T5_32] Phys. Rev. B 78 (2008) 132506. [T5_33] Int. J. Mod. Phys. B 26 (2012) 1250207. [T5_34] New J. Phys. 11 (2009) 025005. [T5_35] EPL 92 (2010) 57011. [T5_36] J. Phys.: Conf. Ser. 150 (2009) 052036.

Table 6: Optimum $T_c$ of Fe$Pn$-superconductors not requiring doping

| System | | Compound | $T_c^{\,opt.}$(K) | Ref. |
|---|---|---|---|---|
| 112 | | $Ca_{1-x}La_xFe(As_{0.9}Sb_{0.1})_2$ | 47 | [T6_1] |
| PtAs-blocking layers; 10 : $l$ : 8 : 5(2 $-x$): 5$x$ : 10 | $l = 3$ | $Ca_{10}(Pt_3As_8)(Fe_{2-x}Pt_xAs_2)_5$ | 13 | [T6_2] |
| | 4 | $Ca_{10}(Pt_4As_8)(Fe_{2-x}Pt_xAs_2)_5$ | 38 | [T6_2] |
| 32522; $m+1$ : $m$ : 3$m-1$ : 2 : 2 | $m = 2$ | $Sr_3Sc_2O_5Fe_2As_2$ | No SC | [T6_3] |
| | 2 | $Ba_3Sc_2O_5Fe_2As_2$ | No SC | [T6_4] |
| | 3 | $Sr_4(Sc, Ti)_3O_8Fe_2As_2$ | 28 | [T6_4] |
| | 3 | $Ba_4Sc_3O_8Fe_2As_2$ | 13 | [T6_4] |
| | 3 | $Ca_4(Sc, Ti)_3O_8Fe_2As_2$ | 33 | [T6_5] |
| | 3 | $Sr_4(Sc, Ti)_3O_8Fe_2As_2$ | 28 | [T6_4] |
| | 3 | $Ca_4(Mg, Ti)_3O_8Fe_2As_2$ | 47 | [T6_6] |
| | 4 | $Ca_5(Mg, Ti)_4O_{11}Fe_2As_2$ | 35 | [T6_7] |
| | 4 | $Ca_5(Sc, Ti)_4O_{11}Fe_2As_2$ | 41 | [T6_5] |
| | 5 | $Ca_6(Sc, Ti)_5O_{14}Fe_2As_2$ | 42 | [T6_5] |
| 42622; 2 : 2 : $n+2$ : $n$ : 3$n$ : 2 : 2 | $n = 2$ | $Ca_4Al_2O_6Fe_2As_2$ | 28 | [T6_8] |
| | 2 | $Fe_2P_2Ca_4Al_2O_6$ | 17 | [T6_8] |
| | 2 | $Ca_4(Al, Ti)_2O_6Fe_2As_2$ | 12-13 | [T6_9] |
| | 2 | $Sr_4V_2O_6Fe_2As_2$ | 37.2 | [T6_10] |
| | 2 | $Sr_4Sc_2O_6Fe_2As_2$ | No SC | [T6_11] |
| | 2 | $Sr_4Sc_2O_6Fe_2P_2$ | 17 | [T6_12] |
| | 2 | $Sr_4(Mg, Ti)_2O_6Fe_2As_2$ | 26 | [T6_13] |
| | 2 | $Ba_4Sc_2O_6Fe_2As_2$ | No SC | [T6_14] |
| | 2 | $Sr_4Cr_2O_6Fe_2As_2$ | No SC | [T6_11, 14] |
| | 3 | $Ca_5(Al, Ti)_3O_9Fe_2As_2$ | 39 | [T6_9] |
| | 4 | $Ca_6(Al, Ti)_4O_{12}Fe_2As_2$ | 36 | [T6_9] |
| | 6 | $Ca_8(Mg, Ti)_6O_{18}Fe_2As_2$ | 40 | [T6_7] |

[T6_1] J. Phys. Soc. Jpn. 82 (2013) 123702. [T6_2] J. Phys. Soc. Jpn. 80 (2011) 093704. [T6_3] Phys. Rev. B 79 (2009) 024516. [T6_4] Appl. Phys. Express 3 (2010) 063102. [T6_5] Appl. Phys. Lett. 97 (2010) 072506. [T6_6] Appl. Phys. Express 3 (2010) 063103. [T6_7] Supercond. Sci. Technol. 24 (2011) 085020. [T6_8] Appl. Phys. Lett. 97 (2010) 172506. [T6_9] Supercond. Sci. Technol. 23 (2010) 115005. [T6_10] Phys. Rev. B 79 (2009) 220512. [T6_11] Supercond. Sci. Technol. 22 (2009) 085001. [T6_12] Supercond. Sci. Technol. 22 (2009) 075008. [T6_13] Supercond. Sci. Technol. 23 (2010) 045001. [T6_14] anorg. allg. Chem. 635 (2009) 2242.

Table 7: Optimum $T_c$ of Fe$Ch$-superconductors not requiring doping

| System | Compound | $T_c^{\text{opt.}}$(K) | Ref. |
|---|---|---|---|
| 245 | $K_2Fe_4Se_5$ | 32 | [T7_1] |
| | $Rb_2Fe_4Se_5$ | 32 | [T7_1] |
| | $Cs_2Fe_4Se_5$ | 29 | [T7_1] |
| | $(Tl, K)_2Fe_4Se_5$ | 28 | [T7_1] |
| | $(Tl, Rb)_2Fe_4Se_5$ | 32 | [T7_1] |
| 1111 | $(Li, Fe)OHFeSe$ | 43 | [T7_1] |
| Intercalated-blocking layers | $LiFe_2Se_2$ | 44 | [T7_2] |
| | $NaFe_2Se_2$ | 45-46 | [T7_3] |
| | $CaFe_2Se_2$ | 40 | [T7_3] |
| | $SrFe_2Se_2$ | 35-38 | [T7_3] |
| | $BaFe_2Se_2$ | 39-40 | [T7_3] |
| | $EuFe_2Se_2$ | 40 | [T7_3] |
| | $YbFe_2Se_2$ | 42 | [T7_3] |
| | $Li_{0.6}(NH_2)_{0.2}(NH_3)_{0.8}Fe_2Se_2$ | 43 | [T7_4] |
| | $(NH_3)_yLi_xFeSe_{0.5}Te_{0.5}$ | 26 | [T7_5] |
| | $(NH_3)_yNa_xFeSe_{0.5}Te_{0.5}$ | 22 | [T7_5] |
| | $(NH_3)_yCa_xFeSe_{0.5}Te_{0.5}$ | 17 | [T7_5] |
| | $(NH_3)_yNa_xFeSe$ | 31.5 | [T7_5] |
| | $Li_x(C_2H_8N_2)_yFe_{2-z}Se_2$ | 45 | [T7_6] |
| | $(NH_3)_yCs_{0.4}FeSe$ | 31 | [T7_7] |
| | $Li_x(C_5H_5N)_yFe_{2-z}Se_2$ | 45 | [T7_8] |
| | $Na_{0.65}Fe_{1.93}Se_2$ | 37 | [T7_9] |

[T7_1] Phys. Rev. Lett. 107 (2011) 137003. [T7_2] Phys. Rev. B 89 (2014) 020507. [T7_3] Sci. Rep. 2 (2012) 426. [T7_4] Nat. Mater. 12 (2013) 15. [T7_5] Phys. Rev. B 89 (2014) 144509. [T7_6] J. Phys. Soc. Jpn. 82 (2013) 123705. [T7_7] Phys. Rev. B 88 (2013) 094521. [T7_8] J. Phys.: Condens. Matter 24 (2012) 382202. [T7_9] Nat. Commun. 5 (2014) 4756.

## Table 8. Comparison among 3 representative superconductors

| | Fe-pnictides | MgB$_2$ | Cuprates |
|---|---|---|---|
| **Parent Material** | **(bad) metal** **($T_N$~150K)** | **metal** | **Mott Insulator** **($T_N$~400K)** |
| **Fermi Level** | **3d 5-bands** | **2-bands** | **3d single band** |
| **Max Tc** | **56K** | **40K** | **~140K** |
| **Impurity** | **robust** | **sensitive** | **sensitive** |
| **SC gap symmetry** | **extended s-wave (+- or ++)** | **s-wave** | **d-wave** |
| **Hc$^2$(0)** | **100-200T >** | **~40T** | **~100T** |
| **Anisotropy** | **2-4 (122)** | **~3.5** | **5-7 (YBCO)** **50-90(Bi -system)** |
| **Critical GB angle** | **8-9(Ba 122)** | | **~5° （YBCO)** |

The primary points

1) A review to know the current status of materials ( bulk and thin films) and pairing mechanism covering a brief background, progress to the most up-to-date information, and future's prospect.

2) Data on materials covers a wide range including pressure effects.

3)  Pairing mechanism deals spin and orbitals.